\documentclass[twocolumn,showpacs,preprintnumbers,amsmath,amssymb]{revtex4}
\usepackage{epsfig}

\usepackage[table]{xcolor}
\usepackage{graphicx}
\usepackage{dcolumn}
\usepackage{bm}
\usepackage{hyperref}
\hypersetup{colorlinks=true,linkcolor=magenta,anchorcolor=green,citecolor=cyan,filecolor=black,menucolor=black,urlcolor=brown}
\usepackage{slashed}

\newcommand{\be}{\begin{equation}}
\newcommand{\ee}{\end{equation}}
\newcommand{\ba}{\begin{eqnarray}}
\newcommand{\ea}{\end{eqnarray}}

\begin{document}

\title{Supersonic  friction of a black hole traversing a self-interacting scalar dark matter cloud}

\author{Alexis Boudon}
\affiliation{Universit\'{e} Paris-Saclay, CNRS, CEA, Institut de physique th\'{e}orique, 91191, Gif-sur-Yvette, France}
\author{Philippe Brax}
\affiliation{Universit\'{e} Paris-Saclay, CNRS, CEA, Institut de physique th\'{e}orique, 91191, Gif-sur-Yvette, France}
\author{Patrick Valageas}
\affiliation{Universit\'{e} Paris-Saclay, CNRS, CEA, Institut de physique th\'{e}orique, 91191, Gif-sur-Yvette, France}

\begin{abstract}

Black Holes (BH) traversing a dark matter cloud made out of a self-interacting scalar soliton are slowed down by two complementary effects. At low subsonic speeds, the BH accretes dark matter and this is the only source of dragging along its motion, if we neglect the backreaction of the cloud self-gravity. The situation changes at larger supersonic speeds where a shock appears. This leads to the emergence of an additional friction term, associated with the gravitational and scalar pressure interactions and with the wake behind the moving BH. This is a long distance effect that can be captured by the hydrodynamical regime of the scalar flow far away from the BH. This dynamical friction term has the same form as the celebrated Chandrasekhar collisionless result, albeit with a well-defined Coulomb logarithm and a prefactor that is smaller by a factor $2/3$. 
The infra-red cut-off is naturally provided by the size of the scalar cloud, which is set by the scalar mass and coupling, whilst the ultra-violet behaviour corresponds to the distance from the BH where the velocity field is significantly perturbed by the BH,
which is determined by pressure effects. As a result, supersonic BH are slowed down by both the accretion drag and the dynamical friction. This effect will be  potentially detectable by future gravitational wave experiments as it influences the phase of the gravitational wave signal from inspiralling binaries. 

\end{abstract}

\date{\today} 
 
\maketitle

\section{Introduction}
\label{sec:introduction}

The $\Lambda$CDM model is the present established cosmological framework providing a comprehensive description of the Universe's large-scale structure and evolution. Central to this model is Cold Dark Matter (CDM), a non-relativistic component that exerts a significant gravitational influence, shaping the formation of galaxies and galaxy clusters. CDM constitutes approximately 27\% of the total energy content of the universe \cite{Planck:2018vyg, DES:2022qpf} and plays a crucial role in the clumping and development of cosmic structures under gravitational effects.

Despite its success in explaining various cosmological observations, the standard CDM paradigm encounters challenges at smaller scales. Direct detection experiments focusing on Weakly Interacting Massive Particles (WIMPs) \cite{Feng:2008mu, Hooper:2009zm, Schwarz:2003du, Chang:2013oia, Garcia:2021iag, Feng:2022rxt}, a prominent CDM candidate, are reaching their limitations \cite{Liu:2017drf, Billard:2021uyg, Roszkowski:2017nbc, Arcadi:2017kky}. Moreover, at galactic scales, tensions arise as CDM predictions deviate from observations, manifesting themselves in the core-cusp problem \cite{Hui:2001wy, deBlok:2009sp, Oh:2010mc, Teyssier:2012ie, Read:2015sta}, the missing satellite problem \cite{Klypin:1999uc, Strigari:2007ma, Guo:2010ap, Penarrubia:2012bb}, and the too-big-to-fail problem \cite{Boylan-Kolchin:2011qkt, Garrison-Kimmel:2014vqa, Kaplinghat:2019svz, Pawlowski:2015qta}. While introducing realistic baryons and baryonic feedback in CDM simulations offers potential solutions, these effects may not be sufficient to explain fully  the observed discrepancies. As a result, researchers have explored alternative approaches, including investigating alternative dark matter scenarios such as Primordial Black Holes (PBHs) \cite{Ivanov:1994pa, Clesse:2015wea, Carr:2016drx, Carr:2020xqk, Green:2020jor}, axions \cite{Duffy:2009ig, Marsh:2015xka, Kim:2008hd, Graham:2015ouw, Irastorza:2018dyq, DiLuzio:2020wdo}, and sterile neutrinos \cite{Dodelson:1993je, Shi:1998km, Boyarsky:2009ix, Kusenko:2009up, Feng:2010gw, Abazajian:2012ys}, or considering modifications to gravity theories.

The Ultralight Dark Matter paradigm \cite{Hu:2000ke, Hui:2016ltb, Knapen:2017xzo, Ferreira:2020fam, Hui:2021tkt}, particularly Fuzzy Dark Matter (FDM), has emerged as a promising alternative. FDM introduces ultralight scalar particles with masses around $10^{-22}$ eV, characterized by large de Broglie wavelengths. These particles form solitonic solutions at the centers of dark matter halos \cite{Goodman:2000tg, Schive:2014dra, Schive:2014hza, Arbey:2001qi, Chavanis:2011zi, Chavanis:2011zm, Marsh:2015wka, Calabrese:2016hmp, Chen:2016unw, Schwabe:2016rze, Veltmaat:2016rxo, Gonzalez-Morales:2016yaf, Robles:2012uy, Bernal:2017oih, Mocz:2017wlg, Mukaida:2016hwd, Vicens:2018kdk, Bar:2018acw, Eby:2018ufi, Bar-Or:2018pxz, Marsh:2018zyw, Chavanis:2018pkx, Emami:2018rxq, Levkov:2018kau, Broadhurst:2019fsl,Hayashi:2019ynr, Bar:2019bqz, Garcia:2023abs}, providing a potential resolution to many of the tensions observed at galactic scales \cite{Hu:2000ke, Hui:2016ltb, Knapen:2017xzo, Ferreira:2020fam, Hui:2021tkt} while recovering the success of the CDM at cosmological scales \cite{Hu:2000ke, Johnson:2008se, Hwang:2009js, Park:2012ru, Hlozek:2014lca, Cembranos:2015oya, Urena-Lopez:2015gur, Urena-Lopez:2019kud}. However, recent observations, including the Lyman-$\alpha$ forest and galaxy rotational curves \cite{Irsic:2017yje, Armengaud:2017nkf, Zhang:2017chj, Bar:2018acw, Bar:2021kti}, constrain the FDM particle mass to be above $10^{-21}$ eV, suggesting that FDM alone may no longer suffice to resolve fully the smaller-scale cosmological discrepancies.

A closely related alternative is self-interacting scalar field dark matter, where scalar dark matter possesses repulsive self-interactions \cite{Goodman:2000tg, Bento:2000ah, Riotto:2000kh, Fregolente:2002nx, Li:2013nal, Suarez:2015fga, Suarez:2016eez, Suarez:2017mav,Brax:2019fzb,Brax:2019npi,Dave:2023wjq}. This model offers  means to reconcile the success of FDM while introducing distinct behaviors at galactic scales. Within this scenario, solitonic solutions no longer remain in equilibrium solely due to the balance between self-gravity and quantum pressure. Instead, repulsive self-interactions introduce an effective pressure, influencing the equilibrium state. In the Thomas-Fermi regime \cite{thomas_1927, Fermi_1927}, the repulsive self-interactions dominate over quantum pressure, resulting in equilibrium being determined solely by self-gravity and the effective pressure \cite{Chavanis:2011zi}. A promising method to detect possible signs of such dark matter involves studying its gravitational effects on black hole observations, for example by using gravitational waves \cite{Eda:2013gg, Kavanagh:2020cfn, Boudon:2023aa} or observational estimates around BHs \cite{Bar:2019pnz, Chakrabarti:2022owq, Ravanal:2023ytp}.

Dynamical friction and mass accretion play significant roles in astrophysics, impacting various astrophysical phenomena. Extensive research on dynamical friction and mass accretion has been conducted for both FDM \cite{Hui:2016ltb, Berezhiani:2019pzd, Lancaster:2019mde, Hartman:2020fbg, Annulli:2020lyc, Wang:2021udl,  Traykova:2021dua, Chowdhury:2021zik, Vicente:2022ivh, Traykova:2023qyv} and CDM, treated as collisionless particles \cite{Chandrasekhar:1943ys}. In this context, these investigations have provided valuable insights and constraints on dark matter models, including applications to the Fornax globular cluster timing problem \cite{Bar:2021jff}, a discrepancy observed in the Fornax galaxy where the expected strong dynamical friction, predicted by the standard CDM model, fails to reproduce the observations of slowly migrating globular clusters towards the galaxy center, and their relevance to gravitational waves where dynamical friction can slow down binary systems and induce phase shifts in gravitational wave emission.

In this paper, we explore the effects of dynamical friction and mass accretion experienced by a Schwarzschild black hole moving within a self-interacting scalar dark matter cloud at supersonic velocities. Our primary focus is on the Thomas-Fermi regime, where self-interactions are significant and the wavelike effects of the scalar field are negligible. This regime results in dark matter dynamics within the solitonic solution behaving more like a gas than FDM, although it retains distinctive characteristics. This study of the supersonic regime complements our previous investigation in the subsonic case \cite{Boudon:2022dxi}, offering relevance to ongoing research on gravitational waves. The implications of mass accretion and dynamical friction on binary systems can be critical, potentially detectable by upcoming gravitational wave detectors such as DECIGO or LISA \cite{Macedo:2013qea, Barausse:2014tra, Cardoso:2019rou, Li:2021pxf,Boudon:2023aa}. Additionally, the application
of such results to the Fornax globular cluster timing problem, where the CDM dynamical friction appears too strong, is of particular interest.

The outline of the paper is as follows. Section~\ref{sec:dark-matter} introduces scalar field 
dark matter with quartic self-interactions, discussing its equations of motion and 
equilibrium solitonic solutions. 
Section~\ref{sec:large-radius-expansions} compares the subsonic and supersonic regimes and 
calculates the large-distance expansions of the dark matter flow for both the upstream and 
downstream regions, including the appearance and location of shock fronts and boundary layers.
Section~\ref{sec:drag-force} describes the relation between these asymptotic expansions and 
the BH accretion rate and derives the drag force exerted on the BH.
Section~\ref{sec:accretion} discusses the accretion rate in comparison with the radial
case and with the classical Hoyle-Lyttleton prediction, and highlights the two regimes obtained at moderate
and high Mach numbers.
Section~\ref{sec:dynamical-friction} compares the magnitudes of the accretion drag and dynamical friction,
while Section~\ref{sec:wake} provides an independent computation of the dynamical friction from
the gravitational force exerted by the BH wake.
Section~\ref{sec:supersonic-flow} presents a numerical computation of the density and velocity fields for a moderate Mach number, to illustrate the behaviour of the system with a bow shock upstream of the BH.
Section~\ref{sec:comparison} compares our results with the behaviours of other systems (collisionless, 
perfect fluid and FDM cases).
Finally, we conclude our study in Section~\ref{sec:conclusion}.

\section{Dark matter scalar field}
\label{sec:dark-matter}

\subsection{Scalar-field action}
\label{sec:scalar-action}

As in our previous work \citep{Boudon:2022dxi}, we consider a scalar-field dark matter 
scenario described by the action
\be
S_\phi = \int d^4x \sqrt{-g} \left[ - \frac{1}{2} g^{\mu\nu} \partial_\mu\phi 
\partial_\nu\phi - V(\phi) \right] ,
\label{eq:Action}
\ee
with a quartic self-interaction,
\be
V(\phi) = \frac{m^2}{2} \phi^2 +  V_{\rm I}(\phi) \;\;\; \mbox{with} \;\;\; 
V_{\rm I}(\phi) = \frac{\lambda_4}{4} \phi^4 .
\label{eq:V_I-def}
\ee
Here $m$ is the mass of the scalar field and $\lambda_4$ its coupling constant, 
which is taken positive. This corresponds to a repulsive self-interaction, which
gives rise to an effective pressure that can balance gravity. This allows the formation
of stable static equilibria, also called boson stars or solitons.
Thus, in this paper we consider the supersonic motion of a BH inside such an extended
soliton, or quasi-static dark matter halo.

The parameters $m$ and $\lambda_4$ determine the characteristic density and radius
\be
\rho_a = \frac{4 m^4}{3\lambda_4} , \;\;\; r_a = \frac{1}{\sqrt{4\pi {\cal G} \rho_a}} .
\label{eq:rhoa-ra-def}
\ee
The dynamics that we study in this paper will only depend on this combination
$\rho_a$ and on the mass and velocity of the BH. Thus, different dark matter models
with the same $\rho_a$ show the same large-scale dynamics.
We refer to \cite{Boudon:2022dxi} for a presentation of the regions in the parameter space
$(m,\lambda_4)$ where our computations apply, for various BH masses.
We briefly recall below the equations of motion of the scalar field in the relativistic
and nonrelativistic regimes.

\subsection{Relativistic regime}

As in \cite{Boudon:2022dxi}, we neglect the gravitational backreaction of the scalar 
cloud and we consider the steady-state limit, that is, the growth and the displacement
of the BH are small as compared with the BH mass and the dark matter halo radius.
Then, working with the isotropic radial coordinate $r$, the static spherically symmetric
metric can be written as
\be
ds^2 = - f(r) \, dt^2 + h(r) \, (dr^2 + r^2 \, d\vec\Omega^2) .
\label{eq:ds2-def}
\ee
Close to the BH, below a transition radius $r_{\rm sg}$, 
the BH gravity dominates and the isotropic metric functions $f(r)$ 
and $h(r)$ read as
\ba
\frac{r_s}{4} < r \ll r_{\rm sg} : && f(r) = \left( \frac{1-r_s/(4r)}{1+r_s/(4r)}
\right)^2 , \nonumber \\
&& h(r) = (1+r_s/(4r))^4 ,
\label{eq:f-h-def}
\ea
where $r_s = 2 {\cal G} M_{\rm BH}$ is the Schwarzschild radius.
In these coordinates, the BH horizon is located at radius $r=r_s/4$.
Far from the BH, beyond the transition radius $r_{\rm sg}$, 
the dark matter self-gravity dominates but is nonrelativistic and we have
\be
r \gg r_{\rm sg} : \;\;\; f = 1 + 2 \Phi_{\rm N} , \;\;\; h = 1 - 2 \Phi_{\rm N} ,
\label{eq:Newtonian-h-f}
\ee
where the Newtonian gravitational potential $\Phi_{\rm N}$ is given by the Poisson
equation, $\nabla^2\Phi_{\rm N} = 4 \pi {\cal G} \rho$.

In the relativistic regime, the scalar-field dynamics are governed by the nonlinear
Klein-Gordon equation
\be
\frac{\partial^2\phi}{\partial t^2} - \sqrt{\frac{f}{h^3}} \nabla \cdot ( \sqrt{f h}
\nabla \phi ) + f \frac{\partial V}{\partial \phi} = 0 .
\label{eq:KG-phi-1}
\ee
For the spherically symmetric and static metric (\ref{eq:ds2-def}) and the
quartic self-interaction (\ref{eq:V_I-def}), one obtains in the large-scalar mass
limit the solution \citep{Brax:2019npi}
\be
\phi = \phi_0(r,\theta) \, {\rm cn}[ \omega(r,\theta) t - {\bf K}(r,\theta)
\beta(r,\theta), k(r,\theta) ] ,
\label{eq:phi-cn-def}
\ee
where ${\rm cn}(u,k)$ is the Jacobi elliptic function \cite{Gradshteyn1965,Byrd-1971}
of argument $u$, modulus $k$, and period $4 {\bf K}$, and ${\bf K}(k)$ is the complete 
elliptic integral of the first kind, defined by 
${\bf K}(k) = \int_0^{\pi/2} \, d\theta/\sqrt{1-k^2\sin^2\theta}$ for $0 \leq k <1$ 
\cite{Gradshteyn1965,Byrd-1971}. 
Here we noted ${\bf K}(r,\theta) \equiv {\bf K}[k(r,\theta)]$. 
Equation (\ref{eq:phi-cn-def}) is the leading-order approximation 
in the limit 
\be
m r_s \gg 1 ,
\label{eq:large-m}
\ee
that is, the Compton wavelength of the scalar field is
much smaller than the BH horizon.
We focus on this regime in this paper.
The expression (\ref{eq:phi-cn-def}) means that the usual trigonometric functions
encountered in the free case, for a quadratic potential $V(\phi)$, are replaced
by the Jacobi elliptic functions in the anharmonic case, when the quartic 
self-interaction (\ref{eq:V_I-def}) is important.
This corresponds to the Duffing equation \cite{Kovacic-2011} associated with
a cubic nonlinearity.
This regime corresponds to a dark matter density $\rho \sim \rho_a$ and a modulus
$k \sim 1$.
At low density, $\rho \ll \rho_a$, the self-interaction potential is small,
we have $k \ll 1$ and the Jacobi elliptic function converges to a trigonometric
function as ${\rm cn}(u,0) = \cos(u)$.

Substituting the expression (\ref{eq:phi-cn-def}) into the Klein-Gordon equation
(\ref{eq:KG-phi-1}) gives the two conditions
\ba
&& (\nabla \beta)^2 = \frac{h}{f} \left( \frac{2 \omega_0}{\pi} \right)^2 
- \frac{h m^2}{(1-2k^2) {\bf K}^2} ,
\label{eq:beta-1}
\\
&& \frac{\lambda_4 \phi_0^2}{m^2} = \frac{2 k^2}{1-2k^2} .
\label{eq:lambda4-k}
\ea

\subsection{Nonrelativistic regime}

In the nonrelativistic weak-gravity regime, it is convenient to write the real scalar 
field $\phi$ in terms of a complex field $\psi$ as \cite{Hui:2016ltb,Brax:2019fzb}
\be
\phi = \frac{1}{\sqrt{2 m}} \left( e^{-i m t} \psi + e^{i m t} \psi^\star \right) .
\label{eq:phi-psi}
\ee
In this regime, where typical frequencies $\dot\psi/\psi$ and momenta $\nabla\psi/\psi$ 
are much smaller than $m$, the complex scalar field $\psi$ obeys the Schr\"odinger equation,
\be
i \, \dot\psi = - \frac{\nabla^2\psi}{2m} + m ( \Phi_{\rm N} + \Phi_{\rm I} ) \psi ,
\label{eq:Schrodinger}
\ee
where $\Phi_{\rm I}$ is the nonrelativistic self-interaction potential, given by
\be
\Phi_{\rm I} = \frac{m |\psi|^2}{\rho_a}  .
\label{eq:Phi_I-psi}
\ee
It is also convenient to express $\psi$ in terms of the amplitude $\rho$ and the phase $s$
by the Madelung transformation \citep{Madelung_1927,Chavanis:2011zi},
\be
\psi = \sqrt{\frac{\rho}{m}} e^{i s} .
\label{eq:Madelung}
\ee
Then, the real and imaginary parts of the Schr\"odinger equation (\ref{eq:Schrodinger}) give
\ba
&& \dot\rho + \nabla \cdot \left( \rho \frac{\nabla s}{m} \right) = 0 ,  
\label{eq:continuity-s} \\
&& \frac{\dot s}{m} + \frac{(\nabla s)^2}{2 m^2} = - ( \Phi_{\rm N} + \Phi_{\rm I} ) , 
\label{eq:Euler-s}
\ea
while the nonrelativistic self-interaction potential reads
\be
\Phi_{\rm I} = \frac{\rho}{\rho_a}  .
\label{eq:Phi_I-rho}
\ee
Defining the curl-free velocity field $\vec v$ by
\be
{\vec v} = \frac{\nabla s}{m} ,  
\label{eq:v-def}
\ee
Eqs.(\ref{eq:continuity-s})-(\ref{eq:Euler-s}) give the usual continuity and Euler equations,
\ba
&& \dot\rho + \nabla \cdot ( \rho \vec v ) = 0 ,  \label{eq:continuity} \\
&& \dot {\vec v} + (\vec v \cdot \nabla) \vec v = - \nabla ( \Phi_{\rm N} 
+ \Phi_{\rm I} ) . 
\label{eq:Euler}
\ea
Thus, in the nonrelativistic regime, we can go from the Klein-Gordon equation to the
Schr\"odinger equation and next to an hydrodynamical picture.
In this regime, the equation (\ref{eq:beta-1}) corresponds to the Bernoulli equation
associated with the integrated form of the Euler equation (\ref{eq:Euler}),
where the velocity reads
\be
{\vec v} = \frac{\pi}{2} \frac{\nabla \beta}{m} .
\ee

In the Hamilton-Jacobi and Euler equations (\ref{eq:Euler-s}) and (\ref{eq:Euler})
we neglected the quantum pressure term
\be
\Phi_{\rm Q} = - \frac{\nabla^2 \sqrt\rho}{2 m^2 \sqrt\rho} .
\label{eq:Phi_Q-def}
\ee
This is because we consider masses much greater than $10^{-22} \, {\rm eV}$, associated 
with fuzzy dark matter scenarios, so that the de Broglie 
wavelength $\lambda_{\rm dB} = 2\pi/mv$ is much smaller than the scales of interest. 
This implies that wave effects, such as interference patterns, are negligible.
However, the dynamics remain different from that of CDM particles because of the
self-interaction, which is relevant up to galactic scales and balances gravity, allowing
for the formation of stable equilibrium configurations often called solitons.
See \cite{Boudon:2022dxi} for a derivation of the regions in the parameter space
$(m,\lambda_4,M_{\rm BH})$ where our approximations are valid.

\subsection{Nonrelativistic dark matter halo}

On large scales, where the BH gravity is negligible as compared with the dark matter
self-gravity, the Euler equation (\ref{eq:Euler}) admits hydrostatic equilibria,
given by $\nabla (\Phi_{\rm N} + \Phi_{\rm I}) = 0$.
This can be integrated as
\be
\Phi_{\rm N} + \Phi_{\rm I} = \alpha , \;\;\; \mbox{with} \;\;\; \alpha = 
\Phi_{\rm N}(R_{\rm sol}) .
\label{eq:hydro-alpha}
\ee
Here we introduced the radius $R_{\rm sol}$ of the spherically symmetric halo, also called
soliton, where the density vanishes.
In the Thomas-Fermi limit (\ref{eq:hydro-alpha}) where the quantum pressure 
(\ref{eq:Phi_Q-def}) is negligible, the solution reads \cite{Chavanis:2011zi,Brax:2019fzb,Brax:2019npi}
\be
r \gg r_{\rm sg} : \;\;\; \rho(r) = \rho_0 \frac{\sin(r/r_a)}{(r/r_a)}  \;\;\;
\mbox{and} \;\;\; R_{\rm sol} = \pi r_a ,
\label{eq:Rsol}
\ee
and the transition radius $r_{\rm sg}$ is given by
\be
r_{\rm sg} = r_s \frac{\rho_a}{\rho_0} \gg r_s .
\label{eq:r_sg-def}
\ee
The bulk density $\rho_0$ is set by the mass of this dark matter halo, 
$M_{\rm sol}= (4/\pi) \rho_0 R_{\rm sol}^3$.
This is the second dark matter parameter, in addition to $\rho_a$, that enters
the dynamics that we study in this paper.
It depends on the formation history of the dark matter halo.
In this regime, the effective pressure associated with the self-interaction $\Phi$
also defines a sound speed $c_s$ given by
\be
c_s^2(\rho) = \frac{\rho}{\rho_a} \ll 1 ,
\ee
which corresponds to a polytropic gas of adiabatic index $\gamma=2$.
From Eq.(\ref{eq:r_sg-def}) we can see that the sound speed in the bulk is also related to
the transition radius as
\be
r_{\rm sg} = \frac{r_s}{c_{s0}^2} , \;\;\; c_{s0}^2 = \frac{\rho_0}{\rho_a} .
\label{eq:rsg-cs0}
\ee

\subsection{Radial accretion}
\label{sec:radial-accretion}

Close to the horizon, the dark matter cannot remain static and falls into the BH.
The case of radial accretion around a motionless BH was studied in \cite{Brax:2019npi}.
Equations (\ref{eq:beta-1}) and (\ref{eq:lambda4-k}) give the phase $\beta$ and the amplitude $\phi_0$ as a function of the modulus $k(r)$.
The latter is next obtained from the continuity equation averaged over the scalar
oscillations, that is, from the condition of constant flux over all radii in the
steady state. 
Then, as for the Bondi problem of the radial accretion of a perfect gas on a BH,
the dark matter profile is determined by the unique transsonic solution that matches 
the quasi-static equilibrium soliton at large radius and the free fall at the BH horizon.
This gives the accretion rate \citep{Brax:2019npi}.
\be
\dot{M}_{\rm BH,radial} = 3 \pi F_\star \rho_a r_s^2 = 3 \pi F_\star \rho_0 r_s^2 /c_{s0}^2 ,
\label{eq:dotM-def}
\ee
where $F_\star \simeq 0.66$.
The result (\ref{eq:dotM-def}) means that the dark matter density near the horizon
is of the order of the characteristic density $\rho_a$ while the radial velocity
is of the order of the speed of light.

This result is much lower than the Bondi accretion 
$\dot M_{\rm Bondi} \sim \rho_0 r_s^2 / c_{s0}^3$ \citep{Bondi:1952ni}.
This is because the stiff polytropic index $\gamma=2$ makes the  repulsive 
self-interaction strong enough to  slow down the infall significantly. 
Moreover, in contrast with the Bondi case with $1<\gamma<5/3$, the sonic radius
$r_c$ where the Mach number $|v_r|/c_s$ reaches unity is located within
the relativistic regime, where the hydrodynamical picture is no longer valid
and one needs to use the Klein-Gordon equation of motion (\ref{eq:KG-phi-1}),
or its large-mass limit (\ref{eq:beta-1})-(\ref{eq:lambda4-k}).

\subsection{Isentropic potential flow}
\label{sec:isentropic}

Introducing as in \cite{Boudon:2022dxi} the dimensionless variables
\be
\hat r = \frac{r}{r_s} , \;\;\; \hat\rho = 2 \frac{\rho}{\rho_a} , \;\;\;
\hat\beta = \frac{\pi}{2m r_s} \beta , \;\;\; \vec v = \hat\nabla \hat\beta ,
\label{eq:x-hat-beta-def}
\ee
the continuity equation (\ref{eq:continuity}) and the Bernoulli equation associated
with the Euler equation (\ref{eq:Euler}) coincide with those of an isentropic
potential flow with a polytropic index $\gamma=2$,
\be
\hat\nabla \cdot (\hat\rho \vec v) = 0 , \;\;\; \frac{v^2}{2} + V + H = 0 ,
\label{eq:continuity-Bernouilli}
\ee
where the external potential $V(\hat r)$ and the enthalpy $H(\hat\rho)$ are given by
\be
V(\hat r) = - \frac{\hat\rho_0}{2} - \frac{v_0^2}{2} - \frac{1}{2\hat r} , \;\;\;
H(\hat r) = \frac{\hat \rho}{2} .
\ee
Here and throughout this paper we work in the BH frame, where the BH is at rest
and the dark matter cloud moves at the uniform velocity $\vec v_0$ far from the BH.
From the Bernouilli equation (\ref{eq:continuity-Bernouilli}) the density can be 
expressed in terms of the velocity by
\be
\hat\rho = \hat\rho_0 + \frac{1}{\hat r} + v_0^2 - v^2 ,
\label{eq:rho-v2}
\ee
and substituting into the continuity equation (\ref{eq:continuity-Bernouilli}) gives
\be
\hat\nabla \cdot \left[ \left( \hat\rho_0+\frac{1}{\hat r} + v_0^2 - (\hat\nabla \hat\beta)^2 \right)
\hat\nabla \hat\beta \right] = 0 .
\label{eq:cubic-beta}
\ee
This equation holds in the nonrelativistic regime, beyond a radius 
$r_{\rm m} \sim 40 r_s$.

\section{Large-distance expansions}
\label{sec:large-radius-expansions}

\subsection{Subsonic and supersonic regimes}
\label{sec:supersonic-regime}

Although it is not possible to obtain the general solution of the nonlinear
equation of motion (\ref{eq:cubic-beta}), we can derive perturbative expansions 
in the large-distance limit. This allows us to understand the main properties 
of the flow and also to obtain analytical results for the BH dynamical friction. 
Indeed, by conservation of mass and momentum in the steady state, the accretion rate 
and the drag force are related to the influx of matter and momentum through any surface 
enclosing the BH, which can be taken to be a sphere of large radius.

\subsubsection{Subsonic regime}
\label{sec:subsonic}

In the subsonic regime, studied in \cite{Boudon:2022dxi}, we obtained at large
distance an expansion of the form
\be
\hat \beta = \hat\beta_{-1} + \hat\beta_0 + \hat\beta_1 + \dots , \;\;\;
\mbox{with} \;\;\; \hat\beta_{n} \sim \hat r^{-n} ,
\label{eq:beta-n}
\ee
where the dots stand for higher order terms over $1/\hat r$ and
\be
\hat\beta_{-1} = v_0 \hat r u ,
\label{eq:beta_-1}
\ee
is the leading-order term associated with the uniform flow $\vec v_0$.
In the subsonic case, we then have
\be
\mbox{subsonic:} \;\;\;\hat\beta_0 = f_0(u) , \;\;\; 
\hat\beta_1 = \frac{f_1(u)}{\hat r} ,
\label{eq:beta0-beta1}
\ee
where we introduced the angular variable $u$, defined as
\be
u = \cos\theta ,
\ee
and the functions $f_n$ are smooth over $-1 < u < 1$.
The first-order correction $f_0$ is generated by the $1/\hat r$ term in the
equation (\ref{eq:cubic-beta}), associated with the BH gravity, coupled to
the zeroth-order uniform flow $v_0 \hat u$. The latter being odd, this gives
an odd correction in $u$. 
The second-order correction $f_1$ contains both odd and even terms.
In particular, the even term is directly related to the mass and momentum influx
at large distance, and thus to the BH mass accretion and dynamical friction. 
The first-order correction $f_0$ is obtained by expanding Eq.(\ref{eq:cubic-beta})
over $1/\hat r$ and collecting the leading-order terms of order $1/\hat r^2$.
This gives the linear differential equation
\be
\frac{\partial^2 \hat\beta_0}{\partial\hat x^2} + \frac{\partial^2 \hat\beta_0}
{\partial\hat y^2} + \left( 1 - \frac{v_0^2}{c_{s0}^2} \right) 
\frac{\partial^2 \hat\beta_0}{\partial\hat z^2} 
= \frac{v_0 u}{\hat\rho_0 \hat r^2} ,
\label{eq:beta0-cs}
\ee
where we work in the cartesian coordinates, $\{\hat x, \hat y, \hat z\}$, with 
$\vec v_0 = v_0 \, \vec e_z$.
As pointed out in \cite{Boudon:2022dxi}, in the subsonic regime, $v_0 < c_{s0}$, 
Eq.(\ref{eq:beta0-cs}) is elliptic, whereas in the supersonic regime that we study 
in this paper, $v_0 > c_{s0}$, Eq.(\ref{eq:beta0-cs}) is hyperbolic.
In the subsonic regime, this gives a flow that is regular over all space
and determined by the boundary conditions at infinity (the uniform velocity 
$\vec v_0$) and at the center (the matching radius $r_{\rm m}$ somewhat above the
Schwarzschild radius).

\subsubsection{Supersonic regime}
\label{sec:supersonic}

As for hydrodynamical flows around moving bodies,
such as airplanes, in the subsonic regime acoustic waves travel faster than the body
and are able to propagate to all points in space (after waiting 
for a long/infinite time as in the steady state). This means that the fluid at 
any point  adapts to the presence of the moving body, the flow is smooth and 
determined by the boundary conditions at infinity and at the surface of the body 
(in our case the Schwarzschild radius).

At supersonic velocities, acoustic waves cannot catch up with the airplane speed and 
are deported downstream, within the Mach cone.
Then, the flow upstream remains unperturbed and the matching to the
boundary conditions on the surface of the airplane is made possible thanks to a shock, 
which originates at the front tip of the plane or somewhat before. 
The shock discontinuity provides the means for the flow to jump to a new pattern 
downstream, which can match the boundary conditions on the plane.

A similar behavior appears in our case, when the BH moves at supersonic speed inside
the dark matter cloud. An additional complication is that it is not possible 
to  apply simple perturbative treatments as in Eq.(\ref{eq:beta-n}) on both sides of 
the shock, with junction conditions on the shock. Indeed, we shall see that
boundary layers, governed by nonlinear effects, appear on both sides of the shock.
Therefore, in the supersonic regime, we must split the large-distance expansions
over four domains: 1) the upstream region far before the shock, 2) the boundary layer 
just before the shock, 3) the boundary layer just after the shock, 
4) the downstream region far behind the shock. 

The far upstream and downstream regimes can again be analysed through large-distance expansions such as (\ref{eq:beta-n}).
As in the subsonic regime, this gives a standard
perturbative approach, where each order $\hat\beta_n$ obeys 
a linear differential equation with a right-hand side that involves 
the lower-order terms $\hat\beta_m$ with $m<n$.
However, the functions $\hat\beta_n$ now take different forms in the upstream 
and downstream regions and they may contain logarithmic contributions in
$\ln(\hat r)$.
The boundary layers require new expansions, which take into account nonlinearities.
The full solution is obtained by matching together these four regions.
This involves two asymptotic matchings, between each boundary layer and either the 
upstream or the downstream bulk flow, and one junction condition along the shock 
between the two boundary layers.
We must also match with the uniform velocity $\vec v_0$ at infinity and
simultaneously determine the location of the shock.
The matching to the radial inflow at the Schwarzschild radius appears 
in a natural fashion as a constant of integration.
We detail this procedure in the next sections.

\subsection{Upstream region}
\label{sec:upstream}

\subsubsection{Large-distance expansion}
\label{sec:upstream-expansion}

We first consider the far upstream region. As explained above, because of 
the hyperbolic nature of the equation of motion (\ref{eq:beta0-cs}), this is
no longer a boundary-value problem (as in the subsonic case) but a Cauchy problem, 
with an initial condition upstream at $\hat z \to -\infty$.
Indeed, Eq.(\ref{eq:beta0-cs}) now takes the form of a wave equation,
\be
\frac{\partial^2\hat\beta_0}{\partial\hat x^2} + \frac{\partial^2\hat\beta_0}
{\partial\hat y^2} - \frac{1}{c_z^2} \frac{\partial^2\hat\beta_0}{\partial\hat z^2} 
= \frac{v_0 u}{\hat\rho_0^2 \hat r^2} ,
\label{eq:delta-beta-z-c2}
\ee
with
\be
v_0 > c_{s0} : \;\;\;   \frac{1}{c_z^2} = \frac{v_0^2}{c_{s0}^2} - 1 , \;\;\; c_z > 0 ,
\label{eq:c2-def}
\ee
where $\hat z$ plays the role of time and $c_z$ the role of the propagation speed.

Far from the boundary layer, the flow is smooth and we can again write
a large-distance expansion as in (\ref{eq:beta-n}). However, we shall see
that the terms $\hat\beta_n$ can include logarithmic factors $\ln(\hat r)$.

\subsubsection{First-order correction}
\label{sec:upstream-first-order}

At large velocities $v_0$, the effective pressure in the soliton associated with 
the term $\hat\rho_0$ becomes negligible and we thus expect to recover the collisionless case.
We describe the behavior of fuzzy dark matter \cite{Hui:2016ltb}, that is, scalar-field dark matter without 
self-interactions, in App.~\ref{app:free}.
We first recall in App.~\ref{app:free-Schrodinger} the results obtained in the Schr\"odinger
picture, from the classic scattering by a Newton or Coulomb potential \cite{mott1965,Hui:2016ltb}.
Then, we show in App.~\ref{app:free-hydro} how this behavior can be recovered from the hydrodynamical
approach that we use in this paper.
From the expression (\ref{eq:psi-free-hydro}) obtained in the free case,
we can expect a logarithmic dependence on the distance 
$\hat r$ in addition to the angular dependence, due to the long-range character of the Newtonian
$1/r$ potential. 
Then, at first order we look for a solution of the form
\be
\hat\beta_0 = a \ln(\hat r) + f_0(u) ,
\label{eq:beta0-a-ln}
\ee
where $a$ is a parameter to be determined. Substituting into 
Eq.(\ref{eq:cubic-beta}) and collecting the terms of order $1/\hat r^2$,
which corresponds to substituting into Eq.(\ref{eq:delta-beta-z-c2}),
we obtain the differential equation
\ba
&& (1-u^2) [ (1+c_z^2) u^2 -1 ] f_0'' + u [ 3+c_z^2-3(1+c_z^2) u^2] f_0'  \nonumber \\
&& = \frac{1+c_z^2}{2 v_0} u + a [ 1-2 (1+c_z^2) u^2 ] .
\label{eq:fu-leading}
\ea
This is a first-order equation in $f_0'$, with the general solution
\be
f_0' = \frac{1-2 v_0 a u}{2 v_0 (1-u^2)} + \frac{b}{(1-u^2) \sqrt{(1+c_z^2)u^2-1}} ,
\label{eq:f-u-a-b}
\ee
where $b$ is an integration constant.
Defining the Mach angle $\theta_c$ by
\be
0 < \theta_c < \frac{\pi}{2} : \;\;\; \sin\theta_c = \frac{c_z}{\sqrt{1+c_z^2}} = 
\frac{c_{s0}}{v_0} = \frac{1}{{\cal M}_0} ,
\label{eq:thetac-def}
\ee
where ${\cal M}_0$ is the Mach number, which also gives
\be
u_c=\cos\theta_c=\frac{1}{\sqrt{1+c_z^2}} , \;\;\; \tan\theta_c = c_z ,
\label{eq:uc-def}
\ee
the second term in Eq.(\ref{eq:f-u-a-b}) is singular on the upstream and downstream Mach cones
$\theta= \pi-\theta_c$ and $\theta=\theta_c$. To avoid the unphysical upstream singularity
at $\pi-\theta_c$, the constant $b$ must be zero.
The angular velocity at order $1/\hat r$ is 
$v_{\theta 1} = (-\sin\theta/\hat r) f_0'(u)$.
To avoid a singularity at $\theta=\pi$, along the $\hat z$-axis upstream,
we must have $a=-1/(2v_0)$. This nonzero value shows that the logarithmic term 
$a \ln(\hat r)$ cannot be ignored and we obtain
\be
f'_{0}(u) = \frac{1}{2 v_0 (1-u)} , \;\;\; f_{0}(u) 
= - \frac{\ln(1-u)}{2 v_0}  .
\label{eq:f0-up-def}
\ee
Here we discarded the irrelevant integration constant in $f_{0}$ because
the velocity potential $\hat\beta$ is defined up to a constant that plays no role,
as only gradients $\hat\nabla\hat\beta$ appear in the equations of motion.
This gives the upstream solution
\be
\hat\beta_{0} = - \frac{\ln(\hat r- \hat z)}{2 v_0}  
= - \frac{\ln [ \hat r (1-u) ])}{2 v_0}
\label{eq:beta0-upstream}
\ee
and
\ba
&& v_{r1} = - \frac{1}{2 v_0 \hat r} , \;\;\; 
v_{\theta1} = - \frac{1+u}{2 v_0 \hat r \sqrt{1-u^2}} , \nonumber \\
&& v^2_{1} = \frac{1}{\hat r} , \;\;\; \hat\rho_{1} = 0 .
\label{eq:v1-rho1-up}
\ea
Thus, at this first order, we actually recover the long-distance 
solution of the collisionless case (\ref{eq:hat-s-ln}).
This is because at this order $1/\hat r$,
the density is not modified by the deflection of the particle trajectories by the BH, 
$\hat\rho_1 = 0$ (but there will be corrections at higher orders). 
This implies that at this order there are no pressure effects, because there are 
no density gradients, and therefore we recover the results 
obtained for the supersonic motion in plasmas
\cite{Tam1966} and isothermal gas \cite{Dokuchaev1964,Ruderman_1971}.

In contrast with the subsonic case, the solution (\ref{eq:beta0-upstream}) is 
neither odd nor even. This is because of the logarithmic term that introduces 
the factor $a u$ in Eq.(\ref{eq:f-u-a-b}). This loss of parity is also
expressed by the bow shock, which obviously breaks parity. This is related 
to the hyperbolic nature of the equation of motion (\ref{eq:beta0-cs}),
which distinguishes between the limits $\hat z\to\pm\infty$, as only the
far upstream region $\hat z \to -\infty$ is associated with the initial condition
of the Cauchy problem.
Although the logarithmic term is expected from the free case described in 
App.~\ref{app:free}, we have seen above (\ref{eq:f0-up-def}) that it is required to obtain
a smooth solution upstream. Without this term, the regularity at $u=-1$ would imply 
$b \neq 0$ in (\ref{eq:f-u-a-b}), which would give an unphysical singularity on the upstream
inverted Mach cone at $\theta=\pi-\theta_c$.  

Two other differences from the subsonic case \citep{Boudon:2022dxi}, 
where $v_{r1}=0$ and $v_{\theta1} > 0$, are that we now have 
$v_{r1} < 0$ and  $v_{\theta1}<0$.
Thus, whereas in the subsonic case the increased pressure due to 
the self-interactions was strong enough to slow down the flow as it moves closer
to the BH (but remains at large distance), in the supersonic case the BH gravity 
is dominant and accelerates the dark matter fluid, with $v^2 > v_0^2$.

\subsubsection{Second-order correction}
\label{sec:second-order-upstream}

The second-order correction $\hat\beta_1$, of order $1/\hat r$,
is obtained by collecting the terms of order $1/\hat r^3$ in Eq.(\ref{eq:cubic-beta})
and using the expression (\ref{eq:beta0-upstream}) for the first-order term
$\hat\beta_0$.
This gives the linear differential equation
\be
\frac{\partial^2\hat\beta_1}{\partial\hat x^2} + \frac{\partial^2\hat\beta_1}
{\partial\hat y^2} - \frac{1}{c_z^2} \frac{\partial^2\hat\beta_1}{\partial\hat z^2} 
= \frac{1}{2\hat\rho_0 v_0 \hat r^3} .
\label{eq:beta1-cz}
\ee
In the upstream supersonic regime, the fields at a point $\vec r$ only depend on the
properties of the flow further upwind. In other words, we must solve
Eq.(\ref{eq:beta1-cz}) using the retarded propagator of the linear wave equation.
Thus we write
\ba
\hat\beta_1 & = & \frac{c_z^2}{2\hat\rho_0 v_0} \int \frac{d\hat x' d\hat y' d\hat z'}
{(\hat r'^2+a^2)^{3/2}} \int \frac{dp_x dp_y d\omega}{(2\pi)^3} \nonumber \\
&& \times \frac{e^{i p_x (\hat x-\hat x')+i p_y (\hat y-\hat y') 
- i \omega (\hat z-\hat z')}}{(\omega+i\epsilon)^2 - c_z^2 (p_x^2-p_y^2)} ,
\ea
where $\epsilon \to 0^+$ and $a\to 0^+$.
Here we used the Fourier-space expression of the retarded propagator and
we regularized the small-scale divergence of the source $1/\hat r^3$ with a
smoothing cutoff $a>0$, by replacing $1/\hat r^3 \to 1/(\hat r^2+a^2)^{3/2}$.
Performing the integrals and taking the limit $a\to 0^+$ we obtain
\be
\hat\beta_1 = - \frac{c_z^2}{2\hat\rho_0 v_0 \hat r} \int_1^\infty \frac{dy}
{ (c_z^2+y^2) \sqrt{\sin^2\theta + y^2 \cos^2\theta} } ,
\label{eq:beta1-int}
\ee
which behaves indeed as $1/\hat r$. 
This expression only applies to the half-plane $\hat z < 0$.
The integral (\ref{eq:beta1-int}) is even in $\hat z$ (i.e. in $u$)
but its first derivative is discontinuous at $u=0$. This means that we must
instead use the analytic continuation of (\ref{eq:beta1-int}) to extend this
result to $u >0$.
Alternatively, we can go back to the differential equation (\ref{eq:beta1-cz})
and substitute the ansatz 
\be 
\hat\beta_1 = \frac{f_1(u)}{\hat r} ,
\label{eq:beta1-f1-up}
\ee
which we know to be correct from the result (\ref{eq:beta1-int}), which ensures
that there are no logarithmic corrections such as $\ln\hat r/\hat r$.
This gives a second-order differential equation over $f_1$,
\ba
&& (1-u^2) [ (1+c_z^2) u^2 -1 ] f_1'' + u [ 5+3c_z^2-5(1+c_z^2) u^2] f_1'  
\nonumber \\
&& + (1+c_z^2) (1-3 u^2) f_1 - \frac{1+c_z^2}{4 v_0^3} = 0 .
\label{eq:f1-up}
\ea
The two integration constants are set by the requirement that both $f_1(-1)$ and 
$f_1(-u_c)$ be finite. We finally obtain the expressions
\ba
&& \hspace{-1cm} -1 \leq u \leq - u_c : \;\;
f_1(u) = \frac{1+c_z^2}{8 c v_0^3 \sqrt{(1+c_z^2) u^2-1}} \nonumber \\
&& \times \ln \left[ \frac{1-(1+c_z^2) u - c_z \sqrt{(1+c_z^2) u^2-1}}
{1-(1+c_z^2) u + c_z \sqrt{(1+c_z^2) u^2-1}} \right] , \;\;\;
\label{eq:f1-ln-up}
\ea
\ba
&& \hspace{-1cm} -u_c \leq u \leq u_c : \;\;
f_1(u) = - \frac{1+c_z^2}{4 c v_0^3 \sqrt{1-(1+c_z^2) u^2}} \nonumber \\
&& \times \left[ \frac{\pi}{2} - {\rm Arctan} \frac{1-(1+c_z^2) u}
{c\sqrt{1-(1+c_z^2) u^2}} \right] ,
\label{eq:f1-arctan-up}
\ea
which agree with the integral expression (\ref{eq:beta1-int}).
One can obtain Eq.(\ref{eq:f1-arctan-up}) from Eq.(\ref{eq:f1-ln-up}) by using
the property ${\rm Arctan}(x) = \frac{i}{2} \ln \frac{1-ix}{1+ix}$.
This provides the analytic continuation from $u<-u_c$ to $u>-u_c$.
To derive Eq.(\ref{eq:f1-arctan-up}) we also used the property
${\rm Arctan}(1/x) = \pi/2-{\rm Arctan}(x)$ for $x>0$, to obtain an expression
that is regular at $u=1/(1+c_z^2)$ (one needs to use the appropriate 
denomination of ${\rm arctan}$ to obtain an expression that is regular over 
the desired range of $u$).
Near the shock, at $u\to u_c$, we obtain the Taylor expansion
\be
u \to u_c^- : \;\;\; 
f_1(u) = - \frac{\pi (1+c_z^2)^{3/4}}{4 \sqrt{2} c v_0^3 \sqrt{u_c-u}} + \dots
\label{eq:f1-uc-up}
\ee
The singularity at $u_c$, where the second-order velocities $v_{r2}$ and $v_{\theta 2}$ diverge,
means that this perturbative approach breaks down near the shock, close to the downwind
Mach cone. 

As explained above, because of the hyperbolicity of the equations of motion
in the supersonic regime, the solution in the upstream domain is fully determined
by the local properties of the fluid (the sound speed $c_s$), the relative velocity
$v_0$ and the long-range gravity of the BH. It is independent of the boundary 
conditions at the Schwarzschild radius and does not contain free integration 
constants.

\subsection{Downstream region}
\label{sec:downstream}

\subsubsection{First-order correction}
\label{sec:downstream-first-order}

In the far downstream region, we again have a large-distance expansion
as in (\ref{eq:beta-n}), with possible logarithmic factors $\ln(\hat r)$. 

The first-order upstream solution (\ref{eq:beta0-upstream}) is singular on the 
$\hat z$-axis downstream, at $\theta=0$ and $u=1$. This is a signature of the fact 
that it does not apply downstream, for $\theta < \theta_s$, where $\theta_s(\hat r)$
is the polar angle of the axisymmetric shock front at radius $\hat r$.
The downstream solution still takes the general form (\ref{eq:beta0-a-ln}),
with $f_0'(u)$ again of the form (\ref{eq:f-u-a-b}).
The velocity potential $\hat\beta$ must be continuous across the shock. 
This implies that the term $a\ln\hat r$ is identical in the upstream and downstream 
functions, whence $a=-1/(2 v_0)$ again.
The solution must now be regular at $u=1$, which determines $b$, while
the second integration constant for $f_0$ is set by the continuity at $u=u_c$.
This gives
\be
\hat\beta_0 = - \frac{1}{2 v_0} \ln\hat r + f_0(u) ,
\label{eq:beta0-f0-downstream}
\ee
with 
\ba
&& f_0(u) = - \frac{\ln(1+u)}{2 v_0} + \frac{1}{2 v_0} \nonumber \\
&& \times \ln \left[ \frac{1+(1+c_z^2)u - c_z \sqrt{(1+c_z^2) u^2-1}}
{-1+(1+c_z^2)u + c_z \sqrt{(1+c_z^2) u^2-1}} \right] , \;\;\;\;\;
\label{eq:f0-down}
\ea
and we obtain
\ba
&& v_{r 1} = - \frac{1}{2 v_0 \hat r} , \nonumber \\
&& v_{\theta 1} = \frac{-1}{2 v_0 \hat r \sqrt{1-u^2}} \left[
1+ u - \frac{2 c_z}{\sqrt{u^2/u_c^2-1}} \right] , \nonumber \\
&& v^2_{1} = \frac{1}{\hat r} - \frac{2 c_z}{\hat r \sqrt{u^2/u_c^2-1}} , 
\nonumber \\
&& \hat\rho_{1} = \frac{2 c_z}{\hat r \sqrt{u^2/u_c^2-1}} . 
\label{eq:v1-rho1-down}
\ea
Thus, we can see that the scalar-field density is increased behind the shock,
by an amount that decreases at large distance as $1/\hat r$, whereas the upstream 
density (\ref{eq:v1-rho1-up}) was not modified at this order.
The radial velocity is continuous through the shock. 
This is consistent with the continuity of
$\hat\beta$ for a shock that has a fixed direction $\theta_s$ at leading order
at large distance.

The jump conditions for an isentropic potential flow across a shock are 
different from the Rankine-Hugoniot jump conditions that apply to the Navier-Stokes
equations (e.g. there is no energy equation and no entropy in our system),
see hydrodynamics textbooks such as \cite{Hirsch:2007}.
The jump conditions are the continuity of the velocity potential $\hat\beta$, 
which also implies the continuity of the tangential velocity $v_t$, 
and the continuity of the transverse mass flow $\hat\rho v_n$, where $v_n$ is the normal velocity.
In our case, at large distance we have $v_t=v_r$ and $v_n=v_\theta$ at leading order.
Thus, we recover the continuity of $v_r$ at this order, whereas the 
condition of continuity of $\hat\rho v_\theta$ at order $1/\hat r$ gives the angle 
of the shock,
\be
\hat r \to \infty : \;\;\; \theta_{s} \to \theta_c ,
\label{eq:thetas-thetac}
\ee
where $\theta_c$ was defined in Eq.(\ref{eq:thetac-def}).
Thus, as expected, at large distance the shock follows the Mach cone.

We can see that the first-order angular velocity $v_{\theta 1}$
and density $\hat\rho_{\rm 1}$ of Eq.(\ref{eq:v1-rho1-down})
diverge at $u_c$, that is, on the downwind Mach cone, as the function
$f_0(u)$ has the expansion
\be
u \to u_c^+ \! : \; f_0(u) = - \frac{\ln(1\!-\!u_c)}{2 v_0}
- \frac{\sqrt{2} (1\!+\!c_z^2)^{3/4}}{c_z v_0} \sqrt{u\!-\!u_c} + \dots
\label{eq:f0-uc-down}
\ee
This signals that the perturbative analysis presented above breaks down close
to the shock and we must take into account nonlinear effects in a boundary layer
just behind the shock. 
This singularity appears at the first-order $f_0$, whereas in the upstream case
it only appeared at the second order $f_1$, in Eq.(\ref{eq:f1-uc-up}).

\subsubsection{Second-order corrections}
\label{sec:downstream-second-order}

As we shall see in Sec.~\ref{sec:boundary-layer} below, the matching conditions 
along the boundary layers and the shock generate 
logarithmic contributions that impact the downstream bulk flow.
Therefore, as compared with the usptream expression (\ref{eq:beta1-f1-up})
the second-order expression of the downstream velocity potential contains an additional
logarithmic term,
\be 
\hat\beta_1 = \frac{f_1(u)+ g_1(u) \ln\hat r}{\hat r} .
\label{eq:beta1-f1-down}
\ee
Substituting this expression into the equation of motion (\ref{eq:cubic-beta}),
with the result (\ref{eq:f0-down}) for the first order, and collecting terms of order
$1/\hat r^3$ and $\ln\hat r/\hat r^3$ gives two coupled linear second-order differential equations
over $f_1$ and $g_1$.
Regularity at $u=1$ determines an integration constant for each of these two functions,
and we obtain
\be
g_{1} =  \frac{C_2}{\sqrt{(1+c_z^2) u^2-1}} ,
\label{eq:g1-down}
\ee
and
\ba
&& \hspace{-0.4cm}  f_1 = - \frac{3 (1+c_z^2)^2 u}{2 v_0^3 [ (1\!+\!c_z^2)u^2\!-\!1]} 
+ \frac{1}{\sqrt{(1\!+\!c_z^2) u^2\!-\!1}} \Biggl\lbrace C_1 + \frac{C_2}{2} \nonumber \\
&& \hspace{-0.4cm} \times \ln \!\! \left[ \frac{ [ (1\!+\!c_z^2)u^2\!-\!1]^2 
[ 1 + (1\!+\!c_z^2) u - c_z \sqrt{(1\!+\!c_z^2)u^2\!-\!1} ] }
{ (1\!+\!u)^2  [ -1 + (1\!+\!c_z^2) u + c_z \sqrt{(1\!+\!c_z^2)u^2\!-\!1} ] } \right] \nonumber \\
&& \hspace{-0.5cm} + \frac{1\!+\!c_z^2}{4 c_z v_0^3} \ln \!\! \left[ \! \frac{ (1\!+\!u)^3 
[ -1 + (1\!+\!c_z^2) u + c_z \sqrt{(1\!+\!c_z^2)u^2\!-\!1} ]^4 }
{ [(1\!+\!c_z^2)u^2\!-\!1]^2 [ 1 \!+\! (1\!+\!c_z^2) u \!-\! c_z \sqrt{(1\!+\!c_z^2)u^2\!-\!1} ]^3 } \! \right] 
\nonumber \\
&& \hspace{-0.4cm} + \frac{(1+c_z^2)^{3/2}}{2 c_z v_0^3} \ln \left[ \frac{\sqrt{1+c_z^2}u+1}{\sqrt{1+c_z^2}u-1} \right]
\Biggl\rbrace ,
\label{eq:f1-down}
\ea
where $C_1$ and $C_2$ are the two remaining integration constants.

\subsection{Shock front and boundary layers}
\label{sec:boundary-layer}

\subsubsection{Large-distance expansions}

We have seen above that the large-distance expansions of the upstream and
downstream bulk flows diverge at $u\to u_c$. Then, close to $u_c$ the first 
or second-order velocity corrections become greater than the zeroth-order 
velocity $v_0$ and the large-distance expansion (\ref{eq:beta-n}) breaks down.
Therefore, on both sides of the shock a boundary layer appears, where 
nonlinearities play a key role and we need to go beyond the expansion 
(\ref{eq:beta-n}).

As described in App.~\ref{sec:Width-boundary-layer},
a careful analysis shows that the boundary layers have a width 
$\Delta u \sim \hat r^{-2/3}$.
This implies that we need to introduce expansions over powers of $\hat r^{-1/3}$
and not only of $\hat r^{-1}$. Moreover, there are again logarithmic contributions.
To start with, we need to specify the location $\theta_{s}(\hat r)$ of the shock
front, which we write as the large-distance expansion
\be
\theta_{s}(\hat r) = \theta_c + \frac{\theta_1}{\hat r^{2/3}} 
+ \frac{\theta_2+\psi_2 \ln\hat r}{\hat r} 
+ \frac{\theta_3+\psi_3 \ln\hat r}{\hat r^{4/3}} + \dots 
\label{eq:theta-shock-series}
\ee
This defines in turn the expansion of $u_s(\hat r) = \cos[\theta_s(\hat r)]$.
The zeroth-order terms $\theta_c$ and $u_c$, defined in 
Eqs.(\ref{eq:thetac-def})-(\ref{eq:uc-def}),
were derived in Eq.(\ref{eq:thetas-thetac}) from the matching of the first-order 
upstream and downstream bulk flows $\hat\beta_0$.

Because the width of the boundary layer is of order $\hat r^{-2/3}$, we introduce
the boundary-layer coordinate
\be
U = \hat r^{2/3} [ u -u_{s}(\hat r) ] .
\label{eq:U-def}
\ee
We can see from Eq.(\ref{eq:f1-uc-up}) that the upstream bulk flow diverges as 
$v_\theta \sim \hat r^{-2} (u_c-u)^{-3/2}$, whereas from Eq.(\ref{eq:f0-uc-down})
the downstream bulk flow diverges as $v_\theta \sim \hat r^{-1} (u-u_c)^{-1/2}$.
Thus, the singularity close to the shock appears at a lower order in $1/\hat r$
on the downstream side.
This asymmetry means that whereas for the upstream boundary layer 
(i.e. just before the shock) we have the expansion
\ba
U < 0 : &&  \hat\beta = v_0 \hat r u - \frac{1}{2 v_0} \ln[ \hat r (1-u) ] 
+  \frac{F_2(U)}{\hat r^{2/3}} \nonumber \\
&& + \frac{F_3(U)}{\hat r} + \dots \;\;\;
\label{eq:F2F3-up}
\ea
for the downstream boundary layer (i.e. just after the shock) we have
\ba
U> 0 : && \hat\beta = v_0 \hat r u - \frac{1}{2 v_0} \ln[ \hat r (1-u_c) ] 
+ \frac{F_1(U)}{\hat r^{1/3}} \nonumber \\
&& +  \frac{F_2(U)}{\hat r^{2/3}} + \frac{F_3(U)+{\cal F}_3(U) \ln\hat r}{\hat r} 
+ \dots \;\;\;
\label{eq:F1F2F3-down}
\ea
In both cases we keep the regular part over $u$ of the bulk flow, up to the order
where the expansion over $1/\hat r$ breaks down.
In the upstream case (\ref{eq:F2F3-up}), this corresponds to the first two terms 
of order $\hat r$ and $\hat r^0$, whereas in the more singular downstream case 
(\ref {eq:F1F2F3-down}) this corresponds to the first term only, of order 
$\hat r$ (and to the constant associated with the second term).
This implies that whereas the boundary-layer expansion in $U$ starts at order 
$\hat r^{-2/3}$ in the upstream case, it starts earlier at order $\hat r^{-1/3}$
in the downstream case.
One can check that there are no logarithmic terms $\ln\hat r$ in the upstream 
boundary layer, as there was no logarithmic term either in the upstream 
second-order bulk flow (\ref{eq:beta1-f1-up}).
However, logarithmic terms appear through nonlinear effects in both the shock
curve (\ref{eq:theta-shock-series}) and the downstream boundary layer
(\ref{eq:F1F2F3-down}).

As compared with standard one-dimensional boundary-layer theory
\cite{orszag1978advanced},
$\hat r^{-1/3}$ plays the role of the small parameter and $U$ is the boundary-layer
coordinate that is stretched to account for its infinitesimal width
$\Delta u \sim \hat r^{-2/3}$. 

In Secs.~\ref{sec:accretion} and ~\ref{sec:dynamical-friction} we will compute 
the accretion rate onto the BH and its dynamical friction.
This involves surface integrals on a sphere of radius $R$, where we take the limit
$R\to \infty$ to use the large-distance expansions described above. 
These integrals give rise to a geometrical area prefactor $\hat r^2$. 
This implies that we must compute velocity and density fields up to order
$1/\hat r^2$, to obtain the constant term that determines the accretion
rate and the dynamical friction. This corresponds to the term of order $1/\hat r$
in the velocity potential $\hat\beta$.
This is why we need to go to order $1/\hat r$ in the bulk flows (\ref{eq:beta-n})
and in the boundary layers (\ref{eq:F2F3-up})-(\ref{eq:F1F2F3-down}).

\subsubsection{Order $\theta_1$ and $F_1$}
\label{sec:order-F1}

We simultaneously compute the boundary-layer expansions and the shock front 
order by order in $\hat r^{-1/3}$.
At zeroth order, there are no boundary layers and we extend the upstream and
downstream bulk flows $\hat\beta_0$ up to the shock front.
As found in Eq.(\ref{eq:thetas-thetac}), the matching condition on the shock front
also determines the zeroth-order term $\theta_c$ in the shock expansion
(\ref{eq:theta-shock-series}).

The next order is associated with the term $\theta_1/\hat r^{2/3}$ in the shock 
expansion (\ref{eq:theta-shock-series}) and with the terms $F_1/\hat r^{1/3}$ 
in the boundary-layer expansions (\ref{eq:F2F3-up})-(\ref{eq:F1F2F3-down}).
We can see that the term $F_1$ is absent in the upstream boundary layer.
As noticed above, this is because the singularity of the upstream bulk flow 
appears at a higher order in $1/\hat r$ than for the dowsntream bulk flow.
Therefore, at this order, we truncate the shock expansion 
(\ref{eq:theta-shock-series}) at the term $\theta_1/\hat r^{2/3}$, 
the upstream bulk flow obtained in Sec.~\ref{sec:upstream} extends down to the
shock $\theta_s$, and there is only one boundary layer behind the shock, given by
the expansion (\ref{eq:F1F2F3-down}) truncated at the term $F_1/\hat r^{1/3}$.

The upstream bulk flow, given by Eqs.(\ref{eq:f0-up-def}) and 
(\ref{eq:f1-arctan-up}), provides the boundary conditions on the shock, at the
angular location $\theta_s$, that is, at $U=0$.
Then, using the downstream boundary-layer expression (\ref{eq:F1F2F3-down}),
the continuity of the velocity potential $\hat\beta$ and of the normal
momentum $\hat p_n = \hat \rho v_n$ give
\be
F_1(0)= 0 , \;\;\; F_1'(0)= - \frac{4 v_0 \theta_1}{9 c_z} .
\label{eq:F10-F1p0}
\ee
Next, substituting the expansion (\ref{eq:F1F2F3-down}) in the equation of motion 
(\ref{eq:cubic-beta}) and collecting the terms of order $\hat r^{-5/3}$, we obtain
the differential equation
\be 
\left( U - \frac{c_z\theta_1}{\sqrt{1+c_z^2}} \right) F_1'' + \frac{1}{2}
\left( 1- \frac{9 c_z^2}{v_0\sqrt{1+c_z^2}} F_1'' \right) F_1' = 0 .
\label{eq:F1-down}
\ee
This differential equation is nonlinear, which shows the importance of nonlinear
effects in the boundary layer that were not captured by the perturbative treatment
in Sec.~\ref{sec:downstream-first-order}.
The solution can be expressed by the parametric representation in terms of the
auxiliary variable $Y$,
\ba
&& U= \frac{c_z\theta_1}{\sqrt{1+c_z^2}} + \frac{16 v_0^3 \theta_1^3 - 729 c_z^3 Y^3}
{243 v_0 c_z \sqrt{1+c_z^2} Y^2} , \nonumber \\
&& F_1 = \frac{-64 v_0^3 \theta_1^3 + 729 c_z^3 Y^3}
{486 v_0 c_z \sqrt{1+c_z^2} Y} , \nonumber \\
&& \mbox{with} \;\; 0 < Y < \frac{4 v_0 \theta_1}{9 c_z} , \;\;\; 0 < U < \infty ,
\ea
where we used the boundary conditions (\ref{eq:F10-F1p0}) to determine two
integration constants. 

Next, we write the expansion of this result at large $U$, in powers of $1/U$,
in the rear of the boundary layer. Expressed next in terms of $u$, this gives the
expansion
\ba
&& \hspace{-0.5cm} \hat\beta = \hat r v_0 u + \left[ 
- \frac{\ln[\hat r (1-u_c)]}{2 v_0} - \frac{8 v_0 \theta_1^{3/2} \sqrt{u-u_c} }
{3^{5/2} c_z^{1/2} (1+c_z^2)^{1/4}} + \dots \right] \nonumber \\
&& + \hat r^{-1} \left[ - \frac{8 c_z v_0 \theta_1^3}{81 (1+c_z^2)}
\frac{1}{u-u_c} + \dots \right] + \dots
\label{eq:beta-F1-large-U-down}
\ea
The dots in the brackets correspond to higher orders in $u-u_c$ that are generated
by higher orders in the boundary-layer expansion (\ref{eq:F1F2F3-down})
(the functions $F_2$, $F_3$, ...) while the dots at the end correspond to higher 
orders in $1/\hat r$.

This must be matched to the expansion of the bulk downstream flow 
(\ref{eq:f0-down})-(\ref{eq:f1-down}) at $u \to u_c^+$,
\ba
&& \hspace{-0.5cm} \hat\beta = \hat r v_0 u + \left[ 
- \frac{\ln[\hat r (1-u_c)]}{2 v_0} - \frac{\sqrt{2} (1+c_z^2)^{3/4} \sqrt{u-u_c} }
{v_0 c_z} + \dots \right] \nonumber \\
&& + \hat r^{-1} \left[ - \frac{3 (1+c_z^2)}{4 v_0^3}
\frac{1}{u-u_c} + \dots \right] + \dots
\label{eq:beta-F1-uc-down}
\ea
Both terms in $\theta_1$ in Eq.(\ref{eq:beta-F1-large-U-down}) match those
in Eq.(\ref{eq:beta-F1-uc-down}) for
\be
\theta_1 = \left( \frac{3}{2} \right)^{5/3}  \frac{(1+c_z^2)^{2/3}}
{v_0^{4/3} c_z^{1/3}}  .
\label{eq:theta1-v0}
\ee
This determines the coefficient $\theta_1$ and the location of the shock
(\ref{eq:theta-shock-series}) at this order.
The fact that we can simultaneously match both $\theta_1$ terms in 
Eq.(\ref{eq:beta-F1-large-U-down}) provides a check of the computation.
This asymptotic matching procedure, where we match the large-$U$ behavior
of the boundary layer with the small $u-u_c$ behavior of the downstream bulk
flow (i.e., the rear of the boundary layer with the behavior of the bulk flow
close to the shock), allows us to obtain the global solution over all space. 

The width $\hat r^{-2/3}$ of the boundary layer is found from Eq.(\ref{eq:F1-down}).
This is the scaling that ensures the balance between the linear and nonlinear
terms in Eq.(\ref{eq:F1-down}), as powers of $\hat r$ cancel out.
This is explained in more details in App.~\ref{sec:Width-boundary-layer}.
The regularization of the divergences at the shock found in the perturbative
treatment of the bulk flow in Sec.~\ref{sec:downstream} is made possible
by this nonlinearity.

\subsubsection{Higher orders $F_2$ and $F_3$}
\label{sec:order-F2-F3}

We use the same method to compute the shock and the boundary layers
at orders $\{\theta_2,\psi_2;F_2\}$ and $\{\theta_3,\psi_3;F_3,{\cal F}_3\}$.
The additional ingredient is that we now have two boundary layers. 
The functional form of $F_2,F_3,{\cal F}_3$ is again obtained by substituting
into the equation of motion (\ref{eq:cubic-beta}), while the integration constants
are determined by the junction conditions.
We now have two asymptotic matchings, between the rear of each boundary layer and the
bulk flow, and simple junction conditions between both boundary layers on the shock.
A detailed computation shows that we need to introduce the logarithmic terms in the expansions 
(\ref{eq:theta-shock-series}) and (\ref{eq:F1F2F3-down}) to satisfy these junction conditions.
Then, this fully determines $\psi_2$ in the shock expansion (\ref{eq:theta-shock-series})
and the integration constant $C_2$ in the bulk downstream solution 
(\ref{eq:g1-down})-(\ref{eq:f1-down}), while the integration constant
$C_1$ in Eq.(\ref{eq:f1-down}) is expressed in terms of $\theta_2$, which remains
undetermined.
We do not give the expressions of these higher-order results here,
as they are somewhat lengthy and not especially illuminating.

\subsubsection{Shock front}
\label{sec:shock-front}

From the analytical solution derived in the previous sections, we obtain the discontinuity of the
density and velocity across the shock,
\ba
&& \Delta \hat\rho = \frac{2^{4/3} c_z^{2/3} v_0^{2/3}}{3^{1/3} (1+c_z^2)^{1/3}} \frac{1}{\hat r^{2/3}} 
+ \dots \nonumber \\
&& \Delta v_r = \frac{3^{1/3} (1+c_z^2)^{5/6}}{2^{1/3} c_z^{2/3} v_0^{5/3}} \frac{1}{\hat r^{4/3}}
+ \dots \nonumber \\
&& \Delta v_\theta = \frac{2^{1/3} (1+c_z^2)^{1/6}}{3^{1/3} c_z^{1/3} v_0^{1/3}} \frac{1}{\hat r^{2/3}} 
+ \dots
\label{eq:Delta-rho-shock}
\ea
The discontinuity of the radial velocity appears at a higher order over $1/\hat r$ because at large
distance the radial velocity is parallel to the shock front and conserved at leading order.
Going back to dimensional units, this gives the large-distance scalings
\ba
&& \Delta \rho \sim \left( \frac{r}{r_{\rm sg}}\right)^{-2/3} \rho_ 0 ,  \nonumber \\
&& \Delta v_{\theta} \sim \left( \frac{r}{r_{\rm sg}}\right)^{-2/3} {\cal M}_0^{-1} v_0
=  \left( \frac{r}{r_{\rm sg}}\right)^{-2/3} c_s . 
\label{eq:shock-Delta-vtheta}
\ea

\section{Drag force on the BH}
\label{sec:drag-force}

\subsection{Relation between the accretion rate and the large-distance expansion}

As explained above, at the order that we need in this paper,
we now have the global solution of the flow at large distance,
except for an unknown parameter $\theta_2$, defined from the expansion
(\ref{eq:theta-shock-series}) of the shock, or equivalently an integration constant
in the downstream solution.
This remaining freedom is due to the fact that so far we have not used the inner boundary
condition close to the BH. 
In fact, this parameter will simply be determined by the accretion rate onto the BH,
which is thus sufficient to describe the boundary condition at the Schwarszchild radius.

In the steady state, the accretion rate onto the BH is given by the flux
of matter through any closed surface that surrounds the BH. Choosing a sphere
of radius $\hat r$, the accretion rate writes
\be 
\dot{\hat M}_{\rm BH} = - 2 \pi \hat r^2 \int_{-1}^1 du \; \hat\rho v_r .
\label{eq:Macc-def}
\ee
Thus, we can relate $\dot{\hat M}_{\rm BH}$ to the large-distance expansion
by computing the radial momentum $\hat\rho v_r$ up to order $1/\hat r^2$.
To handle the fact that we have obtained separate expressions for the scalar 
field profile over four domains, the upstream and downstream bulk flows and the
boundary layers, with two asymptotic matchings in-between, we define the
angular function
\be 
\dot{\hat M}_{\rm BH}(u) = - 2 \pi \hat r^2 \int_{-1}^u du \; \hat\rho v_r ,
\label{eq:Macc-u}
\ee
so that the total accretion rate is $\dot{\hat M}_{\rm BH}(u=1)$.
Then, up to integration constants, we obtain $\dot{\hat M}_{\rm BH}(u)$
in each domain from the appropriate expression of the scalar field flow.
Next, as for the flow, the integration constants are determined by the two
asymptotic matchings at the rear of the two boundary layers and by continuity
at the shock location.
This determines the global function $\dot{\hat M}_{\rm BH}(u)$ and the
total accretion rate $\dot{\hat M}_{\rm BH}(1)$.
We obtain
\ba
&& \dot{\hat M}_{\rm BH} = - \frac{4\pi c_z v_0 \theta_2}{\sqrt{1+c_z^2}} 
- \frac{\pi (20+12 c_z^2 + \sqrt{3}\pi)}{3 v_0} \nonumber \\
&& - \frac{4\pi\sqrt{1+c_z^2}}{3 v_0} \ln\left[ \frac{16 (\sqrt{1+c_z^2}-1)^3 v_0^2}
{3 c_z^4 (1+c_z^2)} \right] \nonumber \\
&& - \frac{2\pi}{9 v_0} \ln\left[ \frac{(\sqrt{1+c_z^2}-1)^{18} v_0^{16}}
{2^{16} 3^8 c_z^{20} (1+c_z^2)^{11}} \right] .
\label{eq:Macc-theta2}
\ea
As expected, we can see that the result (\ref{eq:Macc-theta2}) does not depend on the
radius $\hat r$. All terms with higher powers of $\hat r$ eventually cancel out and the
large-radius limit, $\hat r \to\infty$, gives a finite result.
This agrees with the fact that the matter flux does not depend on the choice of the surface
enclosing the BH, in the stationary regime. 

As announced above, the expression (\ref{eq:Macc-theta2}) relates the remaining large-distance 
unknown parameter $\theta_2$ to the accretion rate $\dot{\hat M}_{\rm BH}$.
By construction the large-distance expansion cannot know about the inner boundary
condition (which is beyond its domain of validity) and cannot determine the accretion rate
$\dot{\hat M}_{\rm BH}$. However, the flow at large distance remains sensitive to the
accretion rate because of the constant-flux condition in the steady state, as explicitly
shown by Eq.(\ref{eq:Macc-theta2}).

\subsection{Accretion drag and dynamical friction}

In \cite{Boudon:2022dxi}, using the Euler equation associated with the Bernoulli equation
(\ref{eq:continuity-Bernouilli}), we showed that the drag force on the BH can be written as
\be
\hat F_z = - 2 \pi \hat r^2  \int_{-1}^1 du ( \hat\rho v_r v_z + \hat P u ) ,
\ee
where we have chosen the surface of integration to be a sphere of radius $\hat r$.
As for the accretion rate in Eq.(\ref{eq:Macc-u}), we define a function $\hat F_z(u)$ 
to compute the drag force in each angular domain, up to integration constants.
The junction conditions and asymptotic matching then provide the global function
and the full dynamical friction is obtained from $\hat F_z(u=1)$. Using Eq.(\ref{eq:Macc-theta2})
to express $\theta_2$ in terms of $\dot{\hat M}_{\rm BH}$, we obtain
\be
\hat F_z = \dot{\hat M}_{\rm BH} v_0 + \frac{2 \pi c_z^2}{3 (1+c_z^2)} \ln \left[
\frac{e v_0^4 c_z \hat r^2}{18 (1+c_z^2)^2} \right] ,
\label{eq:Fz-1}
\ee
where $e=\exp(1)$ is the base of the natural logarithm.
In dimensional units, this reads as
\be
F_z = \dot M_{\rm BH} v_0 + \frac{\pi}{3} \rho_a r_s^2 \frac{c_{s0}^2}{v_0^2} \ln \left[
\frac{e v_0^4 c_z r^2}{18 (1+c_z^2)^2 r_s^2}  \right]  .
\label{eq:Fdyn}
\ee
Thus, our computation recovers in a unified manner two contributions to the total drag force,
\be
F_z = F_{\rm acc} + F_{\rm df} ,
\label{eq:Fz-acc-df}
\ee
where the first term is directly related to the accretion of matter, and therefore of 
momentum, by the BH, whereas the second term is the classical dynamical friction,
associated with the long-range gravitational attraction from the wake behind the BH when 
pressure forces are present.

\section{Two regimes for the BH accretion rate}
\label{sec:accretion}

As explained above, we must derive $\dot M_{\rm BH}$ by other methods
than the large-distance expansion, so as to handle the boundary condition at the horizon.
This requires a fully relativistic treatment.
We do not perform such a numerical computation of the axisymmetric
relativistic flow down to the horizon in this paper, but we present a simple approximation
that should capture the main behaviours.

\subsection{Self-regulated accretion at moderate Mach numbers}

As recalled in Sec.~\ref{sec:radial-accretion}, in the radial case the accretion rate
is given by the expression (\ref{eq:dotM-def}).
We showed in \cite{Boudon:2022dxi} that this accretion rate remains valid in the
subsonic regime, up to $v_0 \lesssim c_{s0}$.
Indeed, below the transition radius $r_{\rm sg} \gg r_s$, the flow quickly becomes
approximately radial and one recovers the radial solution.
This accretion rate is much smaller than the spherical Bondi accretion rate 
\citep{Bondi:1952ni}, $\dot M_{\rm Bondi} \sim \rho_0 r_s^2/c_{s0}^3$, because of the 
steep effective adiabatic index $\gamma=2$. 
In the supersonic regime, one usually expects to recover the Hoyle-Lyttleton accretion rate
\cite{1939PCPS...35..405H, Edgar:2004mk} 
\be
\dot M_{\rm HL} = \frac{4 \pi \rho_0 {\cal G}^2 M_{\rm BH}^2}{v_0^3} 
= \frac{\pi \rho_0 r_s^2}{v_0^3} .
\label{eq:Mdot-HL}
\ee
However, for moderate Mach numbers this accretion rate is of the order of the Bondi
prediction and still much higher than the radial accretion rate (\ref{eq:dotM-def}). 
The latter is the highest possible flux (for radial symmetry) allowed by the effective
pressure associated with the self-interactions \citep{Brax:2019npi}. 
Lower accretion rates are associated with solutions that are fully subsonic 
(which is not physical because of the boundary condition at the BH horizon) 
or fully supersonic. Therefore, in the regime $\dot M_{\rm HL} > \dot M_{\rm BH,radial}$
a bow shock appears that slows down and deflects the dark matter and allows the matching
to the boundary conditions at the BH horizon with their much smaller accretion rate.
This creates a subsonic region behind the shock and around the BH, where the flow becomes 
approximately radial close to the BH horizon and matches the accretion rate
(\ref{eq:dotM-def}).
We shall present in Sec.~\ref{sec:supersonic-flow} below numerical computations
that confirm this behaviour.
In a sense, because the maximum possible accretion rate (\ref{eq:dotM-def}) is much smaller
than the incoming flow (\ref{eq:Mdot-HL}), the BH (dressed by the surrounding scalar cloud with large
self-interactions) acts like an obstacle, such as a solid sphere moving in a fluid or a space shuttle
in the atmosphere.

\subsection{Bondi-Hoyle-Lyttleton accretion at high Mach numbers}

At high velocities, $v_0^3 > c_{s0}^2/(3 F_\star)$, the Hoyle-Lyttleton accretion rate
(\ref{eq:Mdot-HL}) becomes smaller than the value (\ref{eq:dotM-def}), associated with
the highest possible flux. This means that matter can directly fall into the BH
along a fully supersonic solution. Thus, the BH is no longer an obstacle but a sink where matter
can freely fall.
However, on the $z$-axis behind the BH, there is still a wake and a conical shock as 
streamlines coming from all directions converge towards the symmetry axis but cannot cross.
There is also a stagnation point on the $z$-axis behind the BH, where the velocity vanishes, 
because the radial velocity must be negative and of the order of the speed of light close to the 
horizon and positive and close to $v_0$ at large radii. This turning point separates the streamlines
that fall into the BH and those that escape to infinity.
Clearly this region is subsonic, therefore we always have a subsonic region behind the BH.
Thus, we can expect that for high velocities the shock becomes attached to the BH, with
an upstream supersonic regime that extends down to the horizon on the front side of the BH
and a narrow shock cone on the back side that contains a subsonic region.
This agrees with the accretion column of the Hoyle-Lyttleton analysis 
\cite{1939PCPS...35..405H, Edgar:2004mk}.
We discuss in more details this regime in the appendix~\ref{sec:Accretion-column}.
We find that pressure forces do not modify the main properties of the Hoyle-Lyttleton accretion
and for $v_0 > c_{s0}^{2/3}$ we have a narrow accretion column on the rear side of the BH.

Therefore, we have the moderate and high-velocity behaviours
\ba
v_0 < \frac{c_{s0}^{2/3}}{(3 F_\star)^{1/3}} : && \dot M_{\rm BH} 
= \frac{12 \pi F_\star \rho_0 {\cal G}^2 M_{\rm BH}^2}{c_{s0}^2} ,
\label{eq:Mdot-low-v0} \\
v_0 > \frac{c_{s0}^{2/3}}{(3 F_\star)^{1/3}} : && \dot M_{\rm BH} 
=  \frac{4 \pi \rho_0 {\cal G}^2 M_{\rm BH}^2}{v_0^3} ,
\label{eq:Mdot-high-v0}
\ea
which we will use in the following.

\section{Comparison of accretion drag and dynamical friction}
\label{sec:dynamical-friction}

\subsection{Accretion drag}

From Eqs.(\ref{eq:Fdyn}) and (\ref{eq:Mdot-low-v0})-(\ref{eq:Mdot-high-v0}),
the accretion drag on the BH shows the low and high-velocity behaviours
\ba
v_0 < \frac{c_{s0}^{2/3}}{(3 F_\star)^{1/3}} : && F_{\rm acc} 
= \frac{12 \pi F_\star \rho_0 {\cal G}^2 M_{\rm BH}^2 v_0}{c_{s0}^2},
\label{eq:Facc-low-v0} \\
v_0 > \frac{c_{s0}^{2/3}}{(3 F_\star)^{1/3}} : && F_{\rm acc} 
=  \frac{4 \pi \rho_0 {\cal G}^2 M_{\rm BH}^2}{v_0^2} .
\label{eq:Facc-high-v0}
\ea

\subsection{Dynamical friction}
\label{subsection:dynamical-friction}

For $v_0 \gtrsim c_{s0}$ the dynamical friction term in (\ref{eq:Fdyn}) reads
\be
F_{\rm df} = \frac{8\pi\rho_0{\cal G}^2 M_{\rm BH}^2}{3 v_0^2} 
\ln\left(\frac{r_a}{r_{\rm UV}}\right) ,
\label{eq-Fz-df}
\ee
with
\be
r_{\rm UV} \simeq \sqrt{\frac{18}{e}} r_{\rm sg} {\cal M}_0^{-3/2} 
= \sqrt{\frac{18}{e}} r_s v_0^{-3/2} c_{s0}^{-1/2} .
\label{eq:r-UV}
\ee
The effective small-scale cutoff $r_{\rm UV}$ is explicitly obtained from the analytical
computation (\ref{eq:Fdyn}). Thus, the pressure associated with the self-interactions
damps the contributions from small scales to the dynamical friction and in contrast
with the collisionless result the Coulomb logarithm does not show a small-scale
divergence. On the other hand, we still have a large-scale logarithmic divergence, 
as for the seminal computation by Chandrasekhar for a stellar cloud 
\cite{Chandrasekhar:1943ys}. 
One often takes this large-scale cutoff to be the size of the cloud.
In our case, this is not a free parameter as it is given by the soliton radius $R_{\rm sol}=\pi r_a$
defined in Eq.(\ref{eq:Rsol}).

We can check that the radius $r_{\rm UV}$ is always greater than the Schwarzschild radius
as $v_0$ and $c_{s0}$ are smaller than the speed of light. 
The comparison with Eq.(\ref{eq:shock-Delta-vtheta}) shows that the radius $r_{\rm UV}$ is
the radius where the velocity is significantly perturbed by the shock, with a relative discontinuity
$\Delta v_\theta/v_0$ of order unity.
As compared with the free collisionless case, this explains the origin of the small-scale cutoff
in the Coulomb logarithm and why smaller radii do not contribute significantly to the dynamical friction.

Thus, we find that that the accretion drag is negligible at low velocity but of the
same order as the dynamical friction at high velocity,
\ba
v_0 \ll \frac{c_{s0}^{2/3}}{(3 F_\star)^{1/3}} : && F_{\rm acc} \ll F_{\rm df} ,
\nonumber \\
v_0 > \frac{c_{s0}^{2/3}}{(3 F_\star)^{1/3}} : &&  F_{\rm acc} \sim F_{\rm df} ,
\ea
where we used Eqs.(\ref{eq:Mdot-low-v0})-(\ref{eq:Mdot-high-v0}), as also discussed
in the appendix~\ref{sec:Accretion-column}.

\section{Gravitational force from the large-distance BH wake}
\label{sec:wake}

The dynamical friction is often estimated from the gravitational force exerted on the moving obect
by the overdensity created in its wake.
For collisionless systems, this was shown to give back the classical Chandrasekhar result
that was obtained from the deflection of distant orbits \cite{Mulder1983}.
In our case, this neglects pressure effects but it should provide at least a correct order of magnitude 
at high wave numbers.
We focus on the high Mach number regime, where the Mach angle is small and the accretion
proceeds through the accretion column at the rear of the BH, as detailed in 
Appendix~\ref{sec:Accretion-column}.
Then, considering a conical accretion column of Mach angle $\theta_c \ll 1$ at large distance,
its gravitational drag force on the BH reads
\ba
F_g & = & {\cal G} M_{\rm BH} \int dr \, \pi (r\theta_c)^2 \frac{\rho-\rho_0}{r^2} \\
& = & {\cal G} M_{\rm BH}  \pi \theta_c^2 \int dr \, (\rho-\rho_0) ,
\label{eq:Fg-def}
\ea
where $\rho$ is the typical density inside the column at distance $r$.
Let us estimate the contribution from large radii, beyond the Hoyle-Lyttleton radius (\ref{eq:r-HL-b-H}),
where the shock is weak.
Upstream of the shock, pressure effects are small and the streamlines follow the Keplerian orbits
and density (\ref{eq:r-b-orbit})-(\ref{eq:rho-orbit}).
At first order over $\theta_c$ and $1/r$, we obtain at large distance
\ba
&& r= \frac{b}{\theta_c} - \frac{2{\cal G} M_{\rm BH}}{v_0 \theta_c^2} , \;\;
v_r = v_0 - \frac{{\cal G} M_{\rm BH}}{v_0 r} , \nonumber \\
&& v_\theta = - v_0 \theta_c - \frac{2{\cal G} M_{\rm BH}}{v_0 \theta_c r} , \;\;
\rho = \rho_0 .
\ea
We recover that upstream of the shock there is no modification of the density at order $1/r$,
see Eq.(\ref{eq:v1-rho1-up}).
These expressions provide the boundary conditions $\{v_{r_1}, v_{\theta_1}, \rho_1 \}$ 
upstream of the shock.
The junction conditions are the continuity of the longitudinal velocity $v_r$ and of the transverse
momentum $\rho v_\theta$, while the Bernoulli equation (\ref{eq:Bernoulli-column}) remains
satisfied.
Writing $\rho_2 = \rho_1 + \Delta \rho$ and $v_{\theta_2} = v_{\theta_1} + \Delta v_\theta$,
going up to second order over $\Delta \rho$ and $\Delta v_\theta$,
we obtain the solution
\be
\Delta v_{\theta} = \frac{8 {\cal G} M_{\rm BH}}{3 c_{s0} r} , \;\;
\Delta\rho = \rho_0 \frac{8 {\cal G} M_{\rm BH}}{3 c_{s0}^2 r} ,
\label{eq:Delta-vtheta-rho}
\ee
where we used $\theta_c = c_{s0}/v_0$ at first order.
Substituting $\rho-\rho_0=\Delta\rho$ in Eq.(\ref{eq:Fg-def}), we obtain
\be
F_g = \frac{8 \pi \rho_0 {\cal G}^2 M_{\rm BH}^2}{3 v_0^2} \int \frac{dr}{r} .
\label{eq:Fg-8/3}
\ee
Thus, we recover the exact expression (\ref{eq-Fz-df}), with the Coulomb logarithm and the prefactor 
$8\pi/3$, which differs from the standard collisionless result $4\pi$ by a factor $2/3$.
Of course, this computation cannot compute the small-scale cutoff $r_{\rm UV}$ of Eq.(\ref{eq:r-UV}).

The result (\ref{eq:Fg-8/3}) neglects the boundary layers and applies the junction conditions 
at the shock between the upstream and downstream bulk flows.
Thus, we found in (\ref{eq:shock-Delta-vtheta}) that along the shock the density jump actually decays as
$r^{-2/3}$ instead of $r^{-1}$. However, the width of the boundary layer also decreases as
$\Delta u \sim r^{-2/3}$, which gives an angular width $\Delta\theta \sim r^{-2/3}$.
Therefore, the contribution from the boundary layer to the gravitational force takes the form
\be
F_{g,{\rm bl}} \sim {\cal G} M_{\rm BH} \int dr \, 2\pi (r \theta_c) r \Delta\theta \frac{\Delta\rho}{r^2} 
\propto \int dr \, r^{-4/3} .
\ee
Thus, this contribution is finite and does not show a large-distance logarithmic divergence.
It is therefore subdominant at large distance.
This is why the simple computation (\ref{eq:Fg-8/3}) can recover the exact large-distance behaviour
of the dynamical friction (\ref{eq-Fz-df}).

\section{Numerical computation at moderate Mach number}
\label{sec:supersonic-flow}

\begin{figure*}
\begin{center}
\includegraphics[width=\columnwidth]{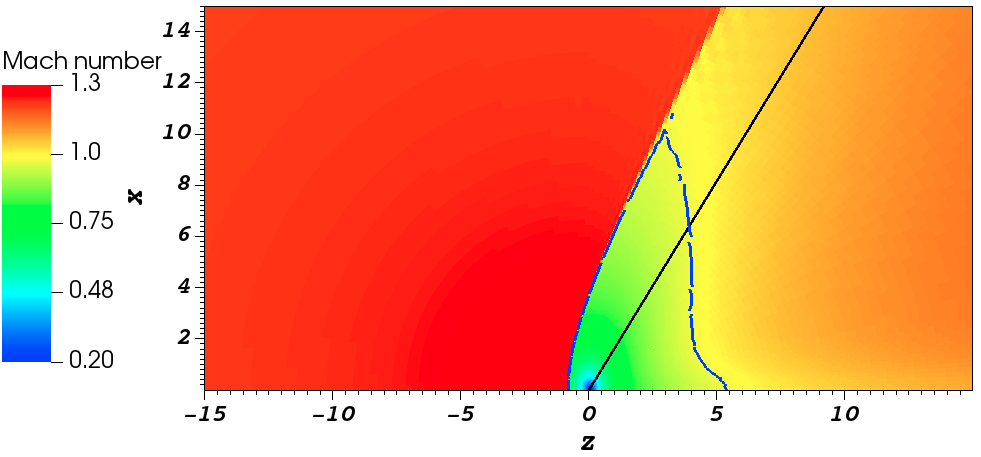}
\includegraphics[width=\columnwidth]{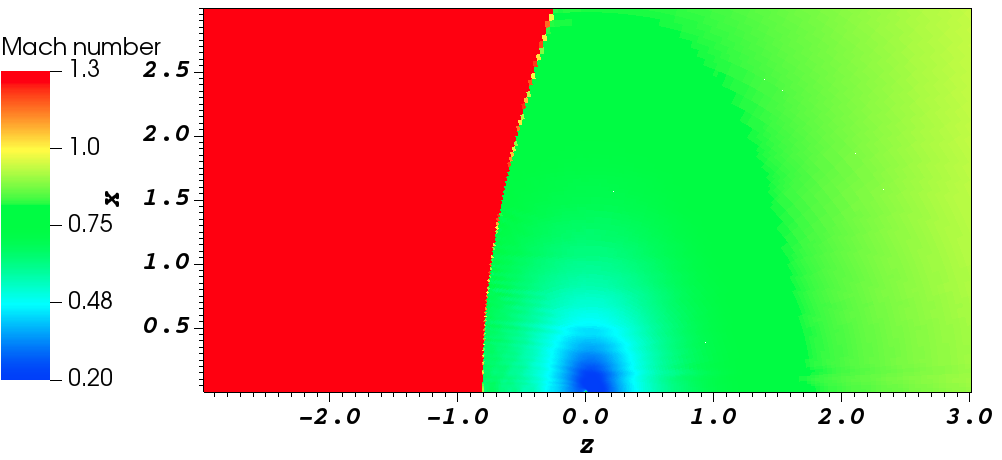}\\
\includegraphics[width=\columnwidth]{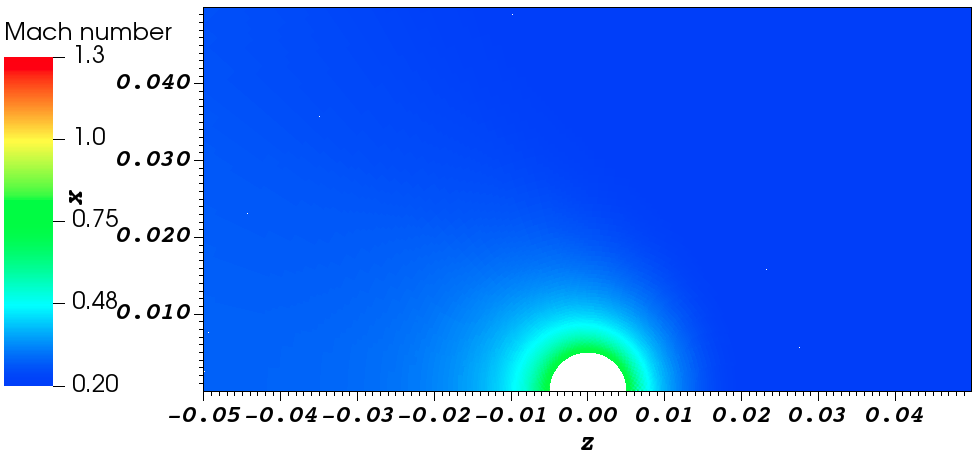}
\includegraphics[width=\columnwidth]{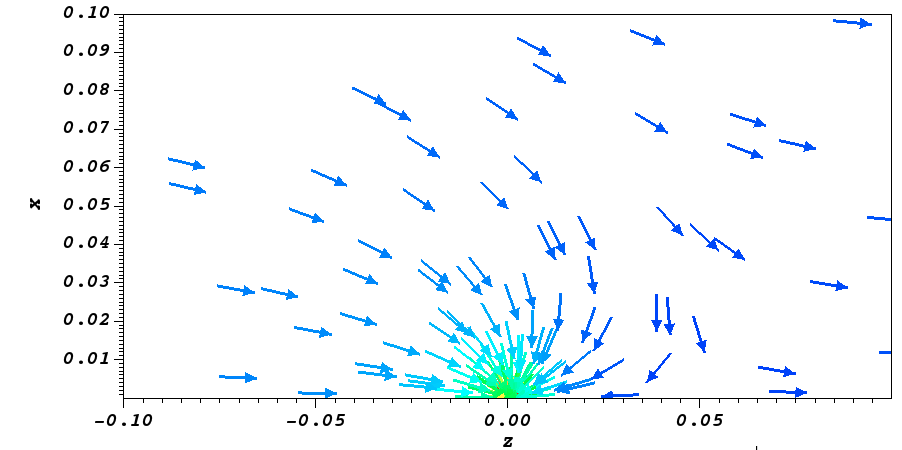}
\end{center}
\caption{
Numerical computation of the scalar dark matter flow around a BH, as viewed in the BH
frame with the dark matter coming from the left at the uniform velocity $\vec v_0$
parallel to the horizontal axis. We take ${\cal M}_0=v_0/c_{s0}=1.2$ and $c_{s0}=0.05$.
The coordinates are in units of the transition radius $r_{\rm sg}$.
The upper and bottom-left panels show maps of the Mach number ${\cal M}=v/c_s$
as we zoom closer to the BH. The lower-right panel shows a map of the velocity field.
}
\label{fig_v1p2}
\end{figure*}

To confirm the behaviour of the system for moderate Mach numbers, 
we perform a numerical computation of the scalar dark matter flow around the BH
using the public AMRVAC code \cite{refId00, KEPPENS2021316}\footnote{https://amrvac.org/index.html}.
This is a parallel adaptive mesh refinement framework aimed at solving partial differential 
equations for use in computational hydrodynamics and astrophysics.
Similar studies for the Bondi-Hoyle accretion of a polytropic gas have been presented 
in \cite{El_Mellah_2015,Xia_2018}, using also the AMRVAC code.
They consider a  polytropic gas of index $\gamma=5/3$. As compared with our case,
they also need to supplement the continuity and Euler equations with the energy equation,
as the entropy is not conserved at the shock. 
In contrast, we do not need this energy equation because our fluid is not a perfect gas
but a scalar field. In particular, the velocity is always curl-free as it is defined as the gradient
of the phase $\beta$, which is the original field of interest, with the amplitude $\sqrt{\rho}$ of the
wavefunction $\psi$.

Thus, we solve the continuity and Euler equations of an isentropic polytropic gas of index
$\gamma=2$, from the mapping described in Sec.~\ref{sec:isentropic}
in the nonrelativistic regime.
The boundary condition at large radii is set by the uniform flow at density $\rho_0$ and
velocity $\vec v_0$. The boundary condition at the inner matching radius $r_{\rm m}$ 
is set by the radial solution (\ref{eq:dotM-def}).
For the initial condition, we take for the density $\rho= \rho_0 \max(1,r_{\rm sg}/(2r))$,
which is an approximation of the radial solution \cite{Brax:2019npi}.
For the initial velocity we take $v_r = - \dot M_{\rm B}/(4\pi\rho r^2) + v_0 \cos\theta$
and $v_\theta = -v_0 \sin\theta$, that is, a simple combination of the uniform flow $\vec v_0$
with the radial infall.
Then, we solve the dynamics over time until the results converge to a steady state.

We use a two-dimensional spherical mesh (thanks to the axisymmetry there is no dependence
on the azimuthal angle $\varphi$), with a uniform stretching in the radial direction.
The upper radius is taken at $50 r_{\rm sg}$ and the lower radius at $0.005 r_{\rm sg}$.
We use a temporally first-order scheme, as we are only interested in a steady state,
and the HLLC flux scheme \cite{HLLC}. We use the dimensionless coordinates and fields 
$r/r_{\rm sg}$, $\rho/\rho_0$, $v/c_{s0}$, to work with quantities of order unity in the transition
region.
We checked that the Bernoulli invariant (\ref{eq:continuity-Bernouilli}) is constant
throughout the computational domain.

We show our results in Fig.~\ref{fig_v1p2} for the case $v_0/c_{s0}=1.2$ and $c_{s0}=0.05$.
We use the visualization tool VisIt \cite{HPV:VisIt} \footnote{https://visit-dav.github.io/visit-website/}.
The maps of the local Mach number ${\cal M}$ in the upper panels clearly show the formation 
of a bow shock, upstream of the BH at a distance
$r \simeq 0.8 r_{\rm sg}$ along the $z$-axis.
After the shock, the velocity decreases while the density and the local sound speed 
increase, which gives a clear drop of the Mach number ${\cal M}=v/c_s$.
This is much more apparent than in the density maps, which are dominated by the radial
increase close to the BH.
The Mach number first keeps decreasing closer to the BH, until a radius $0.2 r_{\rm sg}$,
and next increases as we move closer to the supersonic regime, which is beyond the
computational domain. Indeed, as seen by the lower left panel, the Mach number has not
reached unity at the inner radius of the grid, because we impose the boundary conditions
at a matching radius $r_{\rm m}$ that is still in the nonrelativistic regime. 
However, we can see that close to the BH the system becomes spherically symmetric,
as shown by the spherical color contours of the Mach number.

For a perfect gas of index $\gamma=5/3$, \cite{Foglizzo:aa,El_Mellah_2015,Xia_2018} found
that the shock is attached to the BH (in their case the point mass as they consider
Newtonian gravity) for moderate Mach numbers.
In our case, we can clearly see that the bow shock is detached from the BH horizon
and located at a radius of the order of $r_{\rm sg}$ for $v_0 \sim c_{s0}$.
This is because of the stiffer equation of state, $\gamma = 2$, which also
significantly decreases the radial accretion rate (\ref {eq:dotM-def}) as compared
with the usual Bondi result, as explained in Sec.~\ref{sec:radial-accretion}.
This property allows in turns the flow to become radial closer to the BH and 
confirms the analysis presented in Sec.~\ref{sec:accretion} for moderate Mach
numbers, with the accretion rate (\ref{eq:Mdot-low-v0}).
The pattern of the flow is clearly seen by the velocity field shown in the lower right panel,
with a radial infall close to the BH and a stagnation point behind the BH along the $z$-axis.
At large distance the flow remains close to the uniform velocity $\vec v_0$.

In the upper left panel, the black straight solid line starting from the origin shows 
the Mach angle $\theta_c = \arcsin(c_{s0}/v_0)$ of Eq.(\ref{eq:thetac-def}).
We can see that this agrees with the slope of the shock front at large distance.
The blue solid line shows the sonic line where the Mach number crosses unity ${\cal M}=1$.
Its left part, which follows the shock, actually corresponds to the shock discontinuity
where ${\cal M}$ drops from ${\cal M}_{\rm upstream}>1$ to ${\cal M}_{\rm downstream}<1$.
Therefore, although this line appears in the contour plot computed by VisIt, the Mach
number is not unity on this line but crosses unity in a discontinuous manner.
The right part, which runs from the shock to the $z$-axis in the downstream region,
is a true sonic line, where ${\cal M}=1$. Indeed, whereas the shock efficiently slows down 
the incoming flow not too far from the BH, at large distance in the transverse plane
the shock is weak and the velocity remains close to $\vec v_0$.
Thus, close to the $z-$axis the flow becomes subsonic after the shock, while it remains
supersonic at large distance. Moreover, far downstream behind the BH the flow converges
to the bulk velocity $\vec v_0$ and becomes again supersonic.
Therefore, there is a finite-size region, behind the shock and enclosing the BH, where
the flow is subsonic. This is marked by the sonic line shown by the blue solid line
in the upper left panel. It is within this subsonic region that the flow slows down
and becomes approximately radial closer to the BH.
Below the matching radius $r_{\rm m}$ and somewhat above the BH horizon, not shown in 
the plots, the flow becomes supersonic again and relativistic.
Thus, there are actually two sonic lines. 

As explained in Sec.~\ref{sec:accretion}, at higher velocity we expect a strong asymmetry
down to the horizon, with a shock that is no longer detached and a fully supersonic flow
on the front side of the BH.
However, this high-Mach regime is beyond the reach of our numerical code.
We leave a detailed study of the accretion flow near the BH at these high Mach numbers
to a future work.
This has no impact on the result (\ref{eq:Fdyn}), which follows from the large-distance
expansion, but it would be interesting to check the details of the transition
(\ref{eq:Mdot-low-v0})-(\ref{eq:Mdot-high-v0}).
This regime is discussed in more details in the appendix~\ref{sec:Accretion-column},
adpating the standard Bondi-Hoyle-Lyttleton analysis 
\cite{1939PCPS...35..405H,Bondi-Hoyle-1944,Edgar:2004mk} to our case.

\section{Comparison with other systems}
\label{sec:comparison}

\subsection{Mass accretion}
\label{sec:mass-accretion}

The Bondi and Hoyle-Lyttleton accretion rates for a perfect gas are often computed with 
the expression
\be
\dot{M}_{\rm BHL} = \frac{4\pi \rho_0 \mathcal{G}^2M_{\rm BH}^2}{(c_{s0}^2 + v_0^2)^{3/2}} ,
\ee
which interpolates between the subsonic and supersonic regimes
\cite{Bondi:1952ni, 1939PCPS...35..405H, Edgar:2004mk}.
As explained in Sec.~\ref{sec:accretion}, at low velocities we have a smaller accretion
rate, because of the more efficient self-interactions, whereas at high velocities
we recover the Hoyle-Littleton prediction,
\be
v_0 \ll c_{s0}^{2/3} : \dot M_{\rm BH} \ll \dot{M}_{\rm BHL} , \;\;\;
v_0 \gg c_{s0}^{2/3} : \dot M_{\rm BH} \simeq \dot{M}_{\rm BHL} .
\ee

\subsection{Dynamical friction}

For a collisionless system, when the BH moves at a speed that is much greater than the
velocity dispersion of the cloud particles the classical dynamical friction obtained by
Chandrasekhar \cite{Chandrasekhar:1943ys}  
(and confirmed by numerical simulations \cite{Antonini2011, Binney1987}) reads
\begin{equation}
\text{collisionless:}\quad F_{\rm free} \simeq  \frac{4\pi \rho_0 {\cal G}^2 M_{\rm BH}^2}{v_0^2} 
\ln \left(\frac{b_{\rm max}}{b_{\rm min}} \right) ,
\label{eq:friction-free}
\end{equation}
where $b_{\rm max}$ and $b_{\rm min}$ are large-scale and small-scale cutoffs.
One usually takes $b_{\max} = R$ given by the size of the cloud and 
$b_{\min} = 2 {\cal G} M_{\rm BH}/v_0^2$ the critical impact parameter, associated with bound
orbits if their angular velocity is assumed to vanish when they meet the $z-$axis behind the BH
\cite{Edgar:2004mk}. More generally, $b_{\min}$ corresponds to orbits with a deflection angle of order
unity.

For the perfect gas, one obtains in the supersonic regime \cite{Ruderman_1971,Ostriker:1998fa} 
(also recovered numerically, e.g. \cite{Sanchez-Salcedo_1999,Bernal:2013txa,refId0}) 
\begin{equation}
    \text{Perfect gas:}\quad F_{\rm gas} = \frac{4\pi\rho_0\mathcal{G}^2M_{\rm BH}^2\mathcal{I}}{v_0^2}\,,
\end{equation}
where $\mathcal{I} = \ln(1-1/\mathcal{M}^2)/2 + \ln(b_{\rm max}/b_{\rm min})$. 
In the case $\mathcal{M}\gg 1$, the first term of $\mathcal{I}$ vanishes, and only the second term 
(corresponding to a Coulomb logarithm) remains. Thus, this result is the same as 
\eqref{eq:friction-free}.

Finally, in the case of Fuzzy Dark Matter (FDM), with scalar masses around $10^{-22}$ eV where we can neglect dark matter self-interactions and the de Broglie wavelength is large, the dynamical friction is found to be \cite{Hui:2016ltb}
\begin{equation}
    \text{FDM:}\quad F_{\rm FDM} =\frac{4\pi \rho_0\mathcal{G}^2M_{\rm BH}^2C(\beta, kr)}{v_0^2}\,. 
\end{equation}
Here, $C(\beta, kr)$ is given in terms of confluent hypergeometric functions, 
$\beta={\cal G} M_{\rm BH}m/v_0$ and $k=m v_0$.
For $\beta \ll 1$ and $k r \gg 1$ one obtains $C \sim \ln(kr)$, which gives again
an expression of the form (\ref{eq:friction-free}).

Therefore, in the supersonic regime all these systems give a dynamical friction that is similar
to the Chandrasekhar result (\ref{eq:friction-free}), except that the Coulomb logarithm can vary.
This prefactor is usually difficult to estimate and the infrared and ultraviolet cutoffs are often the result
of an educated guess rather than an explicit derivation.
In our case, the ultraviolet cutoff (\ref{eq:r-UV}) is explicitly obtained from the analytic result
(\ref{eq:Fdyn}). This is because in the steady state momentum conservation allows us to relate
the drag force to the flux of momentum at large distance.
Then, we have seen that $r_{\rm UV}$ is indeed the radius where the incoming velocity of the dark
matter is significantly perturbed.
This plays the role of the critical impact parameter $b_{\rm min} \sim \mathcal{G}M_{\rm BH}/v_0^2$ 
of the collisionless case. However, $r_{\rm UV}$ depends on the physics of the system and on
the behaviour of the self-interactions, as shown by its dependence on $c_{s0}$ in (\ref{eq:r-UV}).
The infrared cutoff is then taken as the size of the cloud, as in other systems, with the 
peculiar property that this is not an additional parameter because for scalar dark matter with
a quartic self-interaction it is independent of the mass of the cloud, see Eq.(\ref{eq:Rsol}).

The radius $r_{\rm UV}$ of Eq.(\ref{eq:r-UV}) is greater than the collisionless critical impact parameter
$b_{\rm min} \sim {\cal G} M_{\rm BH}/v_0^2$,
\be
r_{\rm UV} \sim b_{\rm min} \sqrt{\frac{v_0}{c_{s0}}} .
\label{eq:rUV-bmin}
\ee
This is because of the collective pressure effects, which significantly modify the velocity field
on small scales as compared with the collisionless case, as discussed in 
Sec.~\ref{subsection:dynamical-friction}.
This makes the Coulomb logarithm (\ref{eq-Fz-df}) smaller than for the collisionless case
(\ref{eq:friction-free}).

Apart from the logarithmic term, the prefactor in Eq.(\ref{eq-Fz-df}) is smaller than that in 
Eq.(\ref{eq:friction-free}) by a factor $2/3$. 
This value is confirmed by the computation (\ref {eq:Fg-8/3}) of the gravitational force from
the BH wake at large distance.
It is due to the physics of the dark matter fluid, where the self-interactions play a key role,
as shown by the factor $c_{s0}$ in the velocity and density jumps (\ref{eq:Delta-vtheta-rho}).
In particular, it is determined by the effective polytropic equation of state $\gamma=2$ of an isentropic flow,
with its specific junction conditions.
Thus, for a fixed cloud size the dynamical friction associated with SFDM is smaller than for CDM
by at least a factor $2/3$.

\section{Conclusion}
\label{sec:conclusion}

We have  completed the  study, started in \cite{Boudon:2022dxi}, of the motion of a BH
in a scalar dark matter cloud. Here we focus on the supersonic regime.
This requires a deeper analysis than for the subsonic regime due to  the
appearance of a shock, as is usual for supersonic dynamics. 
Moreover, we found that at large distance this shock front
is surrounded by two boundary layers, which introduce terms that scale as powers of 
$r^{-1/3}$ instead of $r^{-1}$ in the large-distance expansion of the velocity potential.
We have performed these large-distance expansions up to order $r^{-2}$ for the
density and velocity fields.

We showed how the large-distance expansions of the density and velocity fields can be related 
to the BH accretion rate, which is set by the small-scale boundary conditions at the
Schwarzschild radius. This determines the remaining integration constant of the 
large-distance expansion, which is then closed in terms of the incoming velocity $v_0$
and of the accretion rate $\dot M_{\rm BH}$.
However, the value of $\dot M_{\rm BH}$ must be obtained by other means.

The large-distance expansion also provides the expression of the 
drag force experienced by the BH, thanks to the conservation of momentum.
We find that the final result takes a simple form, which can be split as the sum 
of an accretion drag force, due to the accretion of dark matter by the BH, and a dynamical 
friction that takes a form similar to the classical Chandrasekhar result.
However, the amplitude is decreased by a factor $2/3$ and the Coulomb logarithm is finite,
as it does not requite the introduction of small and large scale cutoffs. 
The large-scale radius is set by the size $R_{\rm sol}$ of the dark matter soliton, which
is a function of the combination $m^4/\lambda_4$ of the scalar-field parameters.
The small-scale radius $r_{\rm UV}$ is generated by the dynamics of the flow and
corresponds to the radius where the velocity field is significantly modified, with respect
to the incoming velocity $\vec v_0$.
This radius is always much greater than the BH horizon, for nonrelativistic incoming 
velocities $v_0$ or sound speed $c_{s0}$. 
Thus, the self-interactions provide a significant damping of the dynamical friction. 

For moderate Mach numbers, $v_0 < c_{s0}^{2/3}$, we find that the accretion
rate is still given by the radial prediction, which is much smaller than the 
Bondi-Hoyle-Lyttleton prediction.
This is because of the stiff effective equation of state, with an adiabatic index
$\gamma=2$, which regulates the infall onto the BH near the Schwarzschild radius.
Then, the bow shock is detached from the BH and located upstream of the BH.
Behind the shock there is a subsonic region that encloses the BH and a stagnation point
downstream. In this region the flow becomes approximately radial close to the BH.
Closer to the Schwarzschild radius the flow becomes supersonic again, as in the radial case.
This is confirmed by a numerical simulation for ${\cal M}_0 = 1.2$.

For high Mach numbers, $v_0 > c_{s0}^{2/3}$, we recover the Hoyle-Lyttleton
accretion rate, which is then smaller than the one obtained in the radial case.
Then, as in the classical Hoyle-Lyttleton analysis, most of the accretion takes place
in the accretion column behind the BH and the flow remains strongly asymmetric down
to the Schwarzschild radius. The shock is then attached to the rear of the BH and forms 
the boundary of this narrow accretion column.

The mass accretion and dynamical friction provide invaluable insights about the surrounding environment  of binary BHs, particularly the properties of the dark matter surrounding them. The dynamical friction exerted by the dark matter overdensity formed in the wake of the BHs introduces modifications to the waveforms, leading to a phase shift. The accretion of matter onto the BHs affects gravitational wave emission in a similar manner, but at a different post-Newtonian order. To exploit  fully these effects and utilize them as probes of dark matter environments, improving our knowledge on the different dark matter models and enhancing the sensitivity of gravitational wave detectors is crucial.
Advanced and upcoming detectors, such as Advanced LIGO, LISA and DECIGO, hold promising potential to enhance significantly our capacity to detect and analyze these effects. 
This analysis has been carried out for binary BHs in a companion paper \cite{Boudon:2023aa}. 
Future research in this direction holds great promise in shedding further light on the nature of dark matter and the astrophysical processes that shape the dynamics of binaries in galaxies.

\acknowledgements

This work was granted access to the CCRT High-Performance Computing (HPC) facility under the Grant CCRT2023-valag awarded by the Fundamental Research Division (DRF) of CEA.

\appendix

\section{Free case}
\label{app:free}

In this appendix, we consider the behaviour of a cloud of fuzzy dark matter, that is, a scalar field without
self-interactions,  moving at velocity $\vec v_0$ around a BH.
This provides a reference point for comparison with the case of quartic self-interactions, i.e. the
focus of this paper.

\subsection{Schr\"odinger picture}
\label{app:free-Schrodinger}

Going back to the equation of motion in the form of the Schr\"odinger equation (\ref{eq:Schrodinger})
for the complex scalar field $\psi$, the scattering of the incoming dark matter flux by the
BH reads in the Newtonian regime
\be
i \dot\psi = - \frac{\nabla^2\psi}{2m} + m \Phi_{\rm N} \psi , \;\;\; 
\Phi_{\rm N} = - \frac{{\cal G} M_{\rm BH}}{r} .
\label{eq:Schrod-BH-free}
\ee
There is no self-interaction potential $\Phi_{\rm I}$, as we consider the free case in this appendix,
and we neglect the self-gravity of the dark matter.
This is a classic Coulomb scattering problem and we briefly recall the solution below
\cite{mott1965,Hui:2016ltb}.
We look for a steady-state solution of the form 
\be
\psi(\vec x,t) = e^{-i E t} \hat\psi(\vec x) ,
\ee
which is solution of the time-independent Schr\"odinger equation
\be
\nabla^2\hat\psi + \left( 2 m E + 2 m^2 \frac{{\cal G} M}{r} \right) \hat\psi = 0.
\label{eq:Schrod-scatt-E}
\ee

The background case, without the BH, where the dark matter moves with the uniform velocity 
$\vec v_0 = v_0 \vec e_z$, is given by
\be
\bar{\hat\psi} = \sqrt{\frac{\bar\rho}{m}} \; e^{i k z } , \;\;\; k = \sqrt{2 m E} = m v_0 .
\label{eq:psi-background}
\ee
Factoring out the background, one can check that the Schr\"odinger equation 
(\ref{eq:Schrod-scatt-E}) admits solutions of the form
\be
\hat\psi = \sqrt{\frac{\bar\rho}{m}} \; e^{i k z} F(u) , \;\;\; u=r-z ,
\ee
where $F(u)$ is solution of the differential equation
\be
u F'' + (1-iku) F' + \beta k F = 0 , \;\;\; \beta = \frac{{\cal G} M_{\rm BH} m^2}{k} .
\ee
One can recognize the differential equation satisfied by the confluent hypergeometric equation
$\Phi(\alpha,\gamma;z)$ and we obtain
\be
\psi = \sqrt{\frac{\bar\rho}{m}} \; e^{-i E t + i k z} \Phi(i\beta,1;ik (r-z) ) ,
\label{eq:psi-Coulomb-Phi}
\ee
in agreement with \cite{Hui:2016ltb}.
At large distance, this gives the asymptotic form \cite{mott1965}
\be
\psi \sim \sqrt{\frac{\bar\rho}{m}} \; e^{-i E t + i k z - i \frac{{\cal G} M_{\rm BH} m^2}{k} \ln[ k (r-z) ]} .
\label{eq:psi-Coulomb-asymp}
\ee
The well-known logarithmic divergence of the correction to the background term is due to the
long-range character of the Newton and Coulomb $1/r$ potentials.

\subsection{Hydrodynamical picture}
\label{app:free-hydro}

We now describe how this result can be recovered within the hydrodynamical picture used in this
paper, in the large-scalar mass limit.
Thus, we start from the Hamilton-Jacobi equation (\ref{eq:Euler-s}), where we take
$\Phi_{\rm I}=0$ and $\Phi_{\rm N}$ is given by Eq.(\ref{eq:Schrod-BH-free}).
At zeroth-order over the BH gravity, we recover the background solution (\ref{eq:psi-background})
\be
\bar\psi = \sqrt{\frac{\bar\rho}{m}} \; e^{i \bar s} , \;\;\; 
\bar s = - E t + k z , \;\;\; k = \sqrt{2 m E} = m v_0 .
\label{eq:s-background}
\ee
We factor out the background by introducing $\hat s$ with
\be
s = - E t + k z + \hat s(\vec x) ,
\ee
and we obtain at lowest order
\be
\frac{\partial\hat s}{\partial z} = \frac{{\cal G} M_{\rm BH} m^2}{k r} .
\ee
Looking again for a solution of the form $\hat s(u)$ with $u=r-z$, we obtain
\be
\hat s = - \frac{{\cal G} M_{\rm BH} m^2}{k} \ln[k(r-z)] ,
\label{eq:hat-s-ln}
\ee
and hence
\be
\psi = \sqrt{\frac{\bar\rho}{m}} \; e^{-i E t + i k z - i \frac{{\cal G} M_{\rm BH} m^2}{k} \ln[ k (r-z) ]} .
\label{eq:psi-free-hydro}
\ee
We recover the asymptotic result (\ref{eq:psi-Coulomb-asymp}), with the logarithmic first-order
correction that is characteristic of the $1/r$ Newtonian potential.
This is only the leading order result, upstream of the BH, and higher-order terms are generated
by the nonlinearity of the continuity and Euler equations (\ref{eq:continuity-s})-(\ref{eq:Euler-s}).
We do not investigate this hydrodynamical approach further, as we focus on the case
with nonzero self-interactions in this paper and the free case is more conveniently described
by the linear Schr\"odinger equation with the solution (\ref{eq:psi-Coulomb-Phi}).

We note however that in the large-mass limit the Hamilton-Jacobi equation provides a convenient
tool to include General Relativistic effects on the streamlines, as in Eq.(\ref{eq:beta-1}),
as it can be solved for the exact Schwarzschild metric functions $f$ and $h$, without the need
to use the Newtonian gravity approximation (\ref{eq:Newtonian-h-f}).
This is because we can explicitly solve the geodesics equations in the spherically 
symmetric Schwarzschild metric.

\section{Width of the boundary layer}
\label{sec:Width-boundary-layer}

In this Appendix, we describe how we can obtain the $\hat r^{2/3}$ scaling of the downstream
boundary layer.
At first order in the large-distance expansion, we have obtained in 
Sec.~\ref{sec:downstream-first-order} the downstream bulk flow, given by 
Eq.(\ref{eq:beta0-f0-downstream}). 
The velocity potential has a square-root singularity at the shock location $u_c$, as seen
in Eq.(\ref{eq:f0-uc-down}). This leads to the divergent angular velocity (\ref{eq:v1-rho1-down})
at $u_c$.
Close to $u_c$, this first-order velocity perturbation $v_{\theta 1}$ becomes greater than the
zeroth-order velocity $v_0$. This signals the breakdown of this perturbative expansion.
Therefore, nonlinear terms that have been neglected in this expansion must become important
and regularize the angular velocity. Indeed, at sufficiently large distance we expect the shock to
become increasingly weak and the velocity to converge to $\vec v_0$ (i.e., there is no divergence
in the exact solution).
Thanks to the factor $1/\hat r$ in Eq.(\ref{eq:v1-rho1-down}) for $v_{\theta 1}$, the breakdown
of the perturbative treatment appears at values $u$ that are increasingly close to $u_c$ at large
distance. Therefore, the nonlinear correction to the divergent expression  (\ref{eq:v1-rho1-down})
occurs in an increasingly small region in $u$.
This corresponds to a boundary-layer phenomenon, where nonlinear effects are restricted to
a small region and permit the matching of a bulk solution with an adverse boundary condition.
In our case, $1/\hat r$ plays the role of the small parameter $\epsilon$ of standard one-dimensional
boundary-layer theory.
The scaling of the boundary-layer width is often a power law of $\epsilon$ (here $1/\hat r$), but
the exponent depends on the form of the nonlinear terms.

At lowest order, the shock front is located at $u_c$ and the boundary-layer physics happens
in the transverse direction, parallel to the angular velocity.
Then, as in (\ref{eq:beta0-f0-downstream}), we write for the phase
\be
\hat\beta = v_0 \hat r u - \frac{\ln\hat r}{2 v_0} + f_r(u) ,
\label{eq:beta-f0r}
\ee
but we do not assume that $f_r(u)$ is of order $\hat r^0$ in $\hat r$, which is why we
added the subscript $r$. Instead, the dependence on $\hat r$ is one of the boundary-layer scalings
that we are looking for.
As explained above, at lowest order the boundary-layer physics happens along the angular
direction $\vec e_\theta$; the radial velocity remains finite and it is only the angular velocity
that diverges in (\ref{eq:v1-rho1-down}).
Therefore, derivatives will be dominated by angular derivatives. Besides, as the equation of motion
(\ref{eq:cubic-beta}) only involves $\hat\nabla \hat \beta$, $f_r$ only appears through its derivative,
which we denote $g(u)$,
\be
g(u) = f'_r(u) .
\ee
We also define the transverse coordinate $x$ as
\be
u = u_c + x ,
\ee
and we focus on the boundary layer with $x \ll 1$.
To recover the flow $\vec v_0$ at large distance, the last term in Eq.(\ref{eq:beta-f0r}) must be
subdominant with respect to the first term. This implies $g \ll \hat r$, that is,
$g$ grows more slowly than $\hat r$ at large radii. On the other hand, the 
breakdown of the naive large-distance expansion of Sec.~\ref{sec:downstream-first-order}
means that we have $g \gg 1$.
In the boundary layer, we have strong gradients as the fields evolve on the scale set by the
boundary-layer width $\Delta u$, instead of the radius $\hat r$; $g' \sim g/(\Delta u) \gg g/r$.
Thus, we have
\be
x \ll 1, \;\;\; 1 \ll g \ll \hat r , \;\;\; g' \gg g/\hat r .
\label{eq:inequalities}
\ee
Then, we substitute the ansatz (\ref{eq:beta-f0r}) into the equation of motion (\ref{eq:cubic-beta})
and we use the hierarchies (\ref{eq:inequalities}) to collect the dominant terms. This yields
\be
x g' + \frac{1}{2} g - a g g' = 0  \;\;\; \mbox{with} \;\;\;  a = \frac{3 c_z^2}{2 v_0 \sqrt{1+c_z^2} \hat r} .
\label{eq:ODE-g-a}
\ee
We need to keep these three terms, as we do not know a priori the hierarchy between $x$, $g$ and
$g'$. Other terms in the equation of motion are suppressed with respect to one of these three terms 
by powers of $x$ or $g/\hat r$.

In the large-distance expansion described in Sec.~\ref{sec:downstream-first-order}, we set $a=0$
because of its $1/\hat r$ factor. This gives $x g'+g/2=0$, with the solution $g \propto 1/\sqrt{x}$.
We recover the inverse square-root divergence of $v_{\theta 1}$ in Eq.(\ref{eq:v1-rho1-down})
and of $f_0'$ from Eq.(\ref{eq:f0-uc-down}). 
Clearly, in this approximation, when $x\to 0^+$ the term $a g g'$ becomes greater than 
$g/2$, in spite of the smallness of $a$, and can no longer be neglected.
This nonlinear term will then regularize the behavior at $x \to 0^+$.
It is possible to obtain the exact analytical solution of the nonlinear differential equation
(\ref{eq:ODE-g-a}) in the implicit form
\be
0 < x < \infty , \;\;\; - \frac{b}{2a} < g < 0 , \;\;\; x = \frac{2a}{3} g + \frac{b^3}{12 a^2} g^{-2} ,
\label{eq:x-g-implicit}
\ee
where $b$ is an integration constant.
From this expression we can derive the asymptotic expansions
\ba
&& x \to 0^+ : \;\;\; g = - \frac{b}{2a} + \frac{x}{2 a} + \dots , \nonumber \\
&& x \to \infty : \;\;\; g = - \frac{b^{3/2}}{2 \sqrt{3} a \sqrt{x}} + \dots 
\label{eq:g-asymptotes}
\ea
Thus, we can see that $g(x)$ is now regular at $x \to 0^+$. The divergence of the angular velocity
has been regularized and it remains finite up to the shock front.
On the other hand, the asymptotic matching of the large-$x$ behavior (\ref{eq:g-asymptotes})
with the $u \to u_c^+$ behavior of the bulk flow (\ref{eq:f0-uc-down}) determines the integration
constant $b$,
\be
b = \frac{3 c_z^{2/3} (1+c_z^2)^{1/6}}{2^{1/3} v_0^{4/3} \hat r^{2/3}} .
\ee
The asymptotic scalings (\ref{eq:g-asymptotes}), or the implicit solution (\ref{eq:x-g-implicit}),
show that the width of the boundary layer and the characteristic amplitude of $g$ are
\be
\Delta u = \Delta x \sim b \propto \hat r^{-2/3} , \;\;\; g \sim \frac{b}{a} \propto \hat r^{1/3} .
\label{eq:width-bl}
\ee
These two scalings justify the $\hat r^{2/3}$ in the definition of the boundary-layer coordinate
(\ref{eq:U-def}) and the $\hat r^{-1/3}$ that multiplies the $F_1$ term in Eq.(\ref{eq:F1F2F3-down})
(so that it contributes at the order $\hat r^{1/3}$ in $\partial\hat\beta/\partial u$).
Equation (\ref{eq:ODE-g-a}) corresponds to Eq.(\ref{eq:F1-down}), with
different notations and the neglect of the shock curvature $\theta_1$.

The analysis above closely follows standard one-dimensional boundary layer theory and 
$1/\hat r$ plays the role of a small external parameter.
As we go to higher orders, we need a more systematic approach that can handle our
2D problem, as radial derivatives and the curvature of the shock front start to contribute.
This is done through the expansions (\ref{eq:theta-shock-series})-(\ref{eq:F1F2F3-down}).
Because of the powers $\hat r^{1/3}$ that appear in (\ref{eq:width-bl}), we need a general
expansion in powers of $\hat r^{-1/3}$. In addition, we must pay attention to the fact that
logarithmic terms may appear. In Eqs.(\ref{eq:theta-shock-series}) and (\ref{eq:F1F2F3-down})
we directly wrote those that are nonzero, after the computation is performed.
The results obtained in Sec.~\ref{sec:order-F1} also show that the shock curvature
$\theta_1$ actually already contributes at the lowest order $F_1$. However, this does not change
the scalings (\ref{eq:width-bl}). Instead, this determines the scaling $\hat r^{-2/3}$ for the first
curvature term in the expansion (\ref{eq:theta-shock-series}).
We write a similar expansion (\ref{eq:F2F3-up}) for the upstream boundary layer but as explained
in the main text is starts at order $\hat r^{-2/3}$ instead of $\hat r^{-1/3}$ because the singularity
of the upstream bulk flow at the shock is weaker.

\section{Accretion column}
\label{sec:Accretion-column}

\subsection{Hoyle-Lyttleton accretion rate}

We discuss in this appendix the accretion on the BH in the high Mach number regime
(\ref{eq:Mdot-high-v0}), where it proceeds through an accretion column on the rear side
of the BH, as in the classical Hole-Lyttleton analysis.
We use dimensional coordinates and closely follow the presentation of \cite{Edgar:2004mk},
adapted to our case.
In the hypersonic regime, upstream of the shock front pressure effects are negligible
and the dynamics follows the Keplerian orbits of the collisionless case.
This gives for the streamline of impact parameter $b$ the hyperbolic orbit
\be
r = \frac{b^2 v_0^2}{{\cal G}M_{\rm BH} (1+\cos\theta)+b v_0^2\sin\theta} ,
\label{eq:r-b-orbit}
\ee
with the radial and angular velocities
\be
v_r = \pm \sqrt{v_0^2 + \frac{2{\cal G}M_{\rm BH}}{r} - \frac{b^2 v_0^2}{r^2}} , \;\;\;
v_\theta = - \frac{b v_0}{r} ,
\ee
and the density
\be
\rho = \frac{\rho_0 b^2}{r\sin\theta (2 b - r\sin\theta)} .
\label{eq:rho-orbit}
\ee
At high $v_0$ the accretion column is a roughly conical cylinder around the downstream
$z$-axis behind the shock front, with a narrow angle $\theta_s \ll 1$ that converges at 
large distances to the Mach angle $\theta_c$ of Eq.(\ref{eq:thetac-def}).
From Eq.(\ref{eq:r-b-orbit}), the orbit of impact parameter $b$ crosses the donwstream
$z$-axis $\theta=0$ at the radius
\be
r_1 = \frac{b^2 v_0^2}{2 {\cal G} M_{\rm BH}} ,
\label{eq:r1-b}
\ee
with the velocities
\be
v_{r_1} = v_0 , \;\;\; v_{\theta_1} = - \frac{2 {\cal G} M_{\rm BH}}{b v_0} .
\ee
This gives the boundary conditions upstream of the shock front, where we take
$\theta_s \simeq 0$, while the density reads
\be
\rho_1 = \frac{\rho_0 {\cal G} M_{\rm BH}}{\sin\theta_s b v_0^2} .
\ee
The junction conditions across the shock are the continuity of the tangential velocity
and of the transverse momentum. Taking again $\theta_s \simeq 0$ this gives
just behind the shock
\be
v_{r_2} = v_0 , \;\; v_{\theta_2} = - \frac{c_{s0}^2}{\sin\theta_s v_0} , \;\;
\rho_2 = \frac{\rho_a{\cal G} M_{\rm BH}}{r} 
= \frac{\rho_0 2 {\cal G}^2 M_{\rm BH}^2}{b^2 v_0^2 c_{s0}^2}
\ee
where we used the Bernoulli equation
\be
\frac{v^2}{2} + \frac{\rho}{\rho_a} - \frac{{\cal G} M_{\rm BH}}{r} 
= \frac{v_0^2}{2} + \frac{\rho_0}{\rho_a} .
\label{eq:Bernoulli-column}
\ee
Indeed, in the subsonic regime just behind the BH we have $v^2 \ll c_s^2=\rho/\rho_a$, 
which implies $\rho \simeq \rho_a{\cal G} M_{\rm BH}/r$ for 
$r \lesssim {\cal G} M_{\rm BH}/v_0^2$.
For $\theta_s \sim \theta_c$ defined in Eq.(\ref{eq:thetac-def}) we obtain
\be
\rho_1 \sim \frac{\rho_0 {\cal G} M_{\rm BH}}{b v_0 c_{s0}} , \;\;\;   
v_{\theta_2} \sim - c_{s0} .
\ee
If we neglect $v_{\theta2}$ and assume that the dark matter will be accreted if it is bound to the BH,
$v_0^2/2-{\cal G} M_{\rm BH}/r < 0$, we obtain the Hoyle-Lyttleton radius and impact parameter
\citep{Edgar:2004mk}
\be
r_{\rm HL} = \frac{2{\cal G} M_{\rm BH}}{v_0^2} , \;\;\; b_{\rm HL} = \frac{2{\cal G} M_{\rm BH}}{v_0^2} ,
\label{eq:r-HL-b-H}
\ee
which gives the Hoyle-Lyttleton accretion rate $\dot M_{\rm HL} = \rho_0 v_0 \pi b_{\rm HL}^2$
of Eq.(\ref{eq:Mdot-HL}).

\subsection{Approximate lower bound on the accretion rate}

A more detailed analysis of the accretion column suggests that the accretion rate can be somewhat
smaller \citep{Edgar:2004mk}.
From Eq.(\ref{eq:r1-b}), the mass flux that enters the accretion column through the shock
between radii $r$ and $r+dr$ is
\be
F dr = \rho_0 v_0 2\pi b db = 2\pi \frac{{\cal G} M_{\rm BH} \rho_0}{v_0} dr .
\ee
Denoting $\mu dr$ the mass in the accretion column between radii $r$ and $r+dr$ and 
$v$ the mean longitudinal velocity in the column, the conservation of matter gives
\be
\frac{d}{dr}(\mu v) = F ,
\label{eq:mass-conserv-col}
\ee
whereas the conservation of longitudinal momentum gives
\be
\frac{d}{dr}(\mu v^2) = - \frac{{\cal G}M_{\rm BH} \mu}{r^2} 
+ \frac{d}{dr} (\pi r_\perp^2 P) + F v_0 ,
\label{eq:momentum-conserv-col}
\ee
where the first term on the right-hand side is the gravitational attraction from the BH,
the second term is the pressure force, and the third term is the momentum inflow as
$v_2=v_0$ on the shock.
As the pressure reads $P=\rho^2/(2\rho_a)$, we find that in contrast with the 
Bondi-Hoyle analysis \cite{Bondi-Hoyle-1944,Edgar:2004mk} the pressure is a priori 
of the same order as the gravitational energy, 
$r_\perp^2 P \sim {\cal G} M_{\rm BH}\mu/r \sim \rho_0 {\cal G}^2 M_{\rm BH}^2/v_0^2$.
However, for a conical shock with a constant angle $\theta_s$ we have
$r_\perp \propto r$ while $P \propto \rho^2 \propto r^{-2}$, so that the derivative
of the pressure term vanishes. Therefore, the pressure term is suppressed as compared with
the gravitational term and the analysis can proceed as in \cite{Bondi-Hoyle-1944,Edgar:2004mk}.
The mass conservation equation (\ref{eq:mass-conserv-col}) can be integrated as
\be
\mu v = F (r-r_0) ,
\ee
where $r_0$ is the location of the stagnation point behind the BH.
The momentum conservation equation (\ref{eq:momentum-conserv-col}) can be written as
\be
v \frac{dv}{dr} = - \frac{{\cal G} M_{\rm BH}}{r^2} + \frac{v (v_0-v)}{r-r_0} .
\ee
Then, requiring the velocity to be a monotonic function, from $-1$ close to the BH to $v_0$ at infinity,
implies \cite{Edgar:2004mk}
\be
r_0 > \frac{{\cal G} M_{\rm BH}}{v_0^2} .
\ee
Then the accretion rate is bounded from below by
\be
\dot M_{\rm BH} = \int_0^{r_0} dr F = F r_0 > \frac{2\pi \rho_0 {\cal G}^2 M_{\rm BH}^2}{v_0^3}  ,
\ee
which is twice smaller than the Hoyle-Lyttleton accretion rate (\ref{eq:Mdot-HL}).

\subsection{Velocity threshold for the accretion-column regime}

At radii of the order of the Schwarzschild radius inside the accretion column we have 
$\rho \sim \rho_a$ and $v \sim -1$. As seen from the Bernoulli equation (\ref{eq:Bernoulli-column}), these are the highest
possible density and velocity (at the limit of the Newtonian regime). They are also reached in the radial accretion case.
The solid angle $\Omega$ of the accretion column at a radius of the order of $r_s$ is then related
to the accretion rate by
\be
\Omega r_s^2 \rho_a \sim \dot M_{\rm BH} \sim \rho_0 {\cal G}^2 M_{\rm BH}^2/v_0^3 , 
\ee
which gives
\be
\Omega \sim \frac{c_{s0}^2}{v_0^3} , \;\;\; \mbox{whence} \;\;\;
\Omega \gtrsim 1 \;\;\; \mbox{when} \;\;\; v_0 \lesssim c_{s0}^{2/3} .
\ee
Thus we recover the two regimes (\ref{eq:Mdot-low-v0})-(\ref{eq:Mdot-high-v0}).
For $v_0 > c_{s0}^{2/3}$ the accretion column is narrow behind the BH and the
accretion rate is of the order of the Hoyle-Lyttleton prediction.
For $v_0 < c_{s0}^{2/3}$ the accretion column is large and actually engulfs all sides of the BH.
There is now a bow shock upstream of the BH, as seen in the numerical computation
displayed in Fig.~\ref{fig_v1p2}, and the accretion rate is much smaller than the 
Bondi-Hoyle-Lyttleton prediction, because of the strong impact of the self-interactions
in this subsonic region.

From Eqs.(\ref{eq:r-b-orbit})-(\ref{eq:rho-orbit}) we can see that in the high Mach regime,
$v_0 > c_{s0}^{2/3}$, the density and the velocity at $r\sim r_s$ on the upstream face of the BH,
at $\theta=\pi$, are
\be
v_r \sim 1 , \;\;\; \rho \sim \rho_a \frac{c_{s0}^2}{v_0} < \rho_a c_{s0}^{4/3} \ll \rho_a .
\ee
This confirms that this regime is very asymmetric, with a low infall rate on the upstream face of the BH,
low densities and negligible self-interactions. This allows the matter to fall directly into the BH
while remaining in the supersonic regime and without crossing a shock, over most of the BH surface.
However, most of the accretion rate comes from the narrow accretion column at the back of the BH,
which is associated with an attached shock and a finite-size subsonic region.

\bibliography{supersonic}

\begin{thebibliography}{148}
\expandafter\ifx\csname natexlab\endcsname\relax\def\natexlab#1{#1}\fi
\expandafter\ifx\csname bibnamefont\endcsname\relax
  \def\bibnamefont#1{#1}\fi
\expandafter\ifx\csname bibfnamefont\endcsname\relax
  \def\bibfnamefont#1{#1}\fi
\expandafter\ifx\csname citenamefont\endcsname\relax
  \def\citenamefont#1{#1}\fi
\expandafter\ifx\csname url\endcsname\relax
  \def\url#1{\texttt{#1}}\fi
\expandafter\ifx\csname urlprefix\endcsname\relax\def\urlprefix{URL }\fi
\providecommand{\bibinfo}[2]{#2}
\providecommand{\eprint}[2][]{\url{#2}}

\bibitem[{\citenamefont{Aghanim et~al.}(2020)}]{Planck:2018vyg}
\bibinfo{author}{\bibfnamefont{N.}~\bibnamefont{Aghanim}} \bibnamefont{et~al.}
  (\bibinfo{collaboration}{Planck}), \bibinfo{journal}{Astron. Astrophys.}
  \textbf{\bibinfo{volume}{641}}, \bibinfo{pages}{A6} (\bibinfo{year}{2020}),
  \bibinfo{note}{[Erratum: Astron.Astrophys. 652, C4 (2021)]},
  \eprint{1807.06209}.

\bibitem[{\citenamefont{Doux et~al.}(2022)}]{DES:2022qpf}
\bibinfo{author}{\bibfnamefont{C.}~\bibnamefont{Doux}} \bibnamefont{et~al.}
  (\bibinfo{collaboration}{DES}) (\bibinfo{year}{2022}), \eprint{2203.07128}.

\bibitem[{\citenamefont{Feng et~al.}(2008)\citenamefont{Feng, Tu, and
  Yu}}]{Feng:2008mu}
\bibinfo{author}{\bibfnamefont{J.~L.} \bibnamefont{Feng}},
  \bibinfo{author}{\bibfnamefont{H.}~\bibnamefont{Tu}}, \bibnamefont{and}
  \bibinfo{author}{\bibfnamefont{H.-B.} \bibnamefont{Yu}},
  \bibinfo{journal}{JCAP} \textbf{\bibinfo{volume}{10}}, \bibinfo{pages}{043}
  (\bibinfo{year}{2008}), \eprint{0808.2318}.

\bibitem[{\citenamefont{Hooper}(2010)}]{Hooper:2009zm}
\bibinfo{author}{\bibfnamefont{D.}~\bibnamefont{Hooper}}, in
  \emph{\bibinfo{booktitle}{{Theoretical Advanced Study Institute in Elementary
  Particle Physics}: {The Dawn of the LHC Era}}} (\bibinfo{year}{2010}), pp.
  \bibinfo{pages}{709--764}, \eprint{0901.4090}.

\bibitem[{\citenamefont{Schwarz}(2003)}]{Schwarz:2003du}
\bibinfo{author}{\bibfnamefont{D.~J.} \bibnamefont{Schwarz}},
  \bibinfo{journal}{Annalen Phys.} \textbf{\bibinfo{volume}{12}},
  \bibinfo{pages}{220} (\bibinfo{year}{2003}), \eprint{astro-ph/0303574}.

\bibitem[{\citenamefont{Chang et~al.}(2014)\citenamefont{Chang, Edezhath,
  Hutchinson, and Luty}}]{Chang:2013oia}
\bibinfo{author}{\bibfnamefont{S.}~\bibnamefont{Chang}},
  \bibinfo{author}{\bibfnamefont{R.}~\bibnamefont{Edezhath}},
  \bibinfo{author}{\bibfnamefont{J.}~\bibnamefont{Hutchinson}},
  \bibnamefont{and} \bibinfo{author}{\bibfnamefont{M.}~\bibnamefont{Luty}},
  \bibinfo{journal}{Phys. Rev. D} \textbf{\bibinfo{volume}{89}},
  \bibinfo{pages}{015011} (\bibinfo{year}{2014}), \eprint{1307.8120}.

\bibitem[{\citenamefont{Garcia et~al.}(2022)\citenamefont{Garcia, Kaneta,
  Mambrini, Olive, and Verner}}]{Garcia:2021iag}
\bibinfo{author}{\bibfnamefont{M.~A.~G.} \bibnamefont{Garcia}},
  \bibinfo{author}{\bibfnamefont{K.}~\bibnamefont{Kaneta}},
  \bibinfo{author}{\bibfnamefont{Y.}~\bibnamefont{Mambrini}},
  \bibinfo{author}{\bibfnamefont{K.~A.} \bibnamefont{Olive}}, \bibnamefont{and}
  \bibinfo{author}{\bibfnamefont{S.}~\bibnamefont{Verner}},
  \bibinfo{journal}{JCAP} \textbf{\bibinfo{volume}{03}}, \bibinfo{pages}{016}
  (\bibinfo{year}{2022}), \eprint{2109.13280}.

\bibitem[{\citenamefont{Feng}(2022)}]{Feng:2022rxt}
\bibinfo{author}{\bibfnamefont{J.~L.} \bibnamefont{Feng}}, in
  \emph{\bibinfo{booktitle}{{Les Houches summer school on Dark Matter}}}
  (\bibinfo{year}{2022}), \eprint{2212.02479}.

\bibitem[{\citenamefont{Liu et~al.}(2017)\citenamefont{Liu, Chen, and
  Ji}}]{Liu:2017drf}
\bibinfo{author}{\bibfnamefont{J.}~\bibnamefont{Liu}},
  \bibinfo{author}{\bibfnamefont{X.}~\bibnamefont{Chen}}, \bibnamefont{and}
  \bibinfo{author}{\bibfnamefont{X.}~\bibnamefont{Ji}},
  \bibinfo{journal}{Nature Phys.} \textbf{\bibinfo{volume}{13}},
  \bibinfo{pages}{212} (\bibinfo{year}{2017}), \eprint{1709.00688}.

\bibitem[{\citenamefont{Billard et~al.}(2021)}]{Billard:2021uyg}
\bibinfo{author}{\bibfnamefont{J.}~\bibnamefont{Billard}} \bibnamefont{et~al.}
  (\bibinfo{year}{2021}), \eprint{2104.07634}.

\bibitem[{\citenamefont{Roszkowski et~al.}(2018)\citenamefont{Roszkowski,
  Sessolo, and Trojanowski}}]{Roszkowski:2017nbc}
\bibinfo{author}{\bibfnamefont{L.}~\bibnamefont{Roszkowski}},
  \bibinfo{author}{\bibfnamefont{E.~M.} \bibnamefont{Sessolo}},
  \bibnamefont{and}
  \bibinfo{author}{\bibfnamefont{S.}~\bibnamefont{Trojanowski}},
  \bibinfo{journal}{Rept. Prog. Phys.} \textbf{\bibinfo{volume}{81}},
  \bibinfo{pages}{066201} (\bibinfo{year}{2018}), \eprint{1707.06277}.

\bibitem[{\citenamefont{Arcadi et~al.}(2018)\citenamefont{Arcadi, Dutra, Ghosh,
  Lindner, Mambrini, Pierre, Profumo, and Queiroz}}]{Arcadi:2017kky}
\bibinfo{author}{\bibfnamefont{G.}~\bibnamefont{Arcadi}},
  \bibinfo{author}{\bibfnamefont{M.}~\bibnamefont{Dutra}},
  \bibinfo{author}{\bibfnamefont{P.}~\bibnamefont{Ghosh}},
  \bibinfo{author}{\bibfnamefont{M.}~\bibnamefont{Lindner}},
  \bibinfo{author}{\bibfnamefont{Y.}~\bibnamefont{Mambrini}},
  \bibinfo{author}{\bibfnamefont{M.}~\bibnamefont{Pierre}},
  \bibinfo{author}{\bibfnamefont{S.}~\bibnamefont{Profumo}}, \bibnamefont{and}
  \bibinfo{author}{\bibfnamefont{F.~S.} \bibnamefont{Queiroz}},
  \bibinfo{journal}{Eur. Phys. J. C} \textbf{\bibinfo{volume}{78}},
  \bibinfo{pages}{203} (\bibinfo{year}{2018}), \eprint{1703.07364}.

\bibitem[{\citenamefont{Hui}(2001)}]{Hui:2001wy}
\bibinfo{author}{\bibfnamefont{L.}~\bibnamefont{Hui}}, \bibinfo{journal}{Phys.
  Rev. Lett.} \textbf{\bibinfo{volume}{86}}, \bibinfo{pages}{3467}
  (\bibinfo{year}{2001}), \eprint{astro-ph/0102349}.

\bibitem[{\citenamefont{de~Blok}(2010)}]{deBlok:2009sp}
\bibinfo{author}{\bibfnamefont{W.~J.~G.} \bibnamefont{de~Blok}},
  \bibinfo{journal}{Adv. Astron.} \textbf{\bibinfo{volume}{2010}},
  \bibinfo{pages}{789293} (\bibinfo{year}{2010}), \eprint{0910.3538}.

\bibitem[{\citenamefont{Oh et~al.}(2011)\citenamefont{Oh, Brook, Governato,
  Brinks, Mayer, de~Blok, Brooks, and Walter}}]{Oh:2010mc}
\bibinfo{author}{\bibfnamefont{S.-H.} \bibnamefont{Oh}},
  \bibinfo{author}{\bibfnamefont{C.}~\bibnamefont{Brook}},
  \bibinfo{author}{\bibfnamefont{F.}~\bibnamefont{Governato}},
  \bibinfo{author}{\bibfnamefont{E.}~\bibnamefont{Brinks}},
  \bibinfo{author}{\bibfnamefont{L.}~\bibnamefont{Mayer}},
  \bibinfo{author}{\bibfnamefont{W.~J.~G.} \bibnamefont{de~Blok}},
  \bibinfo{author}{\bibfnamefont{A.}~\bibnamefont{Brooks}}, \bibnamefont{and}
  \bibinfo{author}{\bibfnamefont{F.}~\bibnamefont{Walter}},
  \bibinfo{journal}{Astron. J.} \textbf{\bibinfo{volume}{142}},
  \bibinfo{pages}{24} (\bibinfo{year}{2011}), \eprint{1011.2777}.

\bibitem[{\citenamefont{Teyssier et~al.}(2013)\citenamefont{Teyssier, Pontzen,
  Dubois, and Read}}]{Teyssier:2012ie}
\bibinfo{author}{\bibfnamefont{R.}~\bibnamefont{Teyssier}},
  \bibinfo{author}{\bibfnamefont{A.}~\bibnamefont{Pontzen}},
  \bibinfo{author}{\bibfnamefont{Y.}~\bibnamefont{Dubois}}, \bibnamefont{and}
  \bibinfo{author}{\bibfnamefont{J.}~\bibnamefont{Read}},
  \bibinfo{journal}{Mon. Not. Roy. Astron. Soc.}
  \textbf{\bibinfo{volume}{429}}, \bibinfo{pages}{3068} (\bibinfo{year}{2013}),
  \eprint{1206.4895}.

\bibitem[{\citenamefont{Read et~al.}(2016)\citenamefont{Read, Agertz, and
  Collins}}]{Read:2015sta}
\bibinfo{author}{\bibfnamefont{J.~I.} \bibnamefont{Read}},
  \bibinfo{author}{\bibfnamefont{O.}~\bibnamefont{Agertz}}, \bibnamefont{and}
  \bibinfo{author}{\bibfnamefont{M.~L.~M.} \bibnamefont{Collins}},
  \bibinfo{journal}{Mon. Not. Roy. Astron. Soc.}
  \textbf{\bibinfo{volume}{459}}, \bibinfo{pages}{2573} (\bibinfo{year}{2016}),
  \eprint{1508.04143}.

\bibitem[{\citenamefont{Klypin et~al.}(1999)\citenamefont{Klypin, Kravtsov,
  Valenzuela, and Prada}}]{Klypin:1999uc}
\bibinfo{author}{\bibfnamefont{A.~A.} \bibnamefont{Klypin}},
  \bibinfo{author}{\bibfnamefont{A.~V.} \bibnamefont{Kravtsov}},
  \bibinfo{author}{\bibfnamefont{O.}~\bibnamefont{Valenzuela}},
  \bibnamefont{and} \bibinfo{author}{\bibfnamefont{F.}~\bibnamefont{Prada}},
  \bibinfo{journal}{Astrophys. J.} \textbf{\bibinfo{volume}{522}},
  \bibinfo{pages}{82} (\bibinfo{year}{1999}), \eprint{astro-ph/9901240}.

\bibitem[{\citenamefont{Strigari et~al.}(2007)\citenamefont{Strigari, Bullock,
  Kaplinghat, Diemand, Kuhlen, and Madau}}]{Strigari:2007ma}
\bibinfo{author}{\bibfnamefont{L.~E.} \bibnamefont{Strigari}},
  \bibinfo{author}{\bibfnamefont{J.~S.} \bibnamefont{Bullock}},
  \bibinfo{author}{\bibfnamefont{M.}~\bibnamefont{Kaplinghat}},
  \bibinfo{author}{\bibfnamefont{J.}~\bibnamefont{Diemand}},
  \bibinfo{author}{\bibfnamefont{M.}~\bibnamefont{Kuhlen}}, \bibnamefont{and}
  \bibinfo{author}{\bibfnamefont{P.}~\bibnamefont{Madau}},
  \bibinfo{journal}{Astrophys. J.} \textbf{\bibinfo{volume}{669}},
  \bibinfo{pages}{676} (\bibinfo{year}{2007}), \eprint{0704.1817}.

\bibitem[{\citenamefont{Guo et~al.}(2011)\citenamefont{Guo, White,
  Boylan-Kolchin, De~Lucia, Kauffmann, Lemson, Li, Springel, and
  Weinmann}}]{Guo:2010ap}
\bibinfo{author}{\bibfnamefont{Q.}~\bibnamefont{Guo}},
  \bibinfo{author}{\bibfnamefont{S.}~\bibnamefont{White}},
  \bibinfo{author}{\bibfnamefont{M.}~\bibnamefont{Boylan-Kolchin}},
  \bibinfo{author}{\bibfnamefont{G.}~\bibnamefont{De~Lucia}},
  \bibinfo{author}{\bibfnamefont{G.}~\bibnamefont{Kauffmann}},
  \bibinfo{author}{\bibfnamefont{G.}~\bibnamefont{Lemson}},
  \bibinfo{author}{\bibfnamefont{C.}~\bibnamefont{Li}},
  \bibinfo{author}{\bibfnamefont{V.}~\bibnamefont{Springel}}, \bibnamefont{and}
  \bibinfo{author}{\bibfnamefont{S.}~\bibnamefont{Weinmann}},
  \bibinfo{journal}{Mon. Not. Roy. Astron. Soc.}
  \textbf{\bibinfo{volume}{413}}, \bibinfo{pages}{101} (\bibinfo{year}{2011}),
  \eprint{1006.0106}.

\bibitem[{\citenamefont{Penarrubia et~al.}(2012)\citenamefont{Penarrubia,
  Pontzen, Walker, and Koposov}}]{Penarrubia:2012bb}
\bibinfo{author}{\bibfnamefont{J.}~\bibnamefont{Penarrubia}},
  \bibinfo{author}{\bibfnamefont{A.}~\bibnamefont{Pontzen}},
  \bibinfo{author}{\bibfnamefont{M.~G.} \bibnamefont{Walker}},
  \bibnamefont{and} \bibinfo{author}{\bibfnamefont{S.~E.}
  \bibnamefont{Koposov}}, \bibinfo{journal}{Astrophys. J. Lett.}
  \textbf{\bibinfo{volume}{759}}, \bibinfo{pages}{L42} (\bibinfo{year}{2012}),
  \eprint{1207.2772}.

\bibitem[{\citenamefont{Boylan-Kolchin
  et~al.}(2011)\citenamefont{Boylan-Kolchin, Bullock, and
  Kaplinghat}}]{Boylan-Kolchin:2011qkt}
\bibinfo{author}{\bibfnamefont{M.}~\bibnamefont{Boylan-Kolchin}},
  \bibinfo{author}{\bibfnamefont{J.~S.} \bibnamefont{Bullock}},
  \bibnamefont{and}
  \bibinfo{author}{\bibfnamefont{M.}~\bibnamefont{Kaplinghat}},
  \bibinfo{journal}{Mon. Not. Roy. Astron. Soc.}
  \textbf{\bibinfo{volume}{415}}, \bibinfo{pages}{L40} (\bibinfo{year}{2011}),
  \eprint{1103.0007}.

\bibitem[{\citenamefont{Garrison-Kimmel
  et~al.}(2014)\citenamefont{Garrison-Kimmel, Boylan-Kolchin, Bullock, and
  Kirby}}]{Garrison-Kimmel:2014vqa}
\bibinfo{author}{\bibfnamefont{S.}~\bibnamefont{Garrison-Kimmel}},
  \bibinfo{author}{\bibfnamefont{M.}~\bibnamefont{Boylan-Kolchin}},
  \bibinfo{author}{\bibfnamefont{J.~S.} \bibnamefont{Bullock}},
  \bibnamefont{and} \bibinfo{author}{\bibfnamefont{E.~N.} \bibnamefont{Kirby}},
  \bibinfo{journal}{Mon. Not. Roy. Astron. Soc.}
  \textbf{\bibinfo{volume}{444}}, \bibinfo{pages}{222} (\bibinfo{year}{2014}),
  \eprint{1404.5313}.

\bibitem[{\citenamefont{Kaplinghat et~al.}(2019)\citenamefont{Kaplinghat,
  Valli, and Yu}}]{Kaplinghat:2019svz}
\bibinfo{author}{\bibfnamefont{M.}~\bibnamefont{Kaplinghat}},
  \bibinfo{author}{\bibfnamefont{M.}~\bibnamefont{Valli}}, \bibnamefont{and}
  \bibinfo{author}{\bibfnamefont{H.-B.} \bibnamefont{Yu}},
  \bibinfo{journal}{Mon. Not. Roy. Astron. Soc.}
  \textbf{\bibinfo{volume}{490}}, \bibinfo{pages}{231} (\bibinfo{year}{2019}),
  \eprint{1904.04939}.

\bibitem[{\citenamefont{Pawlowski et~al.}(2015)\citenamefont{Pawlowski, Famaey,
  Merritt, and Kroupa}}]{Pawlowski:2015qta}
\bibinfo{author}{\bibfnamefont{M.~S.} \bibnamefont{Pawlowski}},
  \bibinfo{author}{\bibfnamefont{B.}~\bibnamefont{Famaey}},
  \bibinfo{author}{\bibfnamefont{D.}~\bibnamefont{Merritt}}, \bibnamefont{and}
  \bibinfo{author}{\bibfnamefont{P.}~\bibnamefont{Kroupa}},
  \bibinfo{journal}{Astrophys. J.} \textbf{\bibinfo{volume}{815}},
  \bibinfo{pages}{19} (\bibinfo{year}{2015}), \eprint{1510.08060}.

\bibitem[{\citenamefont{Ivanov et~al.}(1994)\citenamefont{Ivanov, Naselsky, and
  Novikov}}]{Ivanov:1994pa}
\bibinfo{author}{\bibfnamefont{P.}~\bibnamefont{Ivanov}},
  \bibinfo{author}{\bibfnamefont{P.}~\bibnamefont{Naselsky}}, \bibnamefont{and}
  \bibinfo{author}{\bibfnamefont{I.}~\bibnamefont{Novikov}},
  \bibinfo{journal}{Phys. Rev. D} \textbf{\bibinfo{volume}{50}},
  \bibinfo{pages}{7173} (\bibinfo{year}{1994}).

\bibitem[{\citenamefont{Clesse and Garc\'\i{}a-Bellido}(2015)}]{Clesse:2015wea}
\bibinfo{author}{\bibfnamefont{S.}~\bibnamefont{Clesse}} \bibnamefont{and}
  \bibinfo{author}{\bibfnamefont{J.}~\bibnamefont{Garc\'\i{}a-Bellido}},
  \bibinfo{journal}{Phys. Rev. D} \textbf{\bibinfo{volume}{92}},
  \bibinfo{pages}{023524} (\bibinfo{year}{2015}), \eprint{1501.07565}.

\bibitem[{\citenamefont{Carr et~al.}(2016)\citenamefont{Carr, Kuhnel, and
  Sandstad}}]{Carr:2016drx}
\bibinfo{author}{\bibfnamefont{B.}~\bibnamefont{Carr}},
  \bibinfo{author}{\bibfnamefont{F.}~\bibnamefont{Kuhnel}}, \bibnamefont{and}
  \bibinfo{author}{\bibfnamefont{M.}~\bibnamefont{Sandstad}},
  \bibinfo{journal}{Phys. Rev. D} \textbf{\bibinfo{volume}{94}},
  \bibinfo{pages}{083504} (\bibinfo{year}{2016}), \eprint{1607.06077}.

\bibitem[{\citenamefont{Carr and Kuhnel}(2020)}]{Carr:2020xqk}
\bibinfo{author}{\bibfnamefont{B.}~\bibnamefont{Carr}} \bibnamefont{and}
  \bibinfo{author}{\bibfnamefont{F.}~\bibnamefont{Kuhnel}},
  \bibinfo{journal}{Ann. Rev. Nucl. Part. Sci.} \textbf{\bibinfo{volume}{70}},
  \bibinfo{pages}{355} (\bibinfo{year}{2020}), \eprint{2006.02838}.

\bibitem[{\citenamefont{Green and Kavanagh}(2021)}]{Green:2020jor}
\bibinfo{author}{\bibfnamefont{A.~M.} \bibnamefont{Green}} \bibnamefont{and}
  \bibinfo{author}{\bibfnamefont{B.~J.} \bibnamefont{Kavanagh}},
  \bibinfo{journal}{J. Phys. G} \textbf{\bibinfo{volume}{48}},
  \bibinfo{pages}{043001} (\bibinfo{year}{2021}), \eprint{2007.10722}.

\bibitem[{\citenamefont{Duffy and van Bibber}(2009)}]{Duffy:2009ig}
\bibinfo{author}{\bibfnamefont{L.~D.} \bibnamefont{Duffy}} \bibnamefont{and}
  \bibinfo{author}{\bibfnamefont{K.}~\bibnamefont{van Bibber}},
  \bibinfo{journal}{New J. Phys.} \textbf{\bibinfo{volume}{11}},
  \bibinfo{pages}{105008} (\bibinfo{year}{2009}), \eprint{0904.3346}.

\bibitem[{\citenamefont{Marsh}(2016)}]{Marsh:2015xka}
\bibinfo{author}{\bibfnamefont{D.~J.~E.} \bibnamefont{Marsh}},
  \bibinfo{journal}{Phys. Rept.} \textbf{\bibinfo{volume}{643}},
  \bibinfo{pages}{1} (\bibinfo{year}{2016}), \eprint{1510.07633}.

\bibitem[{\citenamefont{Kim and Carosi}(2010)}]{Kim:2008hd}
\bibinfo{author}{\bibfnamefont{J.~E.} \bibnamefont{Kim}} \bibnamefont{and}
  \bibinfo{author}{\bibfnamefont{G.}~\bibnamefont{Carosi}},
  \bibinfo{journal}{Rev. Mod. Phys.} \textbf{\bibinfo{volume}{82}},
  \bibinfo{pages}{557} (\bibinfo{year}{2010}), \bibinfo{note}{[Erratum:
  Rev.Mod.Phys. 91, 049902 (2019)]}, \eprint{0807.3125}.

\bibitem[{\citenamefont{Graham et~al.}(2015)\citenamefont{Graham, Irastorza,
  Lamoreaux, Lindner, and van Bibber}}]{Graham:2015ouw}
\bibinfo{author}{\bibfnamefont{P.~W.} \bibnamefont{Graham}},
  \bibinfo{author}{\bibfnamefont{I.~G.} \bibnamefont{Irastorza}},
  \bibinfo{author}{\bibfnamefont{S.~K.} \bibnamefont{Lamoreaux}},
  \bibinfo{author}{\bibfnamefont{A.}~\bibnamefont{Lindner}}, \bibnamefont{and}
  \bibinfo{author}{\bibfnamefont{K.~A.} \bibnamefont{van Bibber}},
  \bibinfo{journal}{Ann. Rev. Nucl. Part. Sci.} \textbf{\bibinfo{volume}{65}},
  \bibinfo{pages}{485} (\bibinfo{year}{2015}), \eprint{1602.00039}.

\bibitem[{\citenamefont{Irastorza and Redondo}(2018)}]{Irastorza:2018dyq}
\bibinfo{author}{\bibfnamefont{I.~G.} \bibnamefont{Irastorza}}
  \bibnamefont{and} \bibinfo{author}{\bibfnamefont{J.}~\bibnamefont{Redondo}},
  \bibinfo{journal}{Prog. Part. Nucl. Phys.} \textbf{\bibinfo{volume}{102}},
  \bibinfo{pages}{89} (\bibinfo{year}{2018}), \eprint{1801.08127}.

\bibitem[{\citenamefont{Di~Luzio et~al.}(2020)\citenamefont{Di~Luzio,
  Giannotti, Nardi, and Visinelli}}]{DiLuzio:2020wdo}
\bibinfo{author}{\bibfnamefont{L.}~\bibnamefont{Di~Luzio}},
  \bibinfo{author}{\bibfnamefont{M.}~\bibnamefont{Giannotti}},
  \bibinfo{author}{\bibfnamefont{E.}~\bibnamefont{Nardi}}, \bibnamefont{and}
  \bibinfo{author}{\bibfnamefont{L.}~\bibnamefont{Visinelli}},
  \bibinfo{journal}{Phys. Rept.} \textbf{\bibinfo{volume}{870}},
  \bibinfo{pages}{1} (\bibinfo{year}{2020}), \eprint{2003.01100}.

\bibitem[{\citenamefont{Dodelson and Widrow}(1994)}]{Dodelson:1993je}
\bibinfo{author}{\bibfnamefont{S.}~\bibnamefont{Dodelson}} \bibnamefont{and}
  \bibinfo{author}{\bibfnamefont{L.~M.} \bibnamefont{Widrow}},
  \bibinfo{journal}{Phys. Rev. Lett.} \textbf{\bibinfo{volume}{72}},
  \bibinfo{pages}{17} (\bibinfo{year}{1994}), \eprint{hep-ph/9303287}.

\bibitem[{\citenamefont{Shi and Fuller}(1999)}]{Shi:1998km}
\bibinfo{author}{\bibfnamefont{X.-D.} \bibnamefont{Shi}} \bibnamefont{and}
  \bibinfo{author}{\bibfnamefont{G.~M.} \bibnamefont{Fuller}},
  \bibinfo{journal}{Phys. Rev. Lett.} \textbf{\bibinfo{volume}{82}},
  \bibinfo{pages}{2832} (\bibinfo{year}{1999}), \eprint{astro-ph/9810076}.

\bibitem[{\citenamefont{Boyarsky et~al.}(2009)\citenamefont{Boyarsky,
  Ruchayskiy, and Shaposhnikov}}]{Boyarsky:2009ix}
\bibinfo{author}{\bibfnamefont{A.}~\bibnamefont{Boyarsky}},
  \bibinfo{author}{\bibfnamefont{O.}~\bibnamefont{Ruchayskiy}},
  \bibnamefont{and}
  \bibinfo{author}{\bibfnamefont{M.}~\bibnamefont{Shaposhnikov}},
  \bibinfo{journal}{Ann. Rev. Nucl. Part. Sci.} \textbf{\bibinfo{volume}{59}},
  \bibinfo{pages}{191} (\bibinfo{year}{2009}), \eprint{0901.0011}.

\bibitem[{\citenamefont{Kusenko}(2009)}]{Kusenko:2009up}
\bibinfo{author}{\bibfnamefont{A.}~\bibnamefont{Kusenko}},
  \bibinfo{journal}{Phys. Rept.} \textbf{\bibinfo{volume}{481}},
  \bibinfo{pages}{1} (\bibinfo{year}{2009}), \eprint{0906.2968}.

\bibitem[{\citenamefont{Feng}(2010)}]{Feng:2010gw}
\bibinfo{author}{\bibfnamefont{J.~L.} \bibnamefont{Feng}},
  \bibinfo{journal}{Ann. Rev. Astron. Astrophys.}
  \textbf{\bibinfo{volume}{48}}, \bibinfo{pages}{495} (\bibinfo{year}{2010}),
  \eprint{1003.0904}.

\bibitem[{\citenamefont{Abazajian et~al.}(2012)}]{Abazajian:2012ys}
\bibinfo{author}{\bibfnamefont{K.~N.} \bibnamefont{Abazajian}}
  \bibnamefont{et~al.} (\bibinfo{year}{2012}), \eprint{1204.5379}.

\bibitem[{\citenamefont{Hu et~al.}(2000)\citenamefont{Hu, Barkana, and
  Gruzinov}}]{Hu:2000ke}
\bibinfo{author}{\bibfnamefont{W.}~\bibnamefont{Hu}},
  \bibinfo{author}{\bibfnamefont{R.}~\bibnamefont{Barkana}}, \bibnamefont{and}
  \bibinfo{author}{\bibfnamefont{A.}~\bibnamefont{Gruzinov}},
  \bibinfo{journal}{Phys. Rev. Lett.} \textbf{\bibinfo{volume}{85}},
  \bibinfo{pages}{1158} (\bibinfo{year}{2000}), \eprint{astro-ph/0003365}.

\bibitem[{\citenamefont{Hui et~al.}(2017)\citenamefont{Hui, Ostriker, Tremaine,
  and Witten}}]{Hui:2016ltb}
\bibinfo{author}{\bibfnamefont{L.}~\bibnamefont{Hui}},
  \bibinfo{author}{\bibfnamefont{J.~P.} \bibnamefont{Ostriker}},
  \bibinfo{author}{\bibfnamefont{S.}~\bibnamefont{Tremaine}}, \bibnamefont{and}
  \bibinfo{author}{\bibfnamefont{E.}~\bibnamefont{Witten}},
  \bibinfo{journal}{Phys. Rev. D} \textbf{\bibinfo{volume}{95}},
  \bibinfo{pages}{043541} (\bibinfo{year}{2017}), \eprint{1610.08297}.

\bibitem[{\citenamefont{Knapen et~al.}(2017)\citenamefont{Knapen, Lin, and
  Zurek}}]{Knapen:2017xzo}
\bibinfo{author}{\bibfnamefont{S.}~\bibnamefont{Knapen}},
  \bibinfo{author}{\bibfnamefont{T.}~\bibnamefont{Lin}}, \bibnamefont{and}
  \bibinfo{author}{\bibfnamefont{K.~M.} \bibnamefont{Zurek}},
  \bibinfo{journal}{Phys. Rev. D} \textbf{\bibinfo{volume}{96}},
  \bibinfo{pages}{115021} (\bibinfo{year}{2017}), \eprint{1709.07882}.

\bibitem[{\citenamefont{Ferreira}(2021)}]{Ferreira:2020fam}
\bibinfo{author}{\bibfnamefont{E.~G.~M.} \bibnamefont{Ferreira}},
  \bibinfo{journal}{Astron. Astrophys. Rev.} \textbf{\bibinfo{volume}{29}},
  \bibinfo{pages}{7} (\bibinfo{year}{2021}), \eprint{2005.03254}.

\bibitem[{\citenamefont{Hui}(2021)}]{Hui:2021tkt}
\bibinfo{author}{\bibfnamefont{L.}~\bibnamefont{Hui}}, \bibinfo{journal}{Ann.
  Rev. Astron. Astrophys.} \textbf{\bibinfo{volume}{59}}, \bibinfo{pages}{247}
  (\bibinfo{year}{2021}), \eprint{2101.11735}.

\bibitem[{\citenamefont{Goodman}(2000)}]{Goodman:2000tg}
\bibinfo{author}{\bibfnamefont{J.}~\bibnamefont{Goodman}},
  \bibinfo{journal}{New Astron.} \textbf{\bibinfo{volume}{5}},
  \bibinfo{pages}{103} (\bibinfo{year}{2000}), \eprint{astro-ph/0003018}.

\bibitem[{\citenamefont{Schive et~al.}(2014{\natexlab{a}})\citenamefont{Schive,
  Chiueh, and Broadhurst}}]{Schive:2014dra}
\bibinfo{author}{\bibfnamefont{H.-Y.} \bibnamefont{Schive}},
  \bibinfo{author}{\bibfnamefont{T.}~\bibnamefont{Chiueh}}, \bibnamefont{and}
  \bibinfo{author}{\bibfnamefont{T.}~\bibnamefont{Broadhurst}},
  \bibinfo{journal}{Nature Phys.} \textbf{\bibinfo{volume}{10}},
  \bibinfo{pages}{496} (\bibinfo{year}{2014}{\natexlab{a}}),
  \eprint{1406.6586}.

\bibitem[{\citenamefont{Schive et~al.}(2014{\natexlab{b}})\citenamefont{Schive,
  Liao, Woo, Wong, Chiueh, Broadhurst, and Hwang}}]{Schive:2014hza}
\bibinfo{author}{\bibfnamefont{H.-Y.} \bibnamefont{Schive}},
  \bibinfo{author}{\bibfnamefont{M.-H.} \bibnamefont{Liao}},
  \bibinfo{author}{\bibfnamefont{T.-P.} \bibnamefont{Woo}},
  \bibinfo{author}{\bibfnamefont{S.-K.} \bibnamefont{Wong}},
  \bibinfo{author}{\bibfnamefont{T.}~\bibnamefont{Chiueh}},
  \bibinfo{author}{\bibfnamefont{T.}~\bibnamefont{Broadhurst}},
  \bibnamefont{and} \bibinfo{author}{\bibfnamefont{W.~Y.~P.}
  \bibnamefont{Hwang}}, \bibinfo{journal}{Phys. Rev. Lett.}
  \textbf{\bibinfo{volume}{113}}, \bibinfo{pages}{261302}
  (\bibinfo{year}{2014}{\natexlab{b}}), \eprint{1407.7762}.

\bibitem[{\citenamefont{Arbey et~al.}(2001)\citenamefont{Arbey, Lesgourgues,
  and Salati}}]{Arbey:2001qi}
\bibinfo{author}{\bibfnamefont{A.}~\bibnamefont{Arbey}},
  \bibinfo{author}{\bibfnamefont{J.}~\bibnamefont{Lesgourgues}},
  \bibnamefont{and} \bibinfo{author}{\bibfnamefont{P.}~\bibnamefont{Salati}},
  \bibinfo{journal}{Phys. Rev.} \textbf{\bibinfo{volume}{D64}},
  \bibinfo{pages}{123528} (\bibinfo{year}{2001}), \eprint{astro-ph/0105564}.

\bibitem[{\citenamefont{Chavanis}(2011)}]{Chavanis:2011zi}
\bibinfo{author}{\bibfnamefont{P.-H.} \bibnamefont{Chavanis}},
  \bibinfo{journal}{Phys. Rev.} \textbf{\bibinfo{volume}{D84}},
  \bibinfo{pages}{043531} (\bibinfo{year}{2011}), \eprint{1103.2050}.

\bibitem[{\citenamefont{Chavanis and Delfini}(2011)}]{Chavanis:2011zm}
\bibinfo{author}{\bibfnamefont{P.~H.} \bibnamefont{Chavanis}} \bibnamefont{and}
  \bibinfo{author}{\bibfnamefont{L.}~\bibnamefont{Delfini}},
  \bibinfo{journal}{Phys. Rev.} \textbf{\bibinfo{volume}{D84}},
  \bibinfo{pages}{043532} (\bibinfo{year}{2011}), \eprint{1103.2054}.

\bibitem[{\citenamefont{Marsh and Pop}(2015)}]{Marsh:2015wka}
\bibinfo{author}{\bibfnamefont{D.~J.~E.} \bibnamefont{Marsh}} \bibnamefont{and}
  \bibinfo{author}{\bibfnamefont{A.-R.} \bibnamefont{Pop}},
  \bibinfo{journal}{Mon. Not. Roy. Astron. Soc.}
  \textbf{\bibinfo{volume}{451}}, \bibinfo{pages}{2479} (\bibinfo{year}{2015}),
  \eprint{1502.03456}.

\bibitem[{\citenamefont{Calabrese and Spergel}(2016)}]{Calabrese:2016hmp}
\bibinfo{author}{\bibfnamefont{E.}~\bibnamefont{Calabrese}} \bibnamefont{and}
  \bibinfo{author}{\bibfnamefont{D.~N.} \bibnamefont{Spergel}},
  \bibinfo{journal}{Mon. Not. Roy. Astron. Soc.}
  \textbf{\bibinfo{volume}{460}}, \bibinfo{pages}{4397} (\bibinfo{year}{2016}),
  \eprint{1603.07321}.

\bibitem[{\citenamefont{Chen et~al.}(2017)\citenamefont{Chen, Schive, and
  Chiueh}}]{Chen:2016unw}
\bibinfo{author}{\bibfnamefont{S.-R.} \bibnamefont{Chen}},
  \bibinfo{author}{\bibfnamefont{H.-Y.} \bibnamefont{Schive}},
  \bibnamefont{and} \bibinfo{author}{\bibfnamefont{T.}~\bibnamefont{Chiueh}},
  \bibinfo{journal}{Mon. Not. Roy. Astron. Soc.}
  \textbf{\bibinfo{volume}{468}}, \bibinfo{pages}{1338} (\bibinfo{year}{2017}),
  \eprint{1606.09030}.

\bibitem[{\citenamefont{Schwabe et~al.}(2016)\citenamefont{Schwabe, Niemeyer,
  and Engels}}]{Schwabe:2016rze}
\bibinfo{author}{\bibfnamefont{B.}~\bibnamefont{Schwabe}},
  \bibinfo{author}{\bibfnamefont{J.~C.} \bibnamefont{Niemeyer}},
  \bibnamefont{and} \bibinfo{author}{\bibfnamefont{J.~F.}
  \bibnamefont{Engels}}, \bibinfo{journal}{Phys. Rev.}
  \textbf{\bibinfo{volume}{D94}}, \bibinfo{pages}{043513}
  (\bibinfo{year}{2016}), \eprint{1606.05151}.

\bibitem[{\citenamefont{Veltmaat and Niemeyer}(2016)}]{Veltmaat:2016rxo}
\bibinfo{author}{\bibfnamefont{J.}~\bibnamefont{Veltmaat}} \bibnamefont{and}
  \bibinfo{author}{\bibfnamefont{J.~C.} \bibnamefont{Niemeyer}},
  \bibinfo{journal}{Phys. Rev.} \textbf{\bibinfo{volume}{D94}},
  \bibinfo{pages}{123523} (\bibinfo{year}{2016}), \eprint{1608.00802}.

\bibitem[{\citenamefont{Gonz\'alez-Morales
  et~al.}(2017)\citenamefont{Gonz\'alez-Morales, Marsh, Pe\~narrubia, and
  Ure\~na L\'opez}}]{Gonzalez-Morales:2016yaf}
\bibinfo{author}{\bibfnamefont{A.~X.} \bibnamefont{Gonz\'alez-Morales}},
  \bibinfo{author}{\bibfnamefont{D.~J.~E.} \bibnamefont{Marsh}},
  \bibinfo{author}{\bibfnamefont{J.}~\bibnamefont{Pe\~narrubia}},
  \bibnamefont{and} \bibinfo{author}{\bibfnamefont{L.~A.} \bibnamefont{Ure\~na
  L\'opez}}, \bibinfo{journal}{Mon. Not. Roy. Astron. Soc.}
  \textbf{\bibinfo{volume}{472}}, \bibinfo{pages}{1346} (\bibinfo{year}{2017}),
  \eprint{1609.05856}.

\bibitem[{\citenamefont{Robles and Matos}(2012)}]{Robles:2012uy}
\bibinfo{author}{\bibfnamefont{V.~H.} \bibnamefont{Robles}} \bibnamefont{and}
  \bibinfo{author}{\bibfnamefont{T.}~\bibnamefont{Matos}},
  \bibinfo{journal}{Mon. Not. Roy. Astron. Soc.}
  \textbf{\bibinfo{volume}{422}}, \bibinfo{pages}{282} (\bibinfo{year}{2012}),
  \eprint{1201.3032}.

\bibitem[{\citenamefont{Bernal et~al.}(2018)\citenamefont{Bernal,
  Fern{\'a}ndez-Hern{\'a}ndez, Matos, and
  Rodr{\'\i}guez-Meza}}]{Bernal:2017oih}
\bibinfo{author}{\bibfnamefont{T.}~\bibnamefont{Bernal}},
  \bibinfo{author}{\bibfnamefont{L.~M.}
  \bibnamefont{Fern{\'a}ndez-Hern{\'a}ndez}},
  \bibinfo{author}{\bibfnamefont{T.}~\bibnamefont{Matos}}, \bibnamefont{and}
  \bibinfo{author}{\bibfnamefont{M.~A.} \bibnamefont{Rodr{\'\i}guez-Meza}},
  \bibinfo{journal}{Mon. Not. Roy. Astron. Soc.}
  \textbf{\bibinfo{volume}{475}}, \bibinfo{pages}{1447} (\bibinfo{year}{2018}),
  \eprint{1701.00912}.

\bibitem[{\citenamefont{Mocz et~al.}(2017)\citenamefont{Mocz, Vogelsberger,
  Robles, Zavala, Boylan-Kolchin, Fialkov, and Hernquist}}]{Mocz:2017wlg}
\bibinfo{author}{\bibfnamefont{P.}~\bibnamefont{Mocz}},
  \bibinfo{author}{\bibfnamefont{M.}~\bibnamefont{Vogelsberger}},
  \bibinfo{author}{\bibfnamefont{V.~H.} \bibnamefont{Robles}},
  \bibinfo{author}{\bibfnamefont{J.}~\bibnamefont{Zavala}},
  \bibinfo{author}{\bibfnamefont{M.}~\bibnamefont{Boylan-Kolchin}},
  \bibinfo{author}{\bibfnamefont{A.}~\bibnamefont{Fialkov}}, \bibnamefont{and}
  \bibinfo{author}{\bibfnamefont{L.}~\bibnamefont{Hernquist}},
  \bibinfo{journal}{Mon. Not. Roy. Astron. Soc.}
  \textbf{\bibinfo{volume}{471}}, \bibinfo{pages}{4559} (\bibinfo{year}{2017}),
  \eprint{1705.05845}.

\bibitem[{\citenamefont{Mukaida et~al.}(2017)\citenamefont{Mukaida, Takimoto,
  and Yamada}}]{Mukaida:2016hwd}
\bibinfo{author}{\bibfnamefont{K.}~\bibnamefont{Mukaida}},
  \bibinfo{author}{\bibfnamefont{M.}~\bibnamefont{Takimoto}}, \bibnamefont{and}
  \bibinfo{author}{\bibfnamefont{M.}~\bibnamefont{Yamada}},
  \bibinfo{journal}{JHEP} \textbf{\bibinfo{volume}{03}}, \bibinfo{pages}{122}
  (\bibinfo{year}{2017}), \eprint{1612.07750}.

\bibitem[{\citenamefont{Vicens et~al.}(2018)\citenamefont{Vicens, Salvado, and
  Miralda-Escud{\'e}}}]{Vicens:2018kdk}
\bibinfo{author}{\bibfnamefont{J.}~\bibnamefont{Vicens}},
  \bibinfo{author}{\bibfnamefont{J.}~\bibnamefont{Salvado}}, \bibnamefont{and}
  \bibinfo{author}{\bibfnamefont{J.}~\bibnamefont{Miralda-Escud{\'e}}}
  (\bibinfo{year}{2018}), \eprint{1802.10513}.

\bibitem[{\citenamefont{Bar et~al.}(2018)\citenamefont{Bar, Blas, Blum, and
  Sibiryakov}}]{Bar:2018acw}
\bibinfo{author}{\bibfnamefont{N.}~\bibnamefont{Bar}},
  \bibinfo{author}{\bibfnamefont{D.}~\bibnamefont{Blas}},
  \bibinfo{author}{\bibfnamefont{K.}~\bibnamefont{Blum}}, \bibnamefont{and}
  \bibinfo{author}{\bibfnamefont{S.}~\bibnamefont{Sibiryakov}},
  \bibinfo{journal}{Phys. Rev.} \textbf{\bibinfo{volume}{D98}},
  \bibinfo{pages}{083027} (\bibinfo{year}{2018}), \eprint{1805.00122}.

\bibitem[{\citenamefont{Eby et~al.}(2019)\citenamefont{Eby, Mukaida, Takimoto,
  Wijewardhana, and Yamada}}]{Eby:2018ufi}
\bibinfo{author}{\bibfnamefont{J.}~\bibnamefont{Eby}},
  \bibinfo{author}{\bibfnamefont{K.}~\bibnamefont{Mukaida}},
  \bibinfo{author}{\bibfnamefont{M.}~\bibnamefont{Takimoto}},
  \bibinfo{author}{\bibfnamefont{L.~C.~R.} \bibnamefont{Wijewardhana}},
  \bibnamefont{and} \bibinfo{author}{\bibfnamefont{M.}~\bibnamefont{Yamada}},
  \bibinfo{journal}{Phys. Rev.} \textbf{\bibinfo{volume}{D99}},
  \bibinfo{pages}{123503} (\bibinfo{year}{2019}), \eprint{1807.09795}.

\bibitem[{\citenamefont{Bar-Or et~al.}(2019)\citenamefont{Bar-Or, Fouvry, and
  Tremaine}}]{Bar-Or:2018pxz}
\bibinfo{author}{\bibfnamefont{B.}~\bibnamefont{Bar-Or}},
  \bibinfo{author}{\bibfnamefont{J.-B.} \bibnamefont{Fouvry}},
  \bibnamefont{and} \bibinfo{author}{\bibfnamefont{S.}~\bibnamefont{Tremaine}},
  \bibinfo{journal}{Astrophys. J.} \textbf{\bibinfo{volume}{871}},
  \bibinfo{pages}{28} (\bibinfo{year}{2019}), \eprint{1809.07673}.

\bibitem[{\citenamefont{Marsh and Niemeyer}(2018)}]{Marsh:2018zyw}
\bibinfo{author}{\bibfnamefont{D.~J.~E.} \bibnamefont{Marsh}} \bibnamefont{and}
  \bibinfo{author}{\bibfnamefont{J.~C.} \bibnamefont{Niemeyer}}
  (\bibinfo{year}{2018}), \eprint{1810.08543}.

\bibitem[{\citenamefont{Chavanis}(2018)}]{Chavanis:2018pkx}
\bibinfo{author}{\bibfnamefont{P.-H.} \bibnamefont{Chavanis}}
  (\bibinfo{year}{2018}), \eprint{1810.08948}.

\bibitem[{\citenamefont{Emami et~al.}(2018)\citenamefont{Emami, Broadhurst,
  Smoot, Chiueh, and Luu}}]{Emami:2018rxq}
\bibinfo{author}{\bibfnamefont{R.}~\bibnamefont{Emami}},
  \bibinfo{author}{\bibfnamefont{T.}~\bibnamefont{Broadhurst}},
  \bibinfo{author}{\bibfnamefont{G.}~\bibnamefont{Smoot}},
  \bibinfo{author}{\bibfnamefont{T.}~\bibnamefont{Chiueh}}, \bibnamefont{and}
  \bibinfo{author}{\bibfnamefont{H.~N.} \bibnamefont{Luu}}
  (\bibinfo{year}{2018}), \eprint{1806.04518}.

\bibitem[{\citenamefont{Levkov et~al.}(2018)\citenamefont{Levkov, Panin, and
  Tkachev}}]{Levkov:2018kau}
\bibinfo{author}{\bibfnamefont{D.~G.} \bibnamefont{Levkov}},
  \bibinfo{author}{\bibfnamefont{A.~G.} \bibnamefont{Panin}}, \bibnamefont{and}
  \bibinfo{author}{\bibfnamefont{I.~I.} \bibnamefont{Tkachev}},
  \bibinfo{journal}{Phys. Rev. Lett.} \textbf{\bibinfo{volume}{121}},
  \bibinfo{pages}{151301} (\bibinfo{year}{2018}), \eprint{1804.05857}.

\bibitem[{\citenamefont{Broadhurst et~al.}(2019)\citenamefont{Broadhurst,
  de~Martino, Luu, Smoot, and Tye}}]{Broadhurst:2019fsl}
\bibinfo{author}{\bibfnamefont{T.}~\bibnamefont{Broadhurst}},
  \bibinfo{author}{\bibfnamefont{I.}~\bibnamefont{de~Martino}},
  \bibinfo{author}{\bibfnamefont{H.~N.} \bibnamefont{Luu}},
  \bibinfo{author}{\bibfnamefont{G.~F.} \bibnamefont{Smoot}}, \bibnamefont{and}
  \bibinfo{author}{\bibfnamefont{S.~H.~H.} \bibnamefont{Tye}}
  (\bibinfo{year}{2019}), \eprint{1902.10488}.

\bibitem[{\citenamefont{Hayashi and Obata}(2019)}]{Hayashi:2019ynr}
\bibinfo{author}{\bibfnamefont{K.}~\bibnamefont{Hayashi}} \bibnamefont{and}
  \bibinfo{author}{\bibfnamefont{I.}~\bibnamefont{Obata}}
  (\bibinfo{year}{2019}), \eprint{1902.03054}.

\bibitem[{\citenamefont{Bar et~al.}(2019{\natexlab{a}})\citenamefont{Bar, Blum,
  Eby, and Sato}}]{Bar:2019bqz}
\bibinfo{author}{\bibfnamefont{N.}~\bibnamefont{Bar}},
  \bibinfo{author}{\bibfnamefont{K.}~\bibnamefont{Blum}},
  \bibinfo{author}{\bibfnamefont{J.}~\bibnamefont{Eby}}, \bibnamefont{and}
  \bibinfo{author}{\bibfnamefont{R.}~\bibnamefont{Sato}},
  \bibinfo{journal}{Phys. Rev.} \textbf{\bibinfo{volume}{D99}},
  \bibinfo{pages}{103020} (\bibinfo{year}{2019}{\natexlab{a}}),
  \eprint{1903.03402}.

\bibitem[{\citenamefont{Garc\'\i{}a et~al.}(2023)\citenamefont{Garc\'\i{}a,
  Brax, and Valageas}}]{Garcia:2023abs}
\bibinfo{author}{\bibfnamefont{R.~G.} \bibnamefont{Garc\'\i{}a}},
  \bibinfo{author}{\bibfnamefont{P.}~\bibnamefont{Brax}}, \bibnamefont{and}
  \bibinfo{author}{\bibfnamefont{P.}~\bibnamefont{Valageas}}
  (\bibinfo{year}{2023}), \eprint{2304.10221}.

\bibitem[{\citenamefont{Johnson and Kamionkowski}(2008)}]{Johnson:2008se}
\bibinfo{author}{\bibfnamefont{M.~C.} \bibnamefont{Johnson}} \bibnamefont{and}
  \bibinfo{author}{\bibfnamefont{M.}~\bibnamefont{Kamionkowski}},
  \bibinfo{journal}{Phys. Rev.} \textbf{\bibinfo{volume}{D78}},
  \bibinfo{pages}{063010} (\bibinfo{year}{2008}), \eprint{0805.1748}.

\bibitem[{\citenamefont{Hwang and Noh}(2009)}]{Hwang:2009js}
\bibinfo{author}{\bibfnamefont{J.-c.} \bibnamefont{Hwang}} \bibnamefont{and}
  \bibinfo{author}{\bibfnamefont{H.}~\bibnamefont{Noh}},
  \bibinfo{journal}{Phys. Lett.} \textbf{\bibinfo{volume}{B680}},
  \bibinfo{pages}{1} (\bibinfo{year}{2009}), \eprint{0902.4738}.

\bibitem[{\citenamefont{Park et~al.}(2012)\citenamefont{Park, Hwang, and
  Noh}}]{Park:2012ru}
\bibinfo{author}{\bibfnamefont{C.-G.} \bibnamefont{Park}},
  \bibinfo{author}{\bibfnamefont{J.-c.} \bibnamefont{Hwang}}, \bibnamefont{and}
  \bibinfo{author}{\bibfnamefont{H.}~\bibnamefont{Noh}},
  \bibinfo{journal}{Phys. Rev.} \textbf{\bibinfo{volume}{D86}},
  \bibinfo{pages}{083535} (\bibinfo{year}{2012}), \eprint{1207.3124}.

\bibitem[{\citenamefont{Hlozek et~al.}(2015)\citenamefont{Hlozek, Grin, Marsh,
  and Ferreira}}]{Hlozek:2014lca}
\bibinfo{author}{\bibfnamefont{R.}~\bibnamefont{Hlozek}},
  \bibinfo{author}{\bibfnamefont{D.}~\bibnamefont{Grin}},
  \bibinfo{author}{\bibfnamefont{D.~J.~E.} \bibnamefont{Marsh}},
  \bibnamefont{and} \bibinfo{author}{\bibfnamefont{P.~G.}
  \bibnamefont{Ferreira}}, \bibinfo{journal}{Phys. Rev.}
  \textbf{\bibinfo{volume}{D91}}, \bibinfo{pages}{103512}
  (\bibinfo{year}{2015}), \eprint{1410.2896}.

\bibitem[{\citenamefont{Cembranos et~al.}(2016)\citenamefont{Cembranos, Maroto,
  and N{\'u}{\~n}ez~Jare{\~n}o}}]{Cembranos:2015oya}
\bibinfo{author}{\bibfnamefont{J.~A.~R.} \bibnamefont{Cembranos}},
  \bibinfo{author}{\bibfnamefont{A.~L.} \bibnamefont{Maroto}},
  \bibnamefont{and} \bibinfo{author}{\bibfnamefont{S.~J.}
  \bibnamefont{N{\'u}{\~n}ez~Jare{\~n}o}}, \bibinfo{journal}{JHEP}
  \textbf{\bibinfo{volume}{03}}, \bibinfo{pages}{013} (\bibinfo{year}{2016}),
  \eprint{1509.08819}.

\bibitem[{\citenamefont{Ure{\~n}a-L{\'o}pez and
  Gonzalez-Morales}(2016)}]{Urena-Lopez:2015gur}
\bibinfo{author}{\bibfnamefont{L.~A.} \bibnamefont{Ure{\~n}a-L{\'o}pez}}
  \bibnamefont{and} \bibinfo{author}{\bibfnamefont{A.~X.}
  \bibnamefont{Gonzalez-Morales}}, \bibinfo{journal}{JCAP}
  \textbf{\bibinfo{volume}{1607}}, \bibinfo{pages}{048} (\bibinfo{year}{2016}),
  \eprint{1511.08195}.

\bibitem[{\citenamefont{Ure\~na L\'opez}(2019)}]{Urena-Lopez:2019kud}
\bibinfo{author}{\bibfnamefont{L.~A.} \bibnamefont{Ure\~na L\'opez}},
  \bibinfo{journal}{Front. Astron. Space Sci.} \textbf{\bibinfo{volume}{6}},
  \bibinfo{pages}{47} (\bibinfo{year}{2019}).

\bibitem[{\citenamefont{Ir\v{s}i\v{c} et~al.}(2017)\citenamefont{Ir\v{s}i\v{c},
  Viel, Haehnelt, Bolton, and Becker}}]{Irsic:2017yje}
\bibinfo{author}{\bibfnamefont{V.}~\bibnamefont{Ir\v{s}i\v{c}}},
  \bibinfo{author}{\bibfnamefont{M.}~\bibnamefont{Viel}},
  \bibinfo{author}{\bibfnamefont{M.~G.} \bibnamefont{Haehnelt}},
  \bibinfo{author}{\bibfnamefont{J.~S.} \bibnamefont{Bolton}},
  \bibnamefont{and} \bibinfo{author}{\bibfnamefont{G.~D.}
  \bibnamefont{Becker}}, \bibinfo{journal}{Phys. Rev. Lett.}
  \textbf{\bibinfo{volume}{119}}, \bibinfo{pages}{031302}
  (\bibinfo{year}{2017}), \eprint{1703.04683}.

\bibitem[{\citenamefont{Armengaud et~al.}(2017)\citenamefont{Armengaud,
  Palanque-Delabrouille, Y{\`e}che, Marsh, and Baur}}]{Armengaud:2017nkf}
\bibinfo{author}{\bibfnamefont{E.}~\bibnamefont{Armengaud}},
  \bibinfo{author}{\bibfnamefont{N.}~\bibnamefont{Palanque-Delabrouille}},
  \bibinfo{author}{\bibfnamefont{C.}~\bibnamefont{Y{\`e}che}},
  \bibinfo{author}{\bibfnamefont{D.~J.~E.} \bibnamefont{Marsh}},
  \bibnamefont{and} \bibinfo{author}{\bibfnamefont{J.}~\bibnamefont{Baur}},
  \bibinfo{journal}{Mon. Not. Roy. Astron. Soc.}
  \textbf{\bibinfo{volume}{471}}, \bibinfo{pages}{4606} (\bibinfo{year}{2017}),
  \eprint{1703.09126}.

\bibitem[{\citenamefont{Zhang et~al.}(2018)\citenamefont{Zhang, Kuo, Liu, Tsai,
  Cheung, and Chu}}]{Zhang:2017chj}
\bibinfo{author}{\bibfnamefont{J.}~\bibnamefont{Zhang}},
  \bibinfo{author}{\bibfnamefont{J.-L.} \bibnamefont{Kuo}},
  \bibinfo{author}{\bibfnamefont{H.}~\bibnamefont{Liu}},
  \bibinfo{author}{\bibfnamefont{Y.-L.~S.} \bibnamefont{Tsai}},
  \bibinfo{author}{\bibfnamefont{K.}~\bibnamefont{Cheung}}, \bibnamefont{and}
  \bibinfo{author}{\bibfnamefont{M.-C.} \bibnamefont{Chu}},
  \bibinfo{journal}{Astrophys. J.} \textbf{\bibinfo{volume}{863}},
  \bibinfo{pages}{73} (\bibinfo{year}{2018}), \eprint{1708.04389}.

\bibitem[{\citenamefont{Bar et~al.}(2022)\citenamefont{Bar, Blum, and
  Sun}}]{Bar:2021kti}
\bibinfo{author}{\bibfnamefont{N.}~\bibnamefont{Bar}},
  \bibinfo{author}{\bibfnamefont{K.}~\bibnamefont{Blum}}, \bibnamefont{and}
  \bibinfo{author}{\bibfnamefont{C.}~\bibnamefont{Sun}},
  \bibinfo{journal}{Phys. Rev. D} \textbf{\bibinfo{volume}{105}},
  \bibinfo{pages}{083015} (\bibinfo{year}{2022}), \eprint{2111.03070}.

\bibitem[{\citenamefont{Bento et~al.}(2000)\citenamefont{Bento, Bertolami,
  Rosenfeld, and Teodoro}}]{Bento:2000ah}
\bibinfo{author}{\bibfnamefont{M.~C.} \bibnamefont{Bento}},
  \bibinfo{author}{\bibfnamefont{O.}~\bibnamefont{Bertolami}},
  \bibinfo{author}{\bibfnamefont{R.}~\bibnamefont{Rosenfeld}},
  \bibnamefont{and} \bibinfo{author}{\bibfnamefont{L.}~\bibnamefont{Teodoro}},
  \bibinfo{journal}{Phys. Rev. D} \textbf{\bibinfo{volume}{62}},
  \bibinfo{pages}{041302} (\bibinfo{year}{2000}), \eprint{astro-ph/0003350}.

\bibitem[{\citenamefont{Riotto and Tkachev}(2000)}]{Riotto:2000kh}
\bibinfo{author}{\bibfnamefont{A.}~\bibnamefont{Riotto}} \bibnamefont{and}
  \bibinfo{author}{\bibfnamefont{I.}~\bibnamefont{Tkachev}},
  \bibinfo{journal}{Phys. Lett.} \textbf{\bibinfo{volume}{B484}},
  \bibinfo{pages}{177} (\bibinfo{year}{2000}), \eprint{astro-ph/0003388}.

\bibitem[{\citenamefont{Fregolente and Tonasse}(2003)}]{Fregolente:2002nx}
\bibinfo{author}{\bibfnamefont{D.}~\bibnamefont{Fregolente}} \bibnamefont{and}
  \bibinfo{author}{\bibfnamefont{M.~D.} \bibnamefont{Tonasse}},
  \bibinfo{journal}{Phys. Lett. B} \textbf{\bibinfo{volume}{555}},
  \bibinfo{pages}{7} (\bibinfo{year}{2003}), \eprint{hep-ph/0209119}.

\bibitem[{\citenamefont{Li et~al.}(2014)\citenamefont{Li, Rindler-Daller, and
  Shapiro}}]{Li:2013nal}
\bibinfo{author}{\bibfnamefont{B.}~\bibnamefont{Li}},
  \bibinfo{author}{\bibfnamefont{T.}~\bibnamefont{Rindler-Daller}},
  \bibnamefont{and} \bibinfo{author}{\bibfnamefont{P.~R.}
  \bibnamefont{Shapiro}}, \bibinfo{journal}{Phys. Rev.}
  \textbf{\bibinfo{volume}{D89}}, \bibinfo{pages}{083536}
  (\bibinfo{year}{2014}), \eprint{1310.6061}.

\bibitem[{\citenamefont{Su{\'a}rez and Chavanis}(2015)}]{Suarez:2015fga}
\bibinfo{author}{\bibfnamefont{A.}~\bibnamefont{Su{\'a}rez}} \bibnamefont{and}
  \bibinfo{author}{\bibfnamefont{P.-H.} \bibnamefont{Chavanis}},
  \bibinfo{journal}{Phys. Rev.} \textbf{\bibinfo{volume}{D92}},
  \bibinfo{pages}{023510} (\bibinfo{year}{2015}), \eprint{1503.07437}.

\bibitem[{\citenamefont{Su{\'a}rez and Chavanis}(2017)}]{Suarez:2016eez}
\bibinfo{author}{\bibfnamefont{A.}~\bibnamefont{Su{\'a}rez}} \bibnamefont{and}
  \bibinfo{author}{\bibfnamefont{P.-H.} \bibnamefont{Chavanis}},
  \bibinfo{journal}{Phys. Rev.} \textbf{\bibinfo{volume}{D95}},
  \bibinfo{pages}{063515} (\bibinfo{year}{2017}), \eprint{1608.08624}.

\bibitem[{\citenamefont{Su{\'a}rez and Chavanis}(2018)}]{Suarez:2017mav}
\bibinfo{author}{\bibfnamefont{A.}~\bibnamefont{Su{\'a}rez}} \bibnamefont{and}
  \bibinfo{author}{\bibfnamefont{P.-H.} \bibnamefont{Chavanis}},
  \bibinfo{journal}{Phys. Rev.} \textbf{\bibinfo{volume}{D98}},
  \bibinfo{pages}{083529} (\bibinfo{year}{2018}), \eprint{1710.10486}.

\bibitem[{\citenamefont{Brax et~al.}(2019{\natexlab{a}})\citenamefont{Brax,
  Cembranos, and Valageas}}]{Brax:2019fzb}
\bibinfo{author}{\bibfnamefont{P.}~\bibnamefont{Brax}},
  \bibinfo{author}{\bibfnamefont{J.~A.~R.} \bibnamefont{Cembranos}},
  \bibnamefont{and} \bibinfo{author}{\bibfnamefont{P.}~\bibnamefont{Valageas}},
  \bibinfo{journal}{Phys. Rev.} \textbf{\bibinfo{volume}{D100}},
  \bibinfo{pages}{023526} (\bibinfo{year}{2019}{\natexlab{a}}),
  \eprint{1906.00730}.

\bibitem[{\citenamefont{Brax et~al.}(2019{\natexlab{b}})\citenamefont{Brax,
  Valageas, and Cembranos}}]{Brax:2019npi}
\bibinfo{author}{\bibfnamefont{P.}~\bibnamefont{Brax}},
  \bibinfo{author}{\bibfnamefont{P.}~\bibnamefont{Valageas}}, \bibnamefont{and}
  \bibinfo{author}{\bibfnamefont{J.~A.~R.} \bibnamefont{Cembranos}}
  (\bibinfo{year}{2019}{\natexlab{b}}), \eprint{1909.02614}.

\bibitem[{\citenamefont{Dave and Goswami}(2023)}]{Dave:2023wjq}
\bibinfo{author}{\bibfnamefont{B.}~\bibnamefont{Dave}} \bibnamefont{and}
  \bibinfo{author}{\bibfnamefont{G.}~\bibnamefont{Goswami}},
  \bibinfo{journal}{JCAP} \textbf{\bibinfo{volume}{07}}, \bibinfo{pages}{015}
  (\bibinfo{year}{2023}), \eprint{2304.04463}.

\bibitem[{\citenamefont{Thomas}(1927)}]{thomas_1927}
\bibinfo{author}{\bibfnamefont{L.~H.} \bibnamefont{Thomas}},
  \bibinfo{journal}{Mathematical Proceedings of the Cambridge Philosophical
  Society} \textbf{\bibinfo{volume}{23}}, \bibinfo{pages}{542}
  (\bibinfo{year}{1927}).

\bibitem[{\citenamefont{Fermi}(1927)}]{Fermi_1927}
\bibinfo{author}{\bibfnamefont{E.}~\bibnamefont{Fermi}},
  \bibinfo{journal}{Rend. Accad. Naz. Lincei.} \textbf{\bibinfo{volume}{6}},
  \bibinfo{pages}{602} (\bibinfo{year}{1927}).

\bibitem[{\citenamefont{Eda et~al.}(2013)\citenamefont{Eda, Itoh, Kuroyanagi,
  and Silk}}]{Eda:2013gg}
\bibinfo{author}{\bibfnamefont{K.}~\bibnamefont{Eda}},
  \bibinfo{author}{\bibfnamefont{Y.}~\bibnamefont{Itoh}},
  \bibinfo{author}{\bibfnamefont{S.}~\bibnamefont{Kuroyanagi}},
  \bibnamefont{and} \bibinfo{author}{\bibfnamefont{J.}~\bibnamefont{Silk}},
  \bibinfo{journal}{Phys. Rev. Lett.} \textbf{\bibinfo{volume}{110}},
  \bibinfo{pages}{221101} (\bibinfo{year}{2013}), \eprint{1301.5971}.

\bibitem[{\citenamefont{Kavanagh et~al.}(2020)\citenamefont{Kavanagh, Nichols,
  Bertone, and Gaggero}}]{Kavanagh:2020cfn}
\bibinfo{author}{\bibfnamefont{B.~J.} \bibnamefont{Kavanagh}},
  \bibinfo{author}{\bibfnamefont{D.~A.} \bibnamefont{Nichols}},
  \bibinfo{author}{\bibfnamefont{G.}~\bibnamefont{Bertone}}, \bibnamefont{and}
  \bibinfo{author}{\bibfnamefont{D.}~\bibnamefont{Gaggero}},
  \bibinfo{journal}{Phys. Rev. D} \textbf{\bibinfo{volume}{102}},
  \bibinfo{pages}{083006} (\bibinfo{year}{2020}), \eprint{2002.12811}.

\bibitem[{\citenamefont{Boudon et~al.}(2023)\citenamefont{Boudon, Brax,
  Valageas, and Wong}}]{Boudon:2023aa}
\bibinfo{author}{\bibfnamefont{A.}~\bibnamefont{Boudon}},
  \bibinfo{author}{\bibfnamefont{P.}~\bibnamefont{Brax}},
  \bibinfo{author}{\bibfnamefont{P.}~\bibnamefont{Valageas}}, \bibnamefont{and}
  \bibinfo{author}{\bibfnamefont{L.~K.} \bibnamefont{Wong}}
  (\bibinfo{year}{2023}), \eprint{2305.18540},
  \urlprefix\url{https://arxiv.org/pdf/2305.18540.pdf}.

\bibitem[{\citenamefont{Bar et~al.}(2019{\natexlab{b}})\citenamefont{Bar, Blum,
  Lacroix, and Panci}}]{Bar:2019pnz}
\bibinfo{author}{\bibfnamefont{N.}~\bibnamefont{Bar}},
  \bibinfo{author}{\bibfnamefont{K.}~\bibnamefont{Blum}},
  \bibinfo{author}{\bibfnamefont{T.}~\bibnamefont{Lacroix}}, \bibnamefont{and}
  \bibinfo{author}{\bibfnamefont{P.}~\bibnamefont{Panci}},
  \bibinfo{journal}{JCAP} \textbf{\bibinfo{volume}{07}}, \bibinfo{pages}{045}
  (\bibinfo{year}{2019}{\natexlab{b}}), \eprint{1905.11745}.

\bibitem[{\citenamefont{Chakrabarti et~al.}(2022)\citenamefont{Chakrabarti,
  Dave, Dutta, and Goswami}}]{Chakrabarti:2022owq}
\bibinfo{author}{\bibfnamefont{S.}~\bibnamefont{Chakrabarti}},
  \bibinfo{author}{\bibfnamefont{B.}~\bibnamefont{Dave}},
  \bibinfo{author}{\bibfnamefont{K.}~\bibnamefont{Dutta}}, \bibnamefont{and}
  \bibinfo{author}{\bibfnamefont{G.}~\bibnamefont{Goswami}},
  \bibinfo{journal}{JCAP} \textbf{\bibinfo{volume}{09}}, \bibinfo{pages}{074}
  (\bibinfo{year}{2022}), \eprint{2202.11081}.

\bibitem[{\citenamefont{Ravanal et~al.}(2023)\citenamefont{Ravanal, G\'omez,
  and Cruz}}]{Ravanal:2023ytp}
\bibinfo{author}{\bibfnamefont{Y.}~\bibnamefont{Ravanal}},
  \bibinfo{author}{\bibfnamefont{G.}~\bibnamefont{G\'omez}}, \bibnamefont{and}
  \bibinfo{author}{\bibfnamefont{N.}~\bibnamefont{Cruz}}
  (\bibinfo{year}{2023}), \eprint{2306.10204}.

\bibitem[{\citenamefont{Berezhiani et~al.}(2019)\citenamefont{Berezhiani,
  Elder, and Khoury}}]{Berezhiani:2019pzd}
\bibinfo{author}{\bibfnamefont{L.}~\bibnamefont{Berezhiani}},
  \bibinfo{author}{\bibfnamefont{B.}~\bibnamefont{Elder}}, \bibnamefont{and}
  \bibinfo{author}{\bibfnamefont{J.}~\bibnamefont{Khoury}},
  \bibinfo{journal}{JCAP} \textbf{\bibinfo{volume}{10}}, \bibinfo{pages}{074}
  (\bibinfo{year}{2019}), \eprint{1905.09297}.

\bibitem[{\citenamefont{Lancaster et~al.}(2020)\citenamefont{Lancaster,
  Giovanetti, Mocz, Kahn, Lisanti, and Spergel}}]{Lancaster:2019mde}
\bibinfo{author}{\bibfnamefont{L.}~\bibnamefont{Lancaster}},
  \bibinfo{author}{\bibfnamefont{C.}~\bibnamefont{Giovanetti}},
  \bibinfo{author}{\bibfnamefont{P.}~\bibnamefont{Mocz}},
  \bibinfo{author}{\bibfnamefont{Y.}~\bibnamefont{Kahn}},
  \bibinfo{author}{\bibfnamefont{M.}~\bibnamefont{Lisanti}}, \bibnamefont{and}
  \bibinfo{author}{\bibfnamefont{D.~N.} \bibnamefont{Spergel}},
  \bibinfo{journal}{JCAP} \textbf{\bibinfo{volume}{01}}, \bibinfo{pages}{001}
  (\bibinfo{year}{2020}), \eprint{1909.06381}.

\bibitem[{\citenamefont{Hartman et~al.}(2021)\citenamefont{Hartman, Winther,
  and Mota}}]{Hartman:2020fbg}
\bibinfo{author}{\bibfnamefont{S.~T.~H.} \bibnamefont{Hartman}},
  \bibinfo{author}{\bibfnamefont{H.~A.} \bibnamefont{Winther}},
  \bibnamefont{and} \bibinfo{author}{\bibfnamefont{D.~F.} \bibnamefont{Mota}},
  \bibinfo{journal}{Astron. Astrophys.} \textbf{\bibinfo{volume}{647}},
  \bibinfo{pages}{A70} (\bibinfo{year}{2021}), \eprint{2011.00116}.

\bibitem[{\citenamefont{Annulli et~al.}(2020)\citenamefont{Annulli, Cardoso,
  and Vicente}}]{Annulli:2020lyc}
\bibinfo{author}{\bibfnamefont{L.}~\bibnamefont{Annulli}},
  \bibinfo{author}{\bibfnamefont{V.}~\bibnamefont{Cardoso}}, \bibnamefont{and}
  \bibinfo{author}{\bibfnamefont{R.}~\bibnamefont{Vicente}},
  \bibinfo{journal}{Phys. Rev. D} \textbf{\bibinfo{volume}{102}},
  \bibinfo{pages}{063022} (\bibinfo{year}{2020}), \eprint{2009.00012}.

\bibitem[{\citenamefont{Wang and Easther}(2021)}]{Wang:2021udl}
\bibinfo{author}{\bibfnamefont{Y.}~\bibnamefont{Wang}} \bibnamefont{and}
  \bibinfo{author}{\bibfnamefont{R.}~\bibnamefont{Easther}}
  (\bibinfo{year}{2021}), \eprint{2110.03428}.

\bibitem[{\citenamefont{Traykova et~al.}(2021)\citenamefont{Traykova, Clough,
  Helfer, Berti, Ferreira, and Hui}}]{Traykova:2021dua}
\bibinfo{author}{\bibfnamefont{D.}~\bibnamefont{Traykova}},
  \bibinfo{author}{\bibfnamefont{K.}~\bibnamefont{Clough}},
  \bibinfo{author}{\bibfnamefont{T.}~\bibnamefont{Helfer}},
  \bibinfo{author}{\bibfnamefont{E.}~\bibnamefont{Berti}},
  \bibinfo{author}{\bibfnamefont{P.~G.} \bibnamefont{Ferreira}},
  \bibnamefont{and} \bibinfo{author}{\bibfnamefont{L.}~\bibnamefont{Hui}},
  \bibinfo{journal}{Phys. Rev. D} \textbf{\bibinfo{volume}{104}},
  \bibinfo{pages}{103014} (\bibinfo{year}{2021}), \eprint{2106.08280}.

\bibitem[{\citenamefont{Chowdhury et~al.}(2021)\citenamefont{Chowdhury, van~den
  Bosch, Robles, van Dokkum, Schive, Chiueh, and
  Broadhurst}}]{Chowdhury:2021zik}
\bibinfo{author}{\bibfnamefont{D.~D.} \bibnamefont{Chowdhury}},
  \bibinfo{author}{\bibfnamefont{F.~C.} \bibnamefont{van~den Bosch}},
  \bibinfo{author}{\bibfnamefont{V.~H.} \bibnamefont{Robles}},
  \bibinfo{author}{\bibfnamefont{P.}~\bibnamefont{van Dokkum}},
  \bibinfo{author}{\bibfnamefont{H.-Y.} \bibnamefont{Schive}},
  \bibinfo{author}{\bibfnamefont{T.}~\bibnamefont{Chiueh}}, \bibnamefont{and}
  \bibinfo{author}{\bibfnamefont{T.}~\bibnamefont{Broadhurst}},
  \bibinfo{journal}{Astrophys. J.} \textbf{\bibinfo{volume}{916}},
  \bibinfo{pages}{27} (\bibinfo{year}{2021}), \eprint{2105.05268}.

\bibitem[{\citenamefont{Vicente and Cardoso}(2022)}]{Vicente:2022ivh}
\bibinfo{author}{\bibfnamefont{R.}~\bibnamefont{Vicente}} \bibnamefont{and}
  \bibinfo{author}{\bibfnamefont{V.}~\bibnamefont{Cardoso}}
  (\bibinfo{year}{2022}), \eprint{2201.08854}.

\bibitem[{\citenamefont{Traykova et~al.}(2023)\citenamefont{Traykova, Vicente,
  Clough, Helfer, Berti, Ferreira, and Hui}}]{Traykova:2023qyv}
\bibinfo{author}{\bibfnamefont{D.}~\bibnamefont{Traykova}},
  \bibinfo{author}{\bibfnamefont{R.}~\bibnamefont{Vicente}},
  \bibinfo{author}{\bibfnamefont{K.}~\bibnamefont{Clough}},
  \bibinfo{author}{\bibfnamefont{T.}~\bibnamefont{Helfer}},
  \bibinfo{author}{\bibfnamefont{E.}~\bibnamefont{Berti}},
  \bibinfo{author}{\bibfnamefont{P.~G.} \bibnamefont{Ferreira}},
  \bibnamefont{and} \bibinfo{author}{\bibfnamefont{L.}~\bibnamefont{Hui}}
  (\bibinfo{year}{2023}), \eprint{2305.10492}.

\bibitem[{\citenamefont{Chandrasekhar}(1943)}]{Chandrasekhar:1943ys}
\bibinfo{author}{\bibfnamefont{S.}~\bibnamefont{Chandrasekhar}},
  \bibinfo{journal}{Astrophys. J.} \textbf{\bibinfo{volume}{97}},
  \bibinfo{pages}{255} (\bibinfo{year}{1943}).

\bibitem[{\citenamefont{Bar et~al.}(2021)\citenamefont{Bar, Blas, Blum, and
  Kim}}]{Bar:2021jff}
\bibinfo{author}{\bibfnamefont{N.}~\bibnamefont{Bar}},
  \bibinfo{author}{\bibfnamefont{D.}~\bibnamefont{Blas}},
  \bibinfo{author}{\bibfnamefont{K.}~\bibnamefont{Blum}}, \bibnamefont{and}
  \bibinfo{author}{\bibfnamefont{H.}~\bibnamefont{Kim}},
  \bibinfo{journal}{Phys. Rev. D} \textbf{\bibinfo{volume}{104}},
  \bibinfo{pages}{043021} (\bibinfo{year}{2021}), \eprint{2102.11522}.

\bibitem[{\citenamefont{Boudon et~al.}(2022)\citenamefont{Boudon, Brax, and
  Valageas}}]{Boudon:2022dxi}
\bibinfo{author}{\bibfnamefont{A.}~\bibnamefont{Boudon}},
  \bibinfo{author}{\bibfnamefont{P.}~\bibnamefont{Brax}}, \bibnamefont{and}
  \bibinfo{author}{\bibfnamefont{P.}~\bibnamefont{Valageas}}
  (\bibinfo{year}{2022}), \eprint{2204.09401}.

\bibitem[{\citenamefont{Macedo et~al.}(2013)\citenamefont{Macedo, Pani,
  Cardoso, and Crispino}}]{Macedo:2013qea}
\bibinfo{author}{\bibfnamefont{C.~F.~B.} \bibnamefont{Macedo}},
  \bibinfo{author}{\bibfnamefont{P.}~\bibnamefont{Pani}},
  \bibinfo{author}{\bibfnamefont{V.}~\bibnamefont{Cardoso}}, \bibnamefont{and}
  \bibinfo{author}{\bibfnamefont{L.~C.~B.} \bibnamefont{Crispino}},
  \bibinfo{journal}{Astrophys. J.} \textbf{\bibinfo{volume}{774}},
  \bibinfo{pages}{48} (\bibinfo{year}{2013}), \eprint{1302.2646}.

\bibitem[{\citenamefont{Barausse et~al.}(2014)\citenamefont{Barausse, Cardoso,
  and Pani}}]{Barausse:2014tra}
\bibinfo{author}{\bibfnamefont{E.}~\bibnamefont{Barausse}},
  \bibinfo{author}{\bibfnamefont{V.}~\bibnamefont{Cardoso}}, \bibnamefont{and}
  \bibinfo{author}{\bibfnamefont{P.}~\bibnamefont{Pani}},
  \bibinfo{journal}{Phys. Rev. D} \textbf{\bibinfo{volume}{89}},
  \bibinfo{pages}{104059} (\bibinfo{year}{2014}), \eprint{1404.7149}.

\bibitem[{\citenamefont{Cardoso and Maselli}(2020)}]{Cardoso:2019rou}
\bibinfo{author}{\bibfnamefont{V.}~\bibnamefont{Cardoso}} \bibnamefont{and}
  \bibinfo{author}{\bibfnamefont{A.}~\bibnamefont{Maselli}},
  \bibinfo{journal}{Astron. Astrophys.} \textbf{\bibinfo{volume}{644}},
  \bibinfo{pages}{A147} (\bibinfo{year}{2020}), \eprint{1909.05870}.

\bibitem[{\citenamefont{Li et~al.}(2021)\citenamefont{Li, Tang, and
  Wu}}]{Li:2021pxf}
\bibinfo{author}{\bibfnamefont{G.-L.} \bibnamefont{Li}},
  \bibinfo{author}{\bibfnamefont{Y.}~\bibnamefont{Tang}}, \bibnamefont{and}
  \bibinfo{author}{\bibfnamefont{Y.-L.} \bibnamefont{Wu}}
  (\bibinfo{year}{2021}), \eprint{2112.14041}.

\bibitem[{\citenamefont{Gradshteyn and Ryzhik}(1965)}]{Gradshteyn1965}
\bibinfo{author}{\bibfnamefont{I.~S.} \bibnamefont{Gradshteyn}}
  \bibnamefont{and} \bibinfo{author}{\bibfnamefont{I.~M.}
  \bibnamefont{Ryzhik}}, \emph{\bibinfo{title}{Table of integrals, series, and
  products}} (\bibinfo{publisher}{New York Academic Press},
  \bibinfo{year}{1965}), \bibinfo{edition}{4th} ed.,
  \urlprefix\url{http://openlibrary.org/books/OL5955048M}.

\bibitem[{\citenamefont{Byrd and Friedman}(1971)}]{Byrd-1971}
\bibinfo{author}{\bibfnamefont{P.}~\bibnamefont{Byrd}} \bibnamefont{and}
  \bibinfo{author}{\bibfnamefont{M.}~\bibnamefont{Friedman}},
  \emph{\bibinfo{title}{Handbook of Elliptic Integrals for Engineers and
  Scientists}} (\bibinfo{publisher}{Springer, Berlin, Heidelberg},
  \bibinfo{year}{1971}).

\bibitem[{\citenamefont{Kovacic and Brennan}(2011)}]{Kovacic-2011}
\bibinfo{author}{\bibfnamefont{I.}~\bibnamefont{Kovacic}} \bibnamefont{and}
  \bibinfo{author}{\bibfnamefont{M.}~\bibnamefont{Brennan}},
  \emph{\bibinfo{title}{The Duffing Equation: Nonlinear Oscillators and their
  Behaviour}} (\bibinfo{publisher}{Wiley}, \bibinfo{year}{2011}).

\bibitem[{\citenamefont{Madelung}(1927)}]{Madelung_1927}
\bibinfo{author}{\bibfnamefont{E.}~\bibnamefont{Madelung}},
  \bibinfo{journal}{Zeitschrift fur Physik} \textbf{\bibinfo{volume}{40}},
  \bibinfo{pages}{322} (\bibinfo{year}{1927}), ISSN \bibinfo{issn}{1434-601X},
  \urlprefix\url{http://dx.doi.org/10.1007/BF01400372}.

\bibitem[{\citenamefont{Bondi}(1952)}]{Bondi:1952ni}
\bibinfo{author}{\bibfnamefont{H.}~\bibnamefont{Bondi}}, \bibinfo{journal}{Mon.
  Not. Roy. Astron. Soc.} \textbf{\bibinfo{volume}{112}}, \bibinfo{pages}{195}
  (\bibinfo{year}{1952}).

\bibitem[{\citenamefont{Mott et~al.}(1965)\citenamefont{Mott, Massey, and
  of~Monographs~on Physics}}]{mott1965}
\bibinfo{author}{\bibfnamefont{N.~F.} \bibnamefont{Mott}},
  \bibinfo{author}{\bibfnamefont{H.~S.~W.} \bibnamefont{Massey}},
  \bibnamefont{and} \bibinfo{author}{\bibfnamefont{T.~I.~S.}
  \bibnamefont{of~Monographs~on Physics}}, \emph{\bibinfo{title}{The theory of
  atomic collisions}}, vol.~\bibinfo{volume}{35} (\bibinfo{publisher}{Clarendon
  Press Oxford}, \bibinfo{year}{1965}).

\bibitem[{\citenamefont{Tam and W.}(1966)}]{Tam1966}
\bibinfo{author}{\bibnamefont{Tam}} \bibnamefont{and}
  \bibinfo{author}{\bibfnamefont{C.~K.} \bibnamefont{W.}},
  \bibinfo{journal}{Physics of Fluids} \textbf{\bibinfo{volume}{9}},
  \bibinfo{pages}{493} (\bibinfo{year}{1966}), ISSN \bibinfo{issn}{0899-8213},
  \urlprefix\url{https://ui.adsabs.harvard.edu/abs/1966PhFl....9..493T/abstract}.

\bibitem[{\citenamefont{Dokuchaev}(1964)}]{Dokuchaev1964}
\bibinfo{author}{\bibfnamefont{V.~P.} \bibnamefont{Dokuchaev}},
  \bibinfo{journal}{SvA} \textbf{\bibinfo{volume}{8}}, \bibinfo{pages}{23}
  (\bibinfo{year}{1964}), ISSN \bibinfo{issn}{0038-5301},
  \urlprefix\url{https://ui.adsabs.harvard.edu/abs/1964SvA.....8...23D/abstract}.

\bibitem[{\citenamefont{Ruderman and Spiegel}(1971)}]{Ruderman_1971}
\bibinfo{author}{\bibfnamefont{M.~A.} \bibnamefont{Ruderman}} \bibnamefont{and}
  \bibinfo{author}{\bibfnamefont{E.~A.} \bibnamefont{Spiegel}},
  \bibinfo{journal}{The Astrophysical Journal} \textbf{\bibinfo{volume}{165}},
  \bibinfo{pages}{1} (\bibinfo{year}{1971}),
  \urlprefix\url{https://doi.org/10.1086%2F150870}.

\bibitem[{\citenamefont{Hirsch}(2007)}]{Hirsch:2007}
\bibinfo{author}{\bibfnamefont{C.}~\bibnamefont{Hirsch}},
  \emph{\bibinfo{title}{Numerical Computation of Internal and External Flows}}
  (\bibinfo{publisher}{Elsevier}, \bibinfo{year}{2007}),
  \urlprefix\url{https://doi.org/10.1016%2Fb978-0-7506-6594-0.x5037-1}.

\bibitem[{\citenamefont{Orszag and Bender}(1978)}]{orszag1978advanced}
\bibinfo{author}{\bibfnamefont{S.}~\bibnamefont{Orszag}} \bibnamefont{and}
  \bibinfo{author}{\bibfnamefont{C.~M.} \bibnamefont{Bender}},
  \emph{\bibinfo{title}{Advanced mathematical methods for scientists and
  engineers}} (\bibinfo{publisher}{McGraw-Hill New York},
  \bibinfo{year}{1978}).

\bibitem[{\citenamefont{{Hoyle} and {Lyttleton}}(1939)}]{1939PCPS...35..405H}
\bibinfo{author}{\bibfnamefont{F.}~\bibnamefont{{Hoyle}}} \bibnamefont{and}
  \bibinfo{author}{\bibfnamefont{R.~A.} \bibnamefont{{Lyttleton}}},
  \bibinfo{journal}{Proceedings of the Cambridge Philosophical Society}
  \textbf{\bibinfo{volume}{35}}, \bibinfo{pages}{405} (\bibinfo{year}{1939}).

\bibitem[{\citenamefont{Edgar}(2004)}]{Edgar:2004mk}
\bibinfo{author}{\bibfnamefont{R.~G.} \bibnamefont{Edgar}},
  \bibinfo{journal}{New Astron. Rev.} \textbf{\bibinfo{volume}{48}},
  \bibinfo{pages}{843} (\bibinfo{year}{2004}), \eprint{astro-ph/0406166}.

\bibitem[{\citenamefont{Mulder et~al.}(1983)\citenamefont{Mulder, Mulder, and
  A.}}]{Mulder1983}
\bibinfo{author}{\bibfnamefont{W.~A.} \bibnamefont{Mulder}},
  \bibinfo{author}{\bibnamefont{Mulder}}, \bibnamefont{and}
  \bibinfo{author}{\bibfnamefont{W.}~\bibnamefont{A.}},
  \bibinfo{journal}{Astron. Astrophys.} \textbf{\bibinfo{volume}{117}},
  \bibinfo{pages}{9} (\bibinfo{year}{1983}), ISSN \bibinfo{issn}{0004-6361},
  \urlprefix\url{https://ui.adsabs.harvard.edu/abs/1983A&A...117....9M/abstract}.

\bibitem[{\citenamefont{{Keppens, R.} et~al.}(2023)\citenamefont{{Keppens, R.},
  {Popescu Braileanu, B.}, {Zhou, Y.}, {Ruan, W.}, {Xia, C.}, {Guo, Y.},
  {Claes, N.}, and {Bacchini, F.}}}]{refId00}
\bibinfo{author}{\bibnamefont{{Keppens, R.}}},
  \bibinfo{author}{\bibnamefont{{Popescu Braileanu, B.}}},
  \bibinfo{author}{\bibnamefont{{Zhou, Y.}}},
  \bibinfo{author}{\bibnamefont{{Ruan, W.}}},
  \bibinfo{author}{\bibnamefont{{Xia, C.}}},
  \bibinfo{author}{\bibnamefont{{Guo, Y.}}},
  \bibinfo{author}{\bibnamefont{{Claes, N.}}}, \bibnamefont{and}
  \bibinfo{author}{\bibnamefont{{Bacchini, F.}}}, \bibinfo{journal}{A\&A}
  \textbf{\bibinfo{volume}{673}}, \bibinfo{pages}{A66} (\bibinfo{year}{2023}),
  \urlprefix\url{https://doi.org/10.1051/0004-6361/202245359}.

\bibitem[{\citenamefont{Keppens et~al.}(2021)\citenamefont{Keppens, Teunissen,
  Xia, and Porth}}]{KEPPENS2021316}
\bibinfo{author}{\bibfnamefont{R.}~\bibnamefont{Keppens}},
  \bibinfo{author}{\bibfnamefont{J.}~\bibnamefont{Teunissen}},
  \bibinfo{author}{\bibfnamefont{C.}~\bibnamefont{Xia}}, \bibnamefont{and}
  \bibinfo{author}{\bibfnamefont{O.}~\bibnamefont{Porth}},
  \bibinfo{journal}{Computers \& Mathematics with Applications}
  \textbf{\bibinfo{volume}{81}}, \bibinfo{pages}{316} (\bibinfo{year}{2021}),
  ISSN \bibinfo{issn}{0898-1221}, \bibinfo{note}{development and Application of
  Open-source Software for Problems with Numerical PDEs},
  \urlprefix\url{https://www.sciencedirect.com/science/article/pii/S0898122120301279}.

\bibitem[{\citenamefont{Mellah and Casse}(2015)}]{El_Mellah_2015}
\bibinfo{author}{\bibfnamefont{I.~E.} \bibnamefont{Mellah}} \bibnamefont{and}
  \bibinfo{author}{\bibfnamefont{F.}~\bibnamefont{Casse}},
  \bibinfo{journal}{Monthly Notices of the Royal Astronomical Society}
  \textbf{\bibinfo{volume}{454}}, \bibinfo{pages}{2657} (\bibinfo{year}{2015}),
  \urlprefix\url{https://doi.org/10.1093%2Fmnras%2Fstv2184}.

\bibitem[{\citenamefont{Xia et~al.}(2018)\citenamefont{Xia, Teunissen, Mellah,
  Chan{\'{e}}, and Keppens}}]{Xia_2018}
\bibinfo{author}{\bibfnamefont{C.}~\bibnamefont{Xia}},
  \bibinfo{author}{\bibfnamefont{J.}~\bibnamefont{Teunissen}},
  \bibinfo{author}{\bibfnamefont{I.~E.} \bibnamefont{Mellah}},
  \bibinfo{author}{\bibfnamefont{E.}~\bibnamefont{Chan{\'{e}}}},
  \bibnamefont{and} \bibinfo{author}{\bibfnamefont{R.}~\bibnamefont{Keppens}},
  \bibinfo{journal}{The Astrophysical Journal Supplement Series}
  \textbf{\bibinfo{volume}{234}}, \bibinfo{pages}{30} (\bibinfo{year}{2018}),
  \urlprefix\url{https://doi.org/10.3847%2F1538-4365%2Faaa6c8}.

\bibitem[{\citenamefont{Toro et~al.}(1994)\citenamefont{Toro, Spruce, and
  Speares}}]{HLLC}
\bibinfo{author}{\bibfnamefont{E.~F.} \bibnamefont{Toro}},
  \bibinfo{author}{\bibfnamefont{M.}~\bibnamefont{Spruce}}, \bibnamefont{and}
  \bibinfo{author}{\bibfnamefont{W.}~\bibnamefont{Speares}},
  \bibinfo{journal}{Shock Waves} \textbf{\bibinfo{volume}{4}},
  \bibinfo{pages}{25} (\bibinfo{year}{1994}),
  \urlprefix\url{https://doi.org/10.1007/BF01414629}.

\bibitem[{\citenamefont{Childs et~al.}(2012)\citenamefont{Childs, Brugger,
  Whitlock, Meredith, Ahern, Pugmire, Biagas, Miller, Harrison, Weber
  et~al.}}]{HPV:VisIt}
\bibinfo{author}{\bibfnamefont{H.}~\bibnamefont{Childs}},
  \bibinfo{author}{\bibfnamefont{E.}~\bibnamefont{Brugger}},
  \bibinfo{author}{\bibfnamefont{B.}~\bibnamefont{Whitlock}},
  \bibinfo{author}{\bibfnamefont{J.}~\bibnamefont{Meredith}},
  \bibinfo{author}{\bibfnamefont{S.}~\bibnamefont{Ahern}},
  \bibinfo{author}{\bibfnamefont{D.}~\bibnamefont{Pugmire}},
  \bibinfo{author}{\bibfnamefont{K.}~\bibnamefont{Biagas}},
  \bibinfo{author}{\bibfnamefont{M.}~\bibnamefont{Miller}},
  \bibinfo{author}{\bibfnamefont{C.}~\bibnamefont{Harrison}},
  \bibinfo{author}{\bibfnamefont{G.~H.} \bibnamefont{Weber}},
  \bibnamefont{et~al.}, in \emph{\bibinfo{booktitle}{High Performance
  Visualization--Enabling Extreme-Scale Scientific Insight}}
  (\bibinfo{year}{2012}), pp. \bibinfo{pages}{357--372}.

\bibitem[{\citenamefont{Foglizzo and Ruffert}(1997)}]{Foglizzo:aa}
\bibinfo{author}{\bibfnamefont{T.}~\bibnamefont{Foglizzo}} \bibnamefont{and}
  \bibinfo{author}{\bibfnamefont{M.}~\bibnamefont{Ruffert}},
  \bibinfo{journal}{Astron. Astrophys.} \textbf{\bibinfo{volume}{320}},
  \bibinfo{pages}{342} (\bibinfo{year}{1997}), \eprint{astro-ph/9604160},
  \urlprefix\url{https://arxiv.org/pdf/astro-ph/9604160.pdf}.

\bibitem[{\citenamefont{Bondi and Hoyle}(1944)}]{Bondi-Hoyle-1944}
\bibinfo{author}{\bibfnamefont{H.}~\bibnamefont{Bondi}} \bibnamefont{and}
  \bibinfo{author}{\bibfnamefont{F.}~\bibnamefont{Hoyle}},
  \bibinfo{journal}{Monthly Notices of the Royal Astronomical Society}
  \textbf{\bibinfo{volume}{104}}, \bibinfo{pages}{273} (\bibinfo{year}{1944}),
  ISSN \bibinfo{issn}{0035-8711},
  \eprint{https://academic.oup.com/mnras/article-pdf/104/5/273/8072203/mnras104-0273.pdf},
  \urlprefix\url{https://doi.org/10.1093/mnras/104.5.273}.

\bibitem[{\citenamefont{Antonini and Merritt}(2011)}]{Antonini2011}
\bibinfo{author}{\bibfnamefont{F.}~\bibnamefont{Antonini}} \bibnamefont{and}
  \bibinfo{author}{\bibfnamefont{D.}~\bibnamefont{Merritt}},
  \bibinfo{journal}{The Astrophysical Journal} \textbf{\bibinfo{volume}{745}},
  \bibinfo{pages}{83} (\bibinfo{year}{2011}).

\bibitem[{\citenamefont{{Binney} and {Tremaine}}(1987)}]{Binney1987}
\bibinfo{author}{\bibfnamefont{J.}~\bibnamefont{{Binney}}} \bibnamefont{and}
  \bibinfo{author}{\bibfnamefont{S.}~\bibnamefont{{Tremaine}}},
  \emph{\bibinfo{title}{{Galactic dynamics}}} (\bibinfo{year}{1987}).

\bibitem[{\citenamefont{Ostriker}(1999)}]{Ostriker:1998fa}
\bibinfo{author}{\bibfnamefont{E.~C.} \bibnamefont{Ostriker}},
  \bibinfo{journal}{Astrophys. J.} \textbf{\bibinfo{volume}{513}},
  \bibinfo{pages}{252} (\bibinfo{year}{1999}), \eprint{astro-ph/9810324}.

\bibitem[{\citenamefont{S{\'a}nchez-Salcedo and
  Brandenburg}(1999)}]{Sanchez-Salcedo_1999}
\bibinfo{author}{\bibfnamefont{F.~J.} \bibnamefont{S{\'a}nchez-Salcedo}}
  \bibnamefont{and}
  \bibinfo{author}{\bibfnamefont{A.}~\bibnamefont{Brandenburg}},
  \bibinfo{journal}{The Astrophysical Journal} \textbf{\bibinfo{volume}{522}},
  \bibinfo{pages}{L35} (\bibinfo{year}{1999}),
  \urlprefix\url{https://dx.doi.org/10.1086/312215}.

\bibitem[{\citenamefont{Bernal and Sanchez-Salcedo}(2013)}]{Bernal:2013txa}
\bibinfo{author}{\bibfnamefont{C.~G.} \bibnamefont{Bernal}} \bibnamefont{and}
  \bibinfo{author}{\bibfnamefont{F.~J.} \bibnamefont{Sanchez-Salcedo}},
  \bibinfo{journal}{Astrophys. J.} \textbf{\bibinfo{volume}{775}},
  \bibinfo{pages}{72} (\bibinfo{year}{2013}), \eprint{1308.4370}.

\bibitem[{\citenamefont{{Thun, Daniel} et~al.}(2016)\citenamefont{{Thun,
  Daniel}, {Kuiper, Rolf}, {Schmidt, Franziska}, and {Kley, Wilhelm}}}]{refId0}
\bibinfo{author}{\bibnamefont{{Thun, Daniel}}},
  \bibinfo{author}{\bibnamefont{{Kuiper, Rolf}}},
  \bibinfo{author}{\bibnamefont{{Schmidt, Franziska}}}, \bibnamefont{and}
  \bibinfo{author}{\bibnamefont{{Kley, Wilhelm}}}, \bibinfo{journal}{A\&A}
  \textbf{\bibinfo{volume}{589}}, \bibinfo{pages}{A10} (\bibinfo{year}{2016}),
  \urlprefix\url{https://doi.org/10.1051/0004-6361/201527629}.

\end{thebibliography}

\end{document}